\documentclass[12pt]{article}
\pdfoutput=1
\usepackage{subfigure}
\usepackage{amssymb,amsmath}
\usepackage{graphicx}
\usepackage{color}
\usepackage[colorlinks=true
,urlcolor=blue
,citecolor=blue
,linkcolor=blue
,pagecolor=blue
,linktocpage=true
,pdfproducer=medialab
]{hyperref}
\usepackage[a4paper,width=15.2cm]{geometry}
\makeatletter \renewcommand{\@dotsep}{10000} \makeatother
\def\be{\begin{equation}}
\def\ee{\end{equation}}
\def\bea{\begin{eqnarray}}
\def\eea{\end{eqnarray}}
\def\bi{\begin{itemize}}
\def\ei{\end{itemize}}

\def\tst{\tilde t}
\def\ttau{\tilde \tau}

\newcommand\prd[3]{{\it Phys.\ Rev.\ }{\bf D #1} (#2) #3}
\newcommand\prl[3]{{\it Phys.\ Rev.\ Lett.\ }{\bf #1} (#2) #3}

\newcommand\jhep[3]{{\it J. High Energy Phys.\ }{\bf #1} (#2) #3}

\def\tst{\tilde t}
\def\ttau{\tilde \tau}

\newcommand{\beq}{\begin{equation}}
\newcommand{\eeq}{\end{equation}}

\begin{document}
\date{\today}

\begin{center}
{\Large\bf 
Particle Spectroscopy of Supersymmetric SU(5) 
 in Light of 125 GeV Higgs and Muon $g-2$ Data 
} \vspace{1cm}
\end{center}

\begin{center}
{\Large
Nobuchika Okada$^{a,}$\footnote{
Email: okadan@ua.edu},
Shabbar Raza $^{b,}$\footnote{
Email: shabbar@udel.edu. On study leave from:
Department of Physics, FUUAST, Islamabad, Pakistan.}
and
Qaisar Shafi $^{b,}$\footnote{
Email: shafi@bartol.udel.edu. }
}

\vspace{0.75cm}

{\it $^a$
Department of Physics and Astronomy, \\
University of Alabama, Tuscaloosa, Alabama 35487, USA
}\\
{\it  $^b$
Bartol Research Institute, Department of Physics and Astronomy,\\
University of Delaware, Newark, DE 19716, USA 
}

\vspace{1.5cm}
\section*{Abstract}
\end{center}

The discovery of the Higgs boson at the Large Hadron Collider (LHC) 
 has a great impact on the minimal supersymmetric extension of 
 the Standard Model (MSSM).  
In the context of the constrained MSSM (CMSSM) 
 and its extension with non-universal masses 
 for the MSSM Higgs doublets (NUHM2), 
 sparticles with masses $ > 1$ TeV are necessary 
 to reproduce the observed Higgs boson mass of 125-126 GeV. 
On the other hand, there appears to be a significant amount 
 of discrepancy between the measured muon $g-2$
 and the Standard Model prediction. 
 A successful explanation of this discrepancy in the MSSM 
 requires new contributions involving relatively light sparticles with masses $ < 1$ TeV. 
In this paper, we attempt  
 to accommodate the two conflicting requirements 
 in an SU(5) inspired extension of the CMSSM. 
 We assign non-universal but flavor blind soft supersymmetry breaking masses 
 to the scalar components in ${\bar 5}$ and $10$ matter supermultiplets.
The two MSSM Higgs doublets in the $5$, $\bar5$ representations of $SU(5)$
 are also assigned unequal soft mass$^2$ at $M_{GUT}$.  
We identify parameter regions which can simultaneously 
 accommodate the observed Higgs boson mass and the muon $g-2$ data, 
and which are compatible  
with other phenomenological constraints 
 such as neutralino dark matter relic abundance and 
rare B-meson decays. 
Some regions of the allowed parameter space will be explored 
 at the upgraded LHC and by dark matter 
 direct detection experiments.

\newpage

\renewcommand{\thefootnote}{\arabic{footnote}}
\setcounter{footnote}{0}



\section{Introduction}
\label{sec:intro}

By resolving the gauge hierarchy problem, supersymmetry (SUSY) is 
 a prime candidate for new physics beyond the Standard Model (SM). 
The MSSM predicts a plethora of new particles in the TeV mass range.
 Under the assumption of $R$-parity conservation, 
the lightest supersymmetric particle (LSP) neutralino is a very plausible dark matter candidate. 
The electroweak symmetry is radiatively broken 
 by the interplay between the soft SUSY breaking masses 
 and the large top Yukawa coupling. 
The quartic Higgs coupling   
 is determined by the electroweak gauge couplings and 
 as a result, the SM-like Higgs boson mass is predicted 
 in terms of the scalar top quark masses and other soft SUSY breaking  parameters. 
Furthermore, in the MSSM with SUSY broken at the TeV scale, 
 the three SM gauge couplings successfully unify 
 at the grand unified theory (GUT) scale, 
 $M_{\rm GUT} \simeq 2 \times 10^{16}$ GeV. 
This fact strongly supports the GUT paradigm 
 that the three SM gauge interactions are unified 
 within a gauge group  such as SU(5) or SO(10).

The recent discovery of the SM-like Higgs boson 
 by the ATLAS~\cite{ATLAS} and CMS~\cite{CMS} collaborations 
 at the Large Hadron Collider (LHC) has several 
important implications for the MSSM, 
 along with other phenomenological constraints. 
Following this discovery, the sparticle mass spectroscopy has been intensively 
 studied, in particular, in light of the observed Higgs boson 
 of mass around $125$ GeV~\cite{SUSY-Higgs}. 
Since the MSSM tree-level prediction for 
 the SM-like Higgs boson is below the $Z$-boson mass, 
a significant enhancement of the Higgs boson mass is achieved through quantum corrections involving the stops.
In the context of the CMSSM 
 and its extension contains non-universal masses 
 of the MSSM Higgs doublets (NUHM), 
 it has been found that the sfermions are 
 as heavy as $1-10$ TeV~\cite{SUSY-Higgs}.

On the other hand, there is some indication of new physics 
 beyond the SM 
stemming from muon $g-2$ ($a_\mu=(g_\mu-2)/2$), which has been precisely 
 measured by the BNL experiment~\cite{BNL}. 
The data shows a discrepancy from the SM prediction 
 ($a_\mu^{SM}$)~\cite{g-2TH}, namely 
\begin{eqnarray} 
 2.1 \times 10^{-10} \leq \Delta a_\mu \leq 50.1 \times 10^{-10} 
 \; (3\sigma), 
\label{g-2}
\end{eqnarray} 
 where $\Delta a_\mu=a_\mu^{\rm exp} - a_\mu^{SM}$. 
This indicates new physics beyond the SM, 
 and the MSSM with relatively light sleptons, 
 neutralinos and charginos is a primary candidate 
 to account for the discrepancy~\cite{g-2SUSY}. 
Unfortunately, because of the heavy sparticles 
 the CMSSM and NUHM fail to resolve this discrepancy.

In this paper, we attempt to reconcile 
 the muon $g-2$ data with other phenomenological constraints 
 on the MSSM, such as realizing a SM-like Higgs boson mass of
 125-126 GeV, Wilkinson Microwave Anisotropy Probe (WMAP9) \cite{WMAP9} and Planck2013 \cite{Planck2013} compatible  
 neutralino dark matter abundance 
 in a generalized CMSSM inspired by SU(5) GUT~\cite{SU(5)CMSSM}. 
The standard CMSSM assumes universality of scalar masses, 
 the gaugino masses and the $A$-parameters. 
Motivated by SU(5) GUT, we do not require 
 universal masses at $M_{\rm GUT}$ 
 for the ${\bar 5}$ and $10$ matter multiplets 
 (flavor universal masses are assumed). 
The GUT scale boundary conditions 
 for (flavor universal) soft SUSY breaking (SSB) sfermion  masses are as follows: 
\begin{eqnarray}
&&  m_{{\tilde D}^c}=m_{\tilde L}= m_{\bar 5}, 
\nonumber\\
&&  m_{\tilde Q}= m_{{\tilde U}^c} = m_{{\tilde E}^c} = m_{10}.  
\end{eqnarray}
As in the  NUHM2, the masses for the MSSM Higgs doublets are 
 taken as free parameters, since they reside 
 in ${5}$ and ${\bar 5}$ 
 multiplets under SU(5). 
The remaining parameters are the same as in the CMSSM. 
Clearly, the NUHM2 case can be realized 
 by setting $m_{\bar 5}= m_{10}$. 
With this set of free parameters, we perform a random parameter 
 scan under a variety of phenomenological constraints 
 and identify parameter regions which satisfy 
 the imposed constraints as well as the SUSY explanation 
 of the muon $g-2$ data. 
For a similar discussion but with non-universal 
 gaugino mass models, see \cite{NUGM}. 
See also \cite{GMSB} for a discussion with the gauge mediation.

\section{Phenomenological constraints and scanning procedure}
\label{sec:scan}
We employ the ISAJET~7.84 package~\cite{ISAJET} 
 to perform random scans over the parameter space 
 in the generalized CMSSM inspired by SU(5) GUT. 
In this package, the weak scale values of gauge and third 
 generation Yukawa couplings are evolved to 
 $M_{\rm GUT}$ via the MSSM renormalization group equations (RGEs)
 in the $\overline{DR}$ regularization scheme.
We do not strictly enforce the unification condition
 $g_3=g_1=g_2$ at $M_{\rm GUT}$, since a few percent deviation
 from unification can be assigned to unknown GUT-scale threshold
 corrections~\cite{Hisano:1992jj}.
With the boundary conditions given at $M_{\rm GUT}$, 
 all the SSB parameters, along with the gauge and Yukawa couplings, 
 are evolved back to the weak scale $M_{\rm Z}$.

In evaluating Yukawa couplings the SUSY threshold 
 corrections~\cite{Pierce:1996zz} are taken into account 
 at the common scale $M_{\rm SUSY}= \sqrt{m_{\tst_L}m_{\tst_R}}$. 
The entire parameter set is iteratively run between 
 $M_{\rm Z}$ and $M_{\rm GUT}$ using the full 2-loop RGEs
 until a stable solution is obtained.
To better account for leading-log corrections, one-loop step-beta
 functions are adopted for gauge and Yukawa couplings, and
 the SSB parameters $m_i$ are extracted from RGEs at appropriate scales
 $m_i=m_i(m_i)$.
The RGE-improved 1-loop effective potential is minimized
 at an optimized scale $M_{\rm SUSY}$, which effectively
 accounts for the leading 2-loop corrections.
Full 1-loop radiative corrections are incorporated
 for all sparticle masses.

The requirement of radiative electroweak symmetry breaking
 (REWSB)~\cite{Ibanez:1982fr} puts an important theoretical
 constraint on the parameter space.
Another important constraint comes from limits on the cosmological
 abundance of stable charged particles~\cite{Nakamura:2010zzi}.
This excludes regions in the parameter space where charged
 SUSY particles, such as $\ttau_1$ or $\tst_1$,
 become the LSP.
We accept only those solutions for which one of the neutralinos
 is the LSP and saturates the dark matter relic abundance bound
 observed by WMAP9 and Planck2013.

We have performed Markov-chain Monte Carlo (MCMC) scans 
 for the following parameter range:
\begin{align}
0\leq  m_{\bar 5}  \leq 2\, \rm{TeV} \nonumber \\
0\leq  m_{10}  \leq 10\, \rm{TeV} \nonumber \\
0\leq  M_{1/2}  \leq 2\, \rm{TeV} \nonumber \\
0\leq \mu \leq 2 \, \rm{TeV} \nonumber \\
0\leq m_{A} \leq 2 \, \rm{TeV} \nonumber \\
A_{0}=-6, -5, -4, -3, 0, 4 \, \rm{TeV} \nonumber \\ 
\tan\beta=10, 30, 50,
 \label{parameterRange}
\end{align}
with  $\mu > 0$ and  $m_t = 173.3\, {\rm GeV}$  \cite{:2009ec}.
Note that our results are not too sensitive to one 
 or two sigma variation in the value of $m_t$  \cite{bartol2}.
We use $m_b^{\overline{DR}}(M_{\rm Z})=2.83$ GeV 
 which is hard-coded into ISAJET.

In scanning the parameter space, we employ the Metropolis-Hastings
 algorithm as described in \cite{Belanger:2009ti}. 
The data points collected all satisfy the requirement of REWSB, 
 with the neutralino in each case being the LSP. 
After collecting the data, we impose the mass bounds on 
 all the particles \cite{Nakamura:2010zzi} and 
 use the IsaTools package~\cite{bsg, bmm} and Ref.~\cite{mamoudi}
 to implement the following phenomenological constraints: 
\begin{eqnarray} 
m_h  = 124-127~{\rm GeV}~~&\cite{ATLAS, CMS}& 
\\
0.8\times 10^{-9} \leq{\rm BR}(B_s \rightarrow \mu^+ \mu^-) 
  \leq 6.2 \times10^{-9} \;(2\sigma)~~&\cite{BsMuMu}& 
\\ 
2.99 \times 10^{-4} \leq 
  {\rm BR}(b \rightarrow s \gamma) 
  \leq 3.87 \times 10^{-4} \; (2\sigma)~~&\cite{Amhis:2012bh}&  
\\
0.15 \leq \frac{
 {\rm BR}(B_u\rightarrow\tau \nu_{\tau})_{\rm MSSM}}
 {{\rm BR}(B_u\rightarrow \tau \nu_{\tau})_{\rm SM}}
        \leq 2.41 \; (3\sigma)~~&\cite{Asner:2010qj}&  
\\
 0.0913 \leq \Omega_{\rm CDM}h^2 (\rm WMAP9) \leq 0.1363   \; (5\sigma)~~&\cite{WMAP9}&
\\ 
 2.1 \times 10^{-10} \leq \Delta a_{\mu} 
  \leq 50.1 \times 10^{-10} \; (3\sigma)~~&\cite{BNL}&
\end{eqnarray}

\section{Results}
\label{sec:results}
\begin{figure}
\centering
\subfiguretopcaptrue

\subfigure{
\includegraphics[width=7cm]{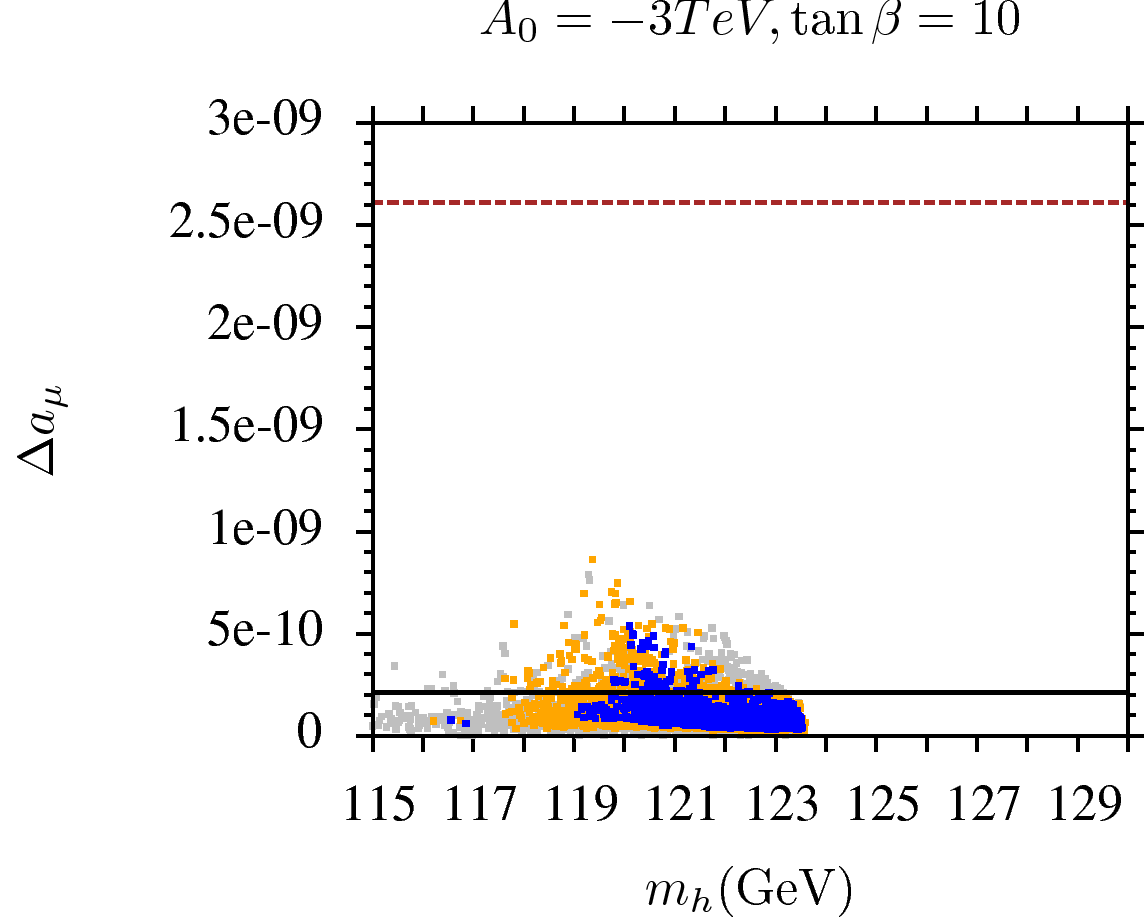}
}
\subfigure{
\includegraphics[width=7cm]{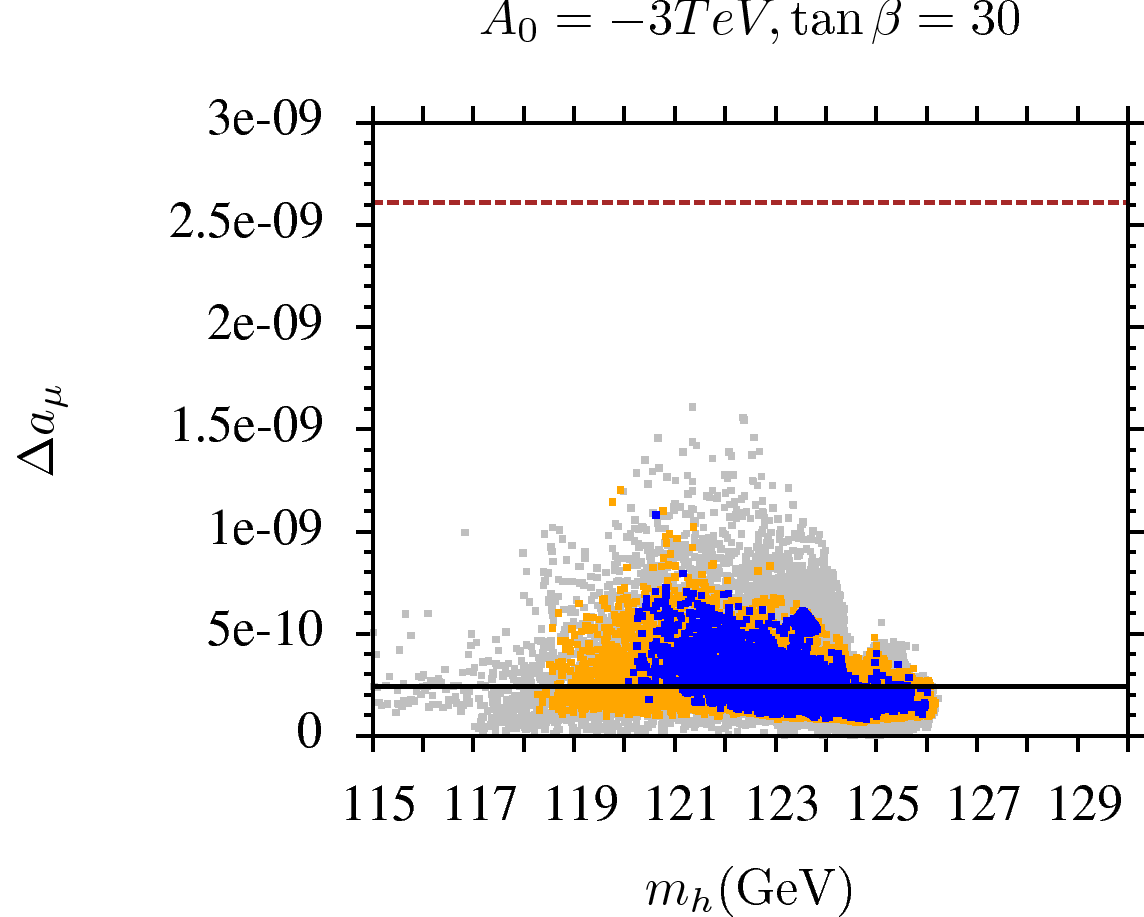}
}
\subfigure{
\includegraphics[width=7cm]{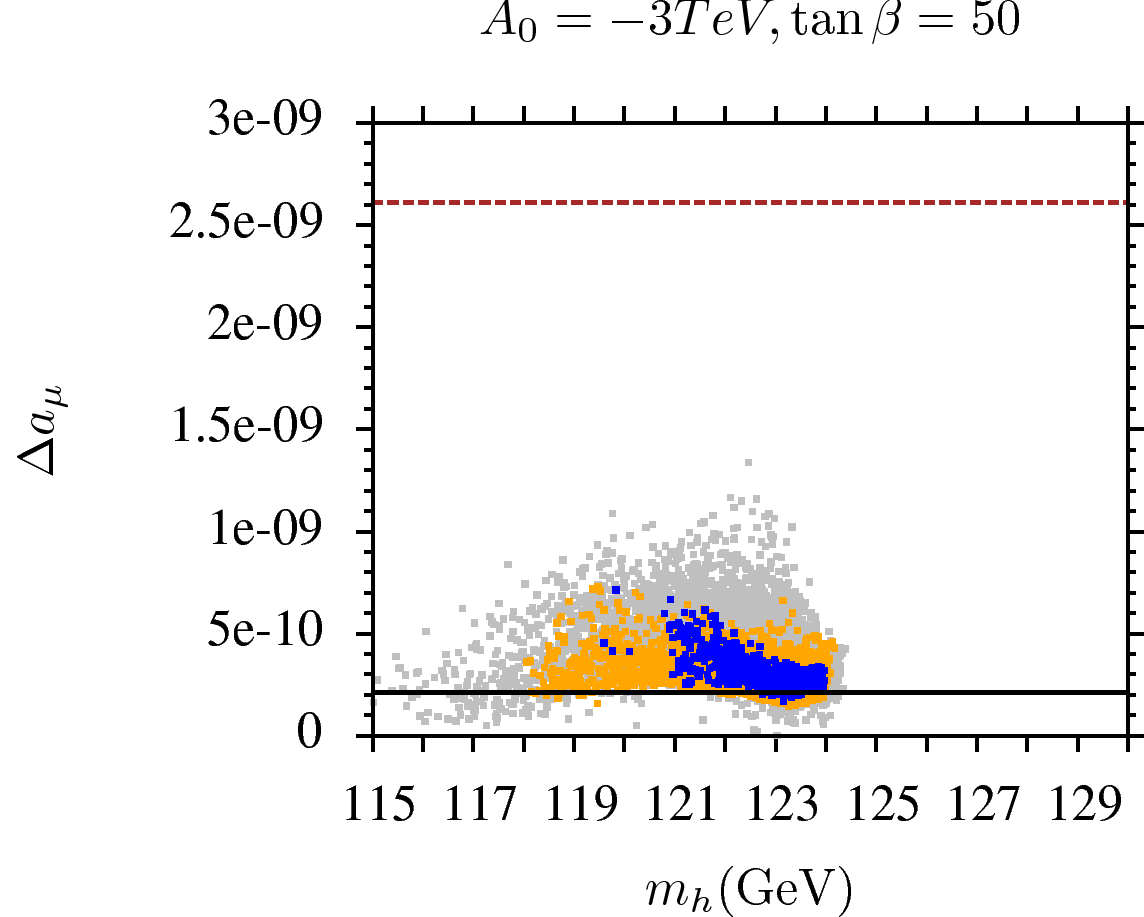}
}
\caption{
Plots in ($\Delta a_\mu$, $m_h$)-plane. 
Gray points satisfy the requirements of REWSB and neutralino LSP. 
Orange points satisfy mass bounds and B-physics bounds. 
Blue points are subset of the orange points and represent 
 solutions satisfying the WMAP9 bounds at 5$\sigma$  
 and $m_{\tilde g}$, $m_{\tilde d_R} \geq 1$ TeV.
}
\label{An3TeVg2mh}
\end{figure}

\begin{figure}
\centering
\subfiguretopcaptrue

\subfigure{
\includegraphics[width=7cm]{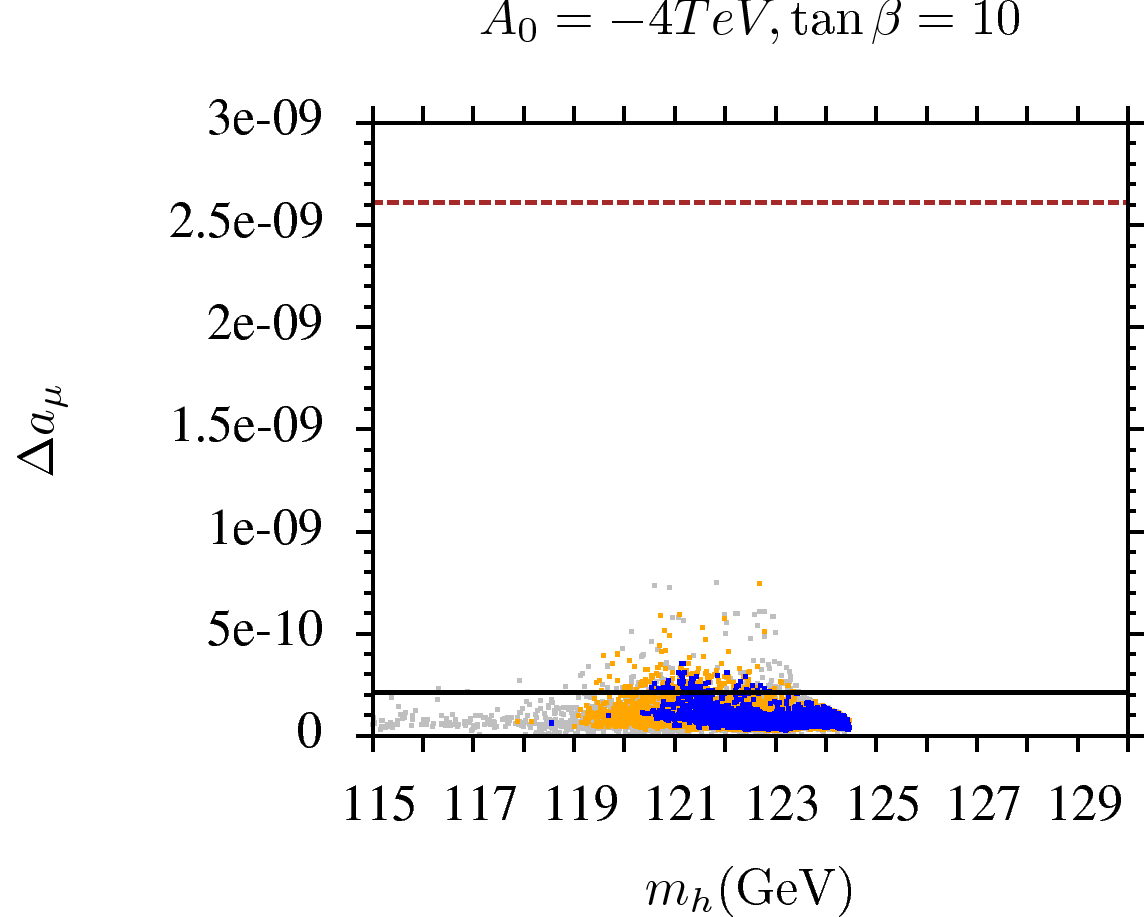}
}
\subfigure{
\includegraphics[width=7cm]{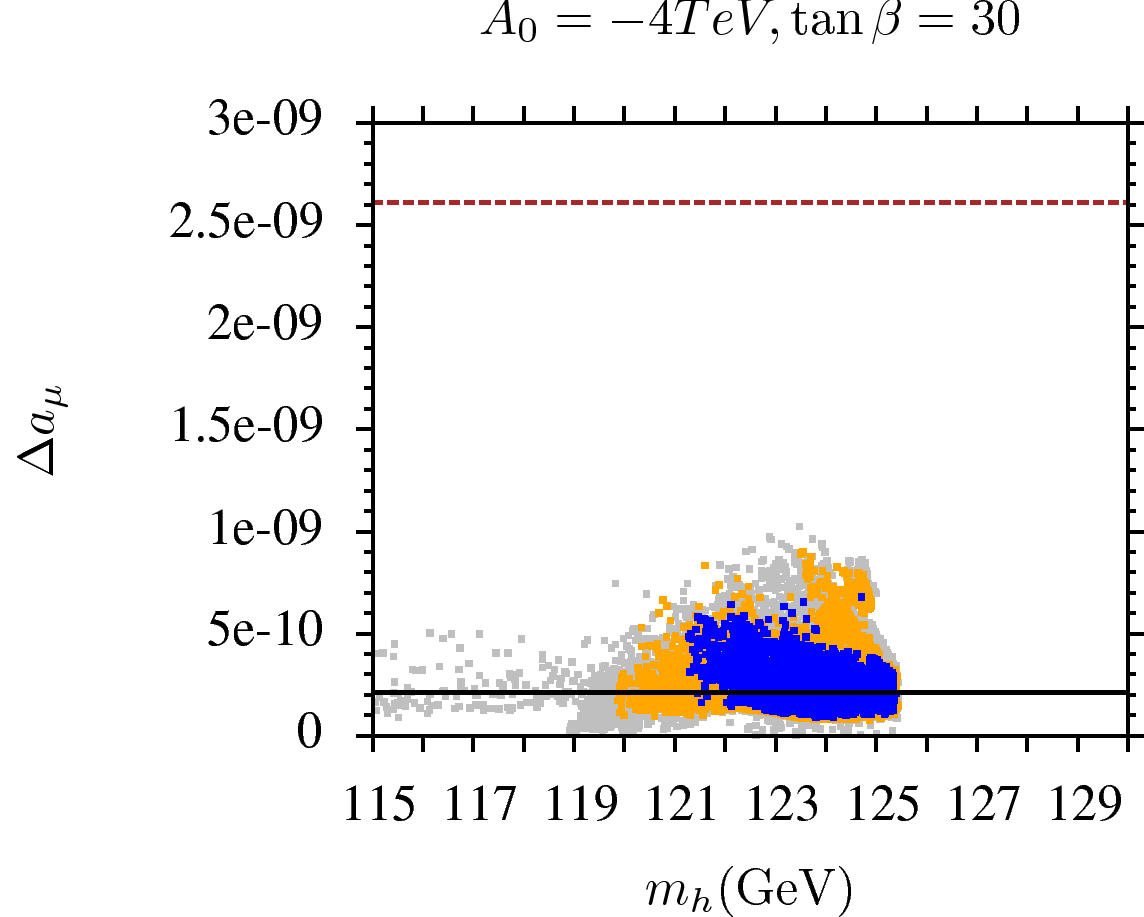}
}
\subfigure{
\includegraphics[width=7cm]{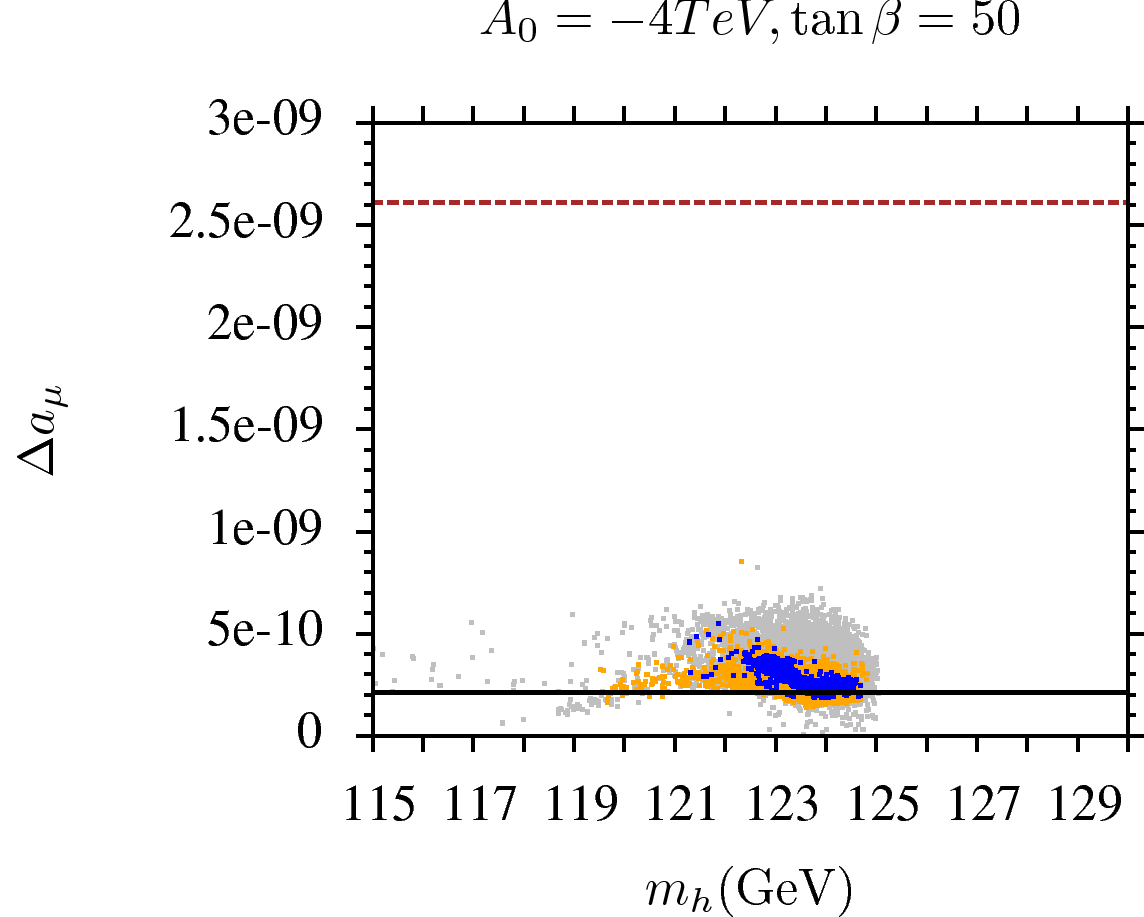}
}
\caption{
Color coding same as in Fig.~\ref{An3TeVg2mh}, 
 with $A_0=-4$ TeV. 
}
\label{An4TeVg2mh}
\end{figure}

\begin{figure}
\centering
\subfiguretopcaptrue

\subfigure{
\includegraphics[width=7cm]{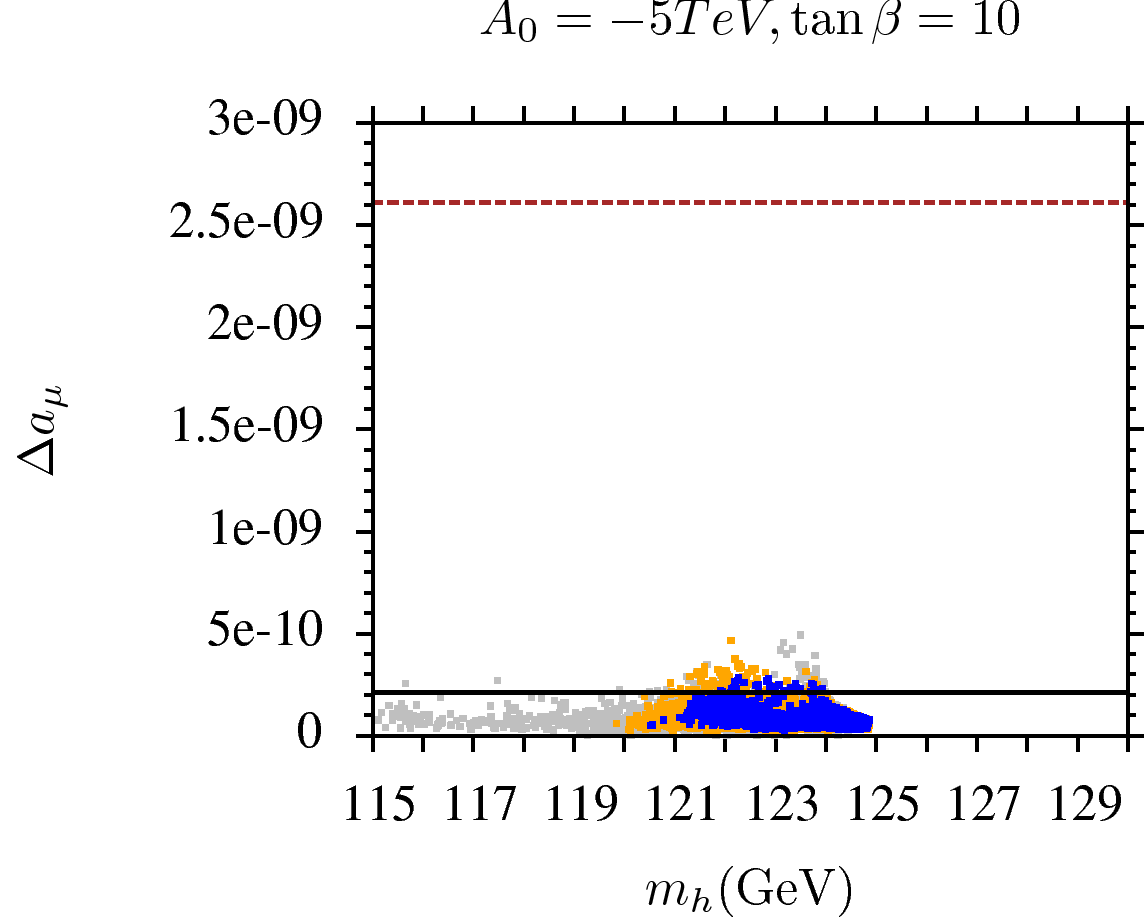}
}
\subfigure{
\includegraphics[width=7cm]{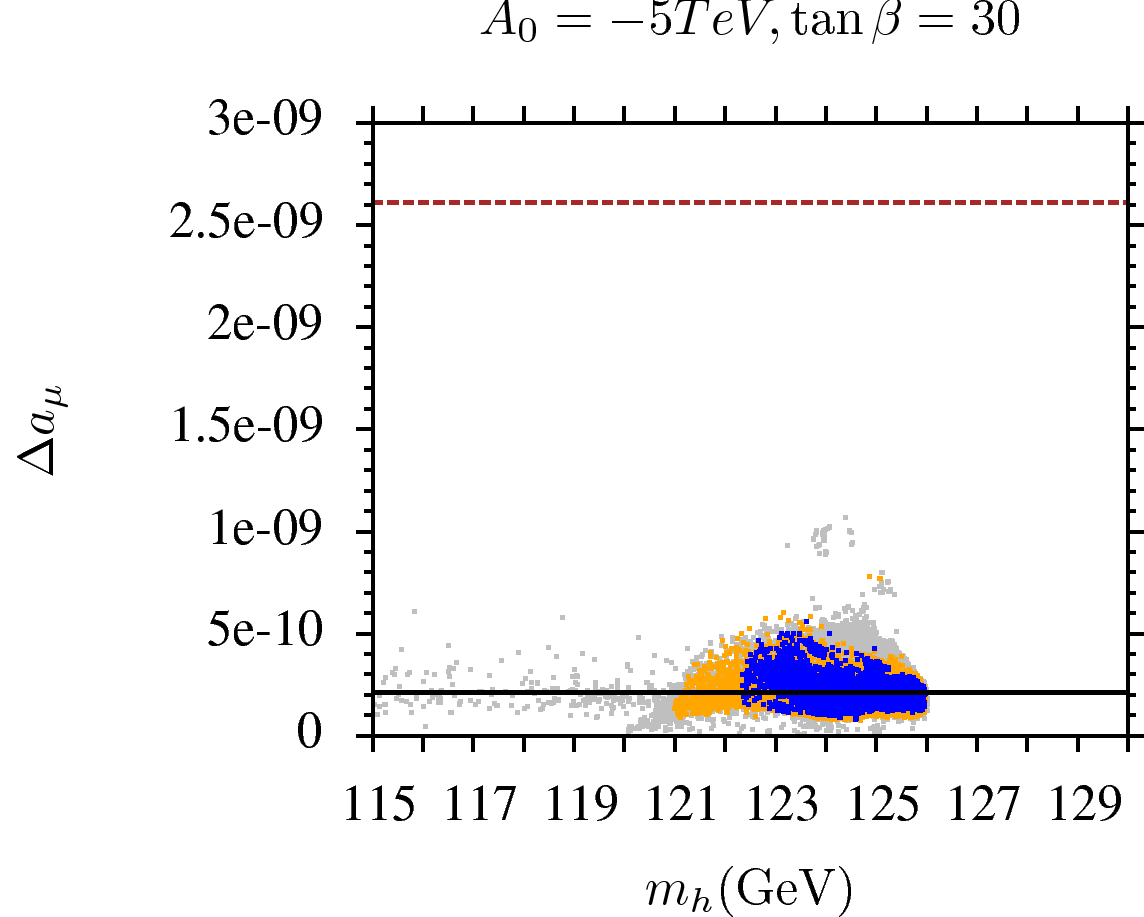}
}
\subfigure{
\includegraphics[width=7cm]{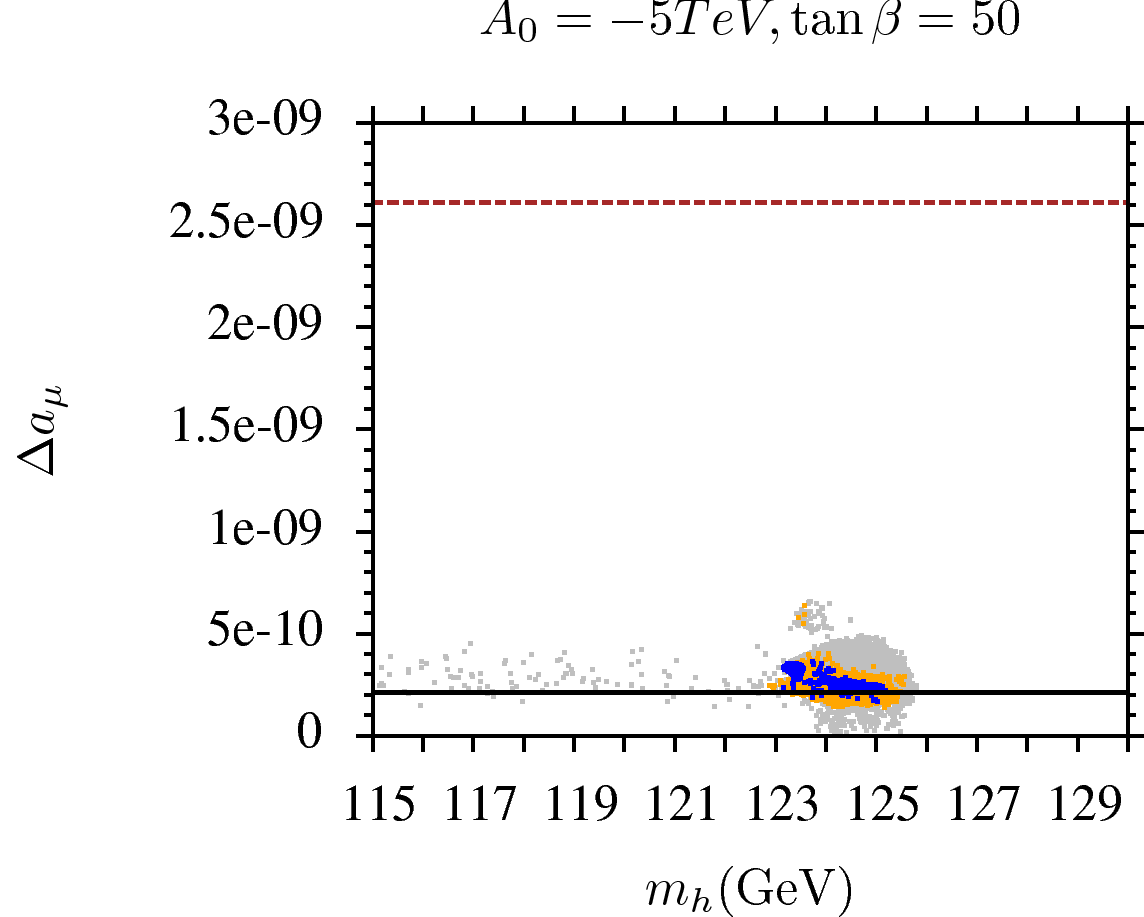}
}
\caption{
Color coding same as in Fig.~\ref{An3TeVg2mh}, 
 for $A_0=-5$ TeV. 
}
\label{An5TeVg2mh}
\end{figure}

\begin{figure}
\centering
\subfiguretopcaptrue

\subfigure{
\includegraphics[width=7cm]{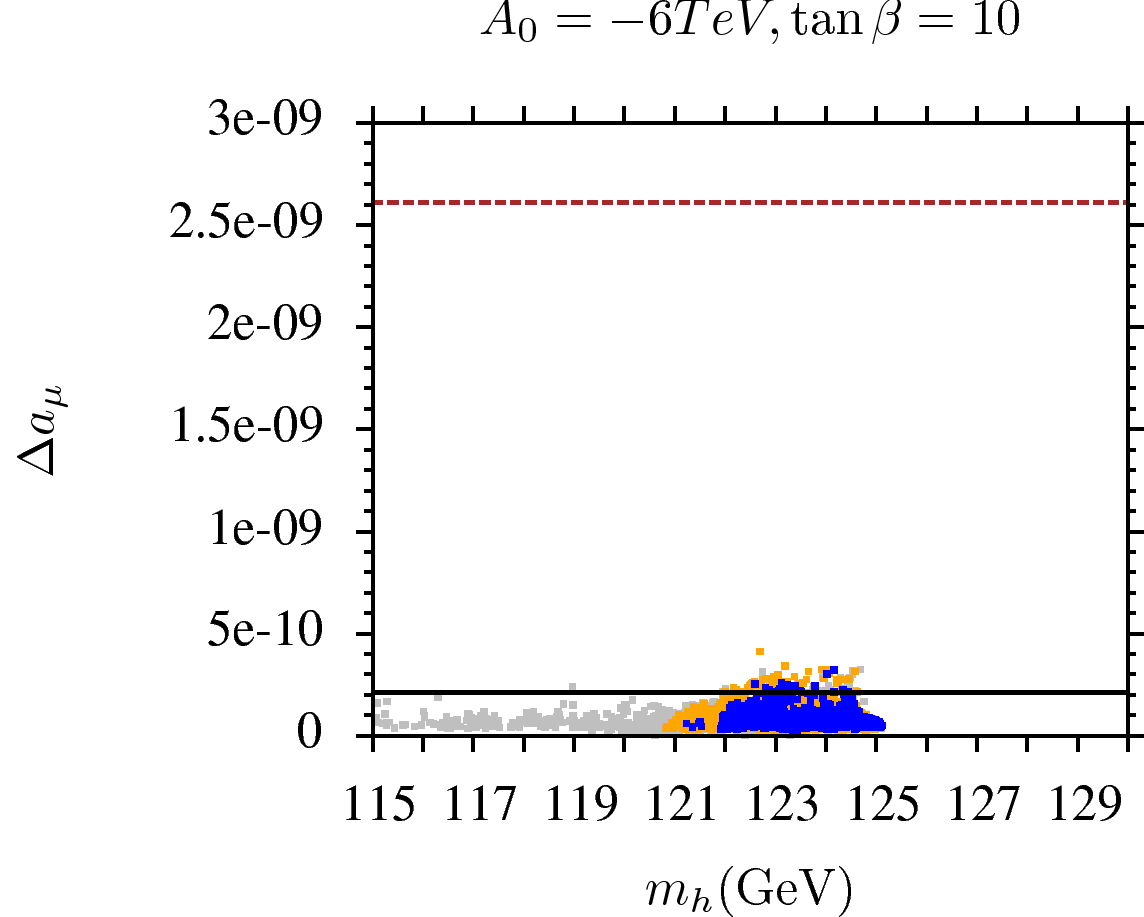}
}
\subfigure{
\includegraphics[width=7cm]{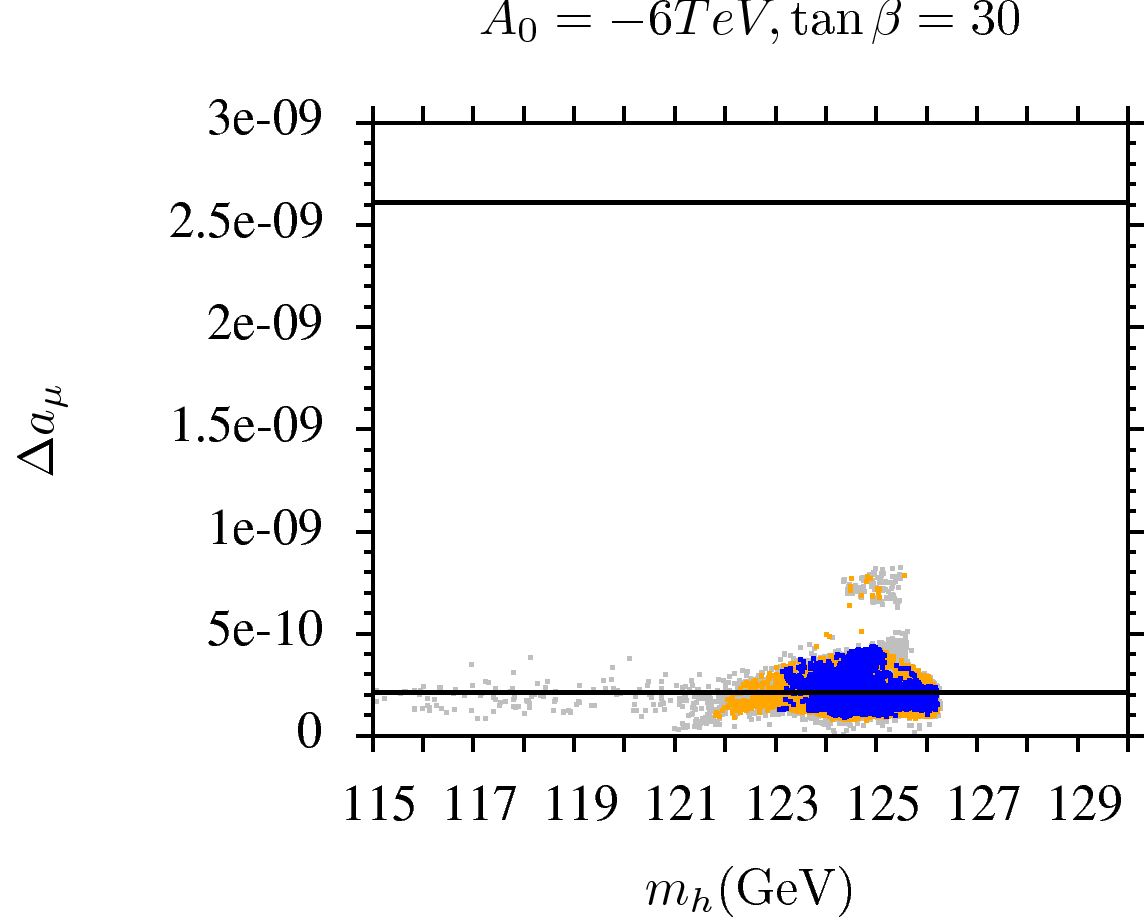}
}
\subfigure{
\includegraphics[width=7cm]{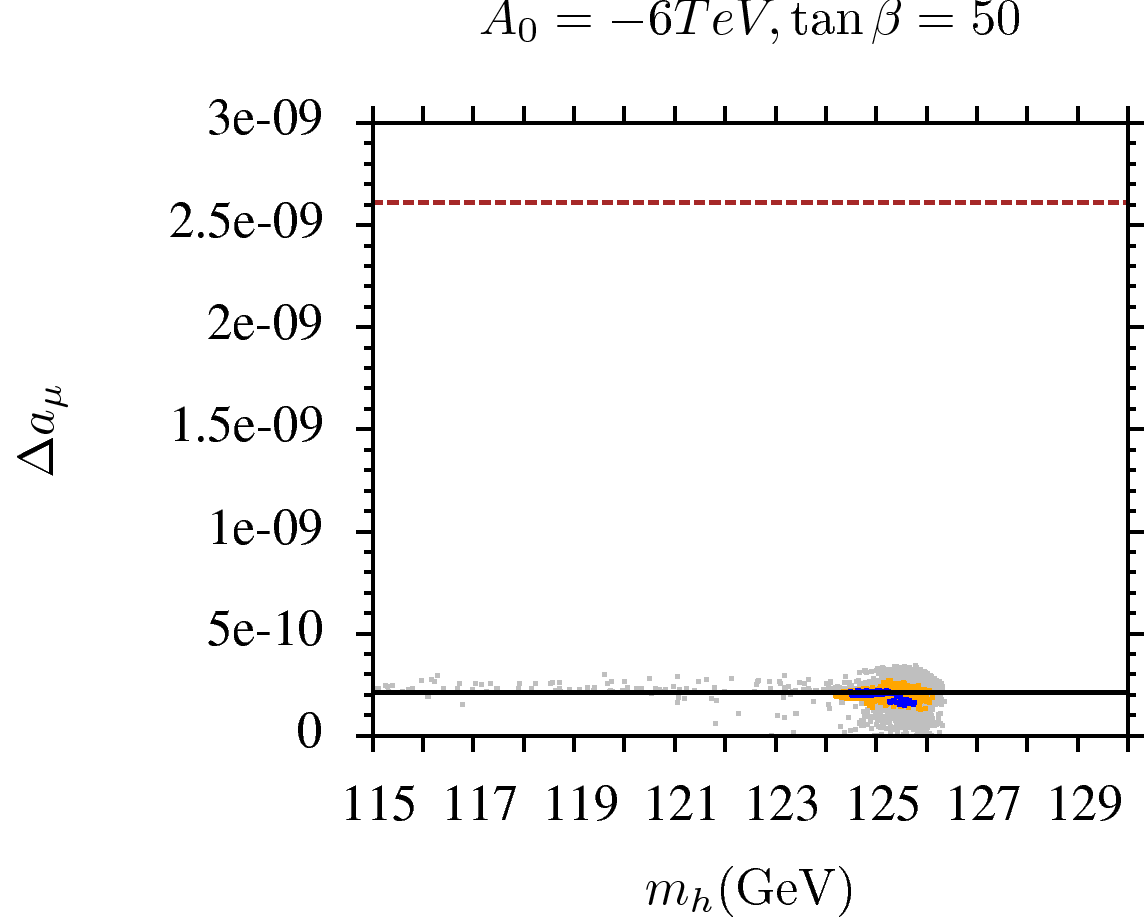}
}
\caption{
Color coding same as in Fig.~\ref{An3TeVg2mh}, 
 for $A_0=-6$ TeV. 
}
\label{An6TeVg2mh}
\end{figure}

\begin{figure}
\centering
\subfiguretopcaptrue

\subfigure{
\includegraphics[width=7cm]{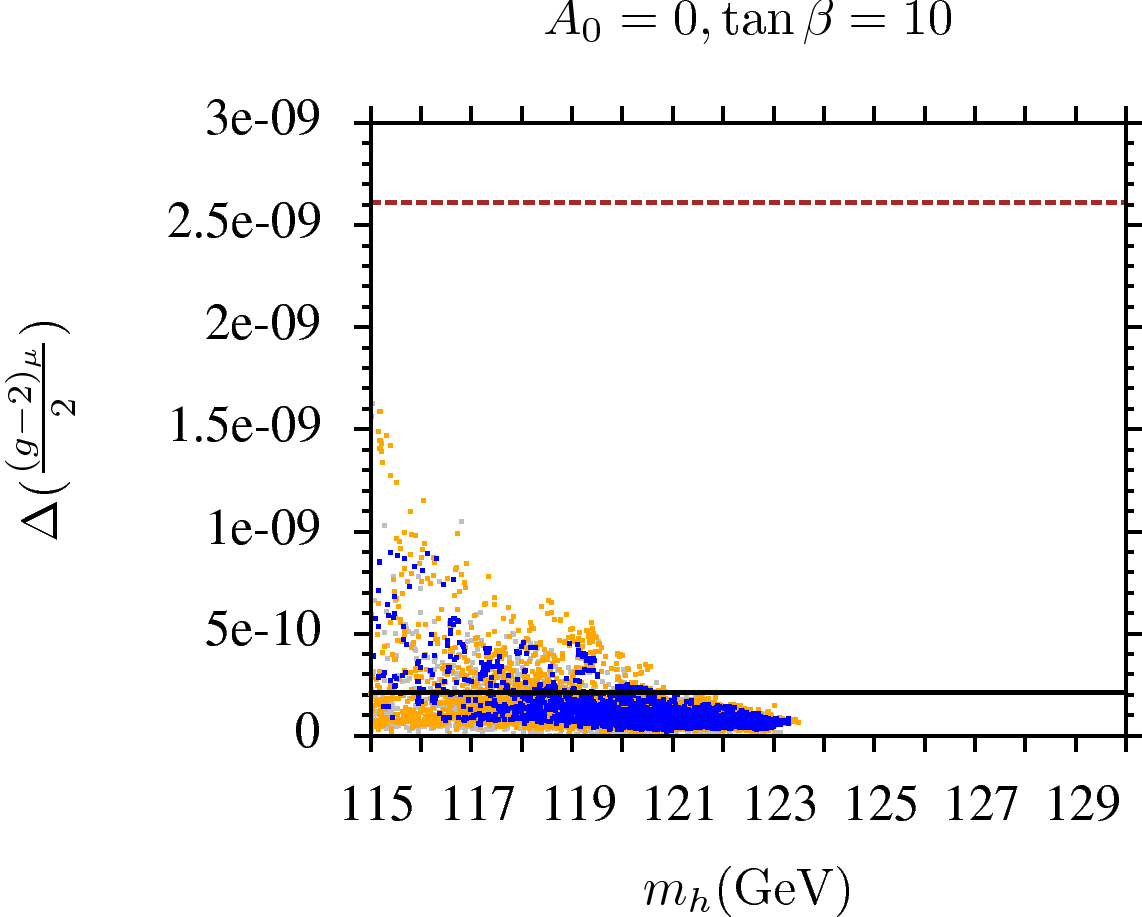}
}
\subfigure{
\includegraphics[width=7cm]{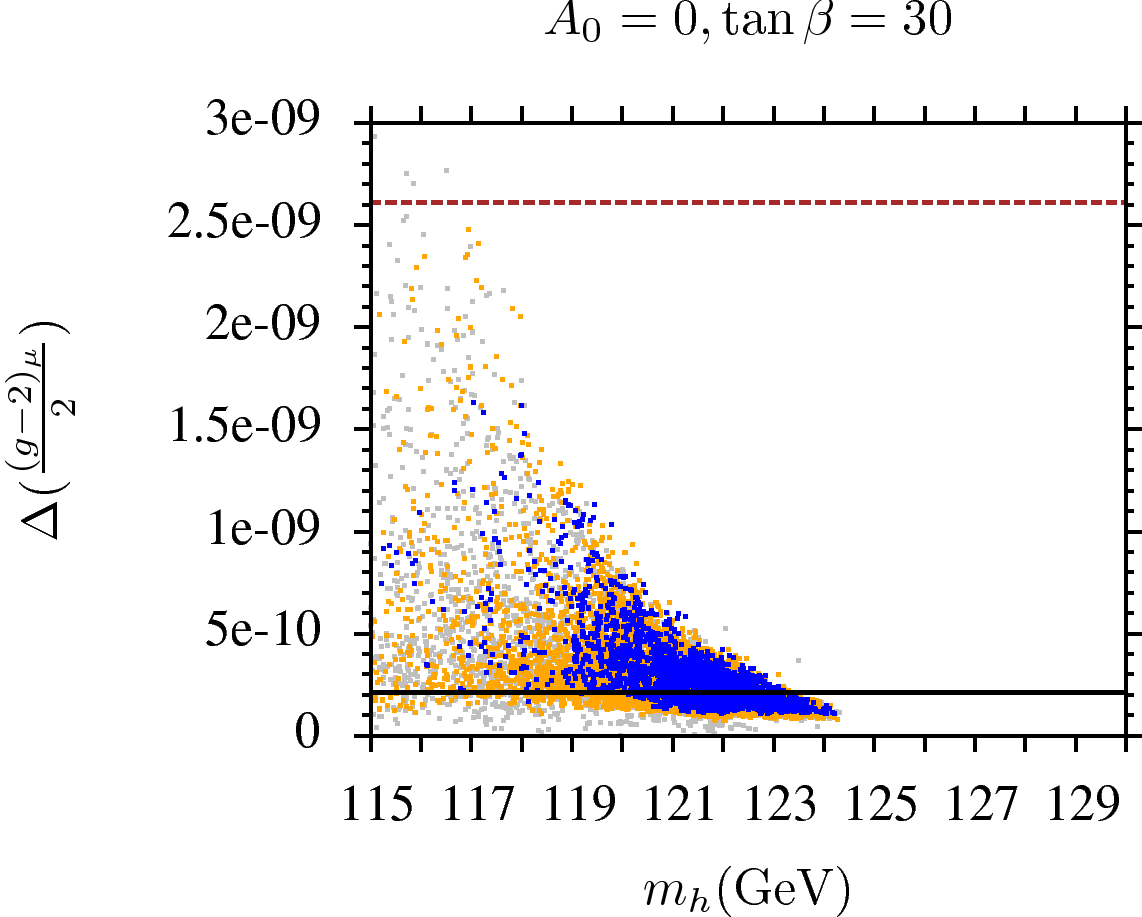}
}
\subfigure{
\includegraphics[width=7cm]{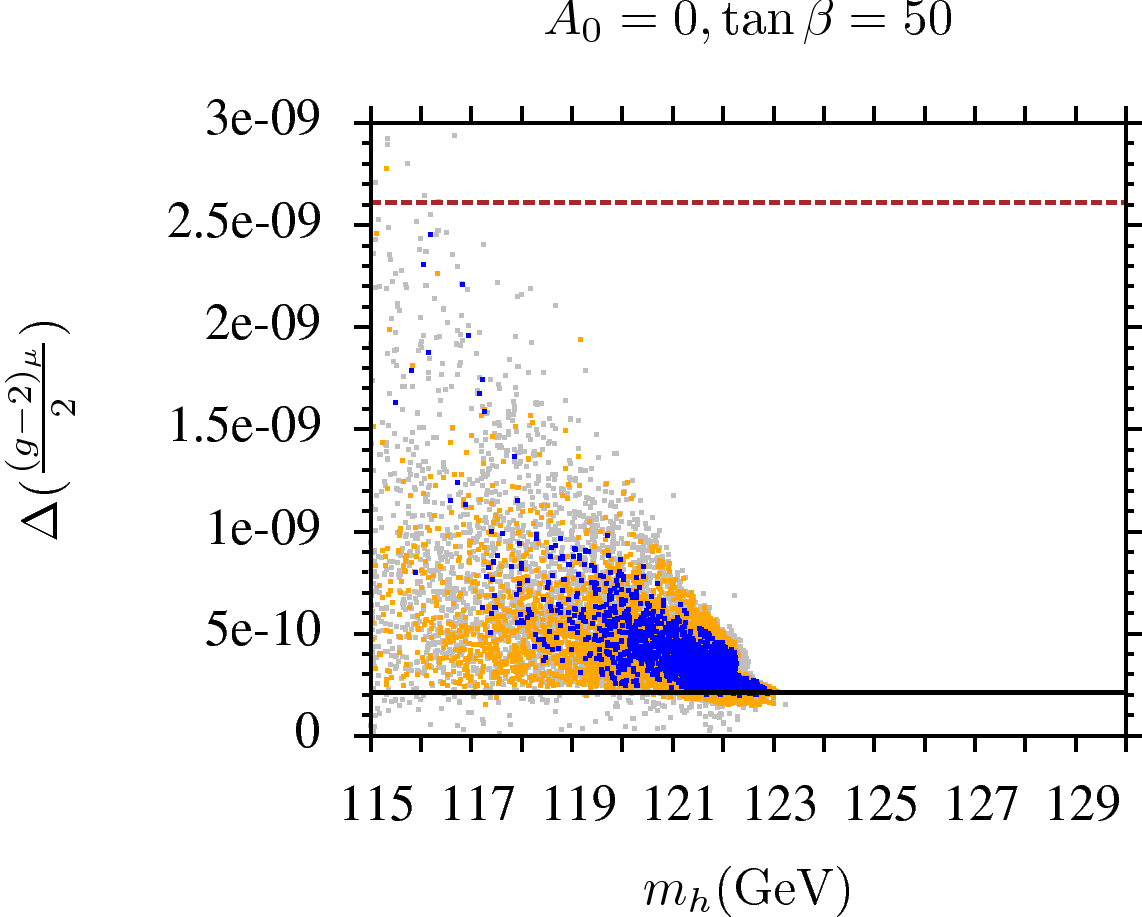}
}
\caption{
Color coding same as in Fig.~\ref{An3TeVg2mh},
for $A_0=0$ TeV. 
}
\label{A0TeVg2mh}
\end{figure}

\begin{figure}
\centering
\subfiguretopcaptrue

\subfigure{
\includegraphics[width=7cm]{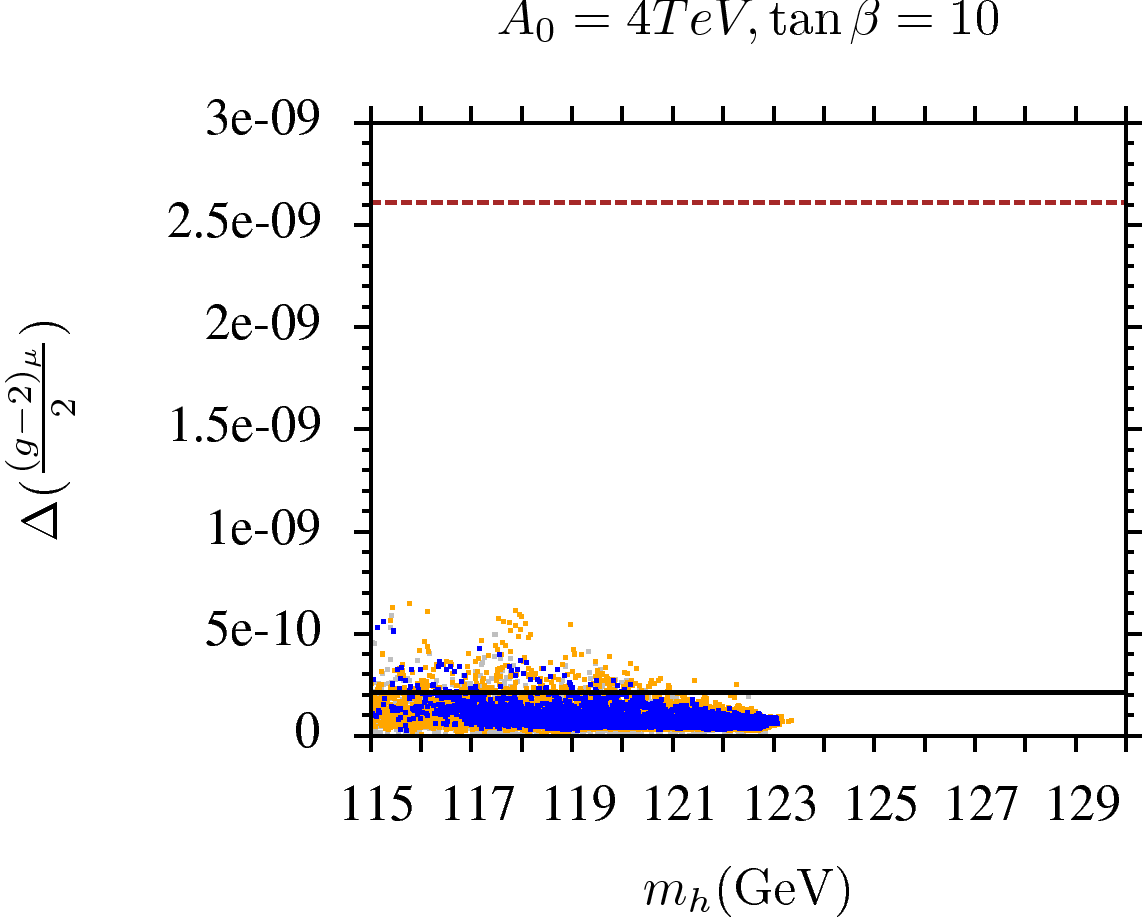}
}
\subfigure{
\includegraphics[width=7cm]{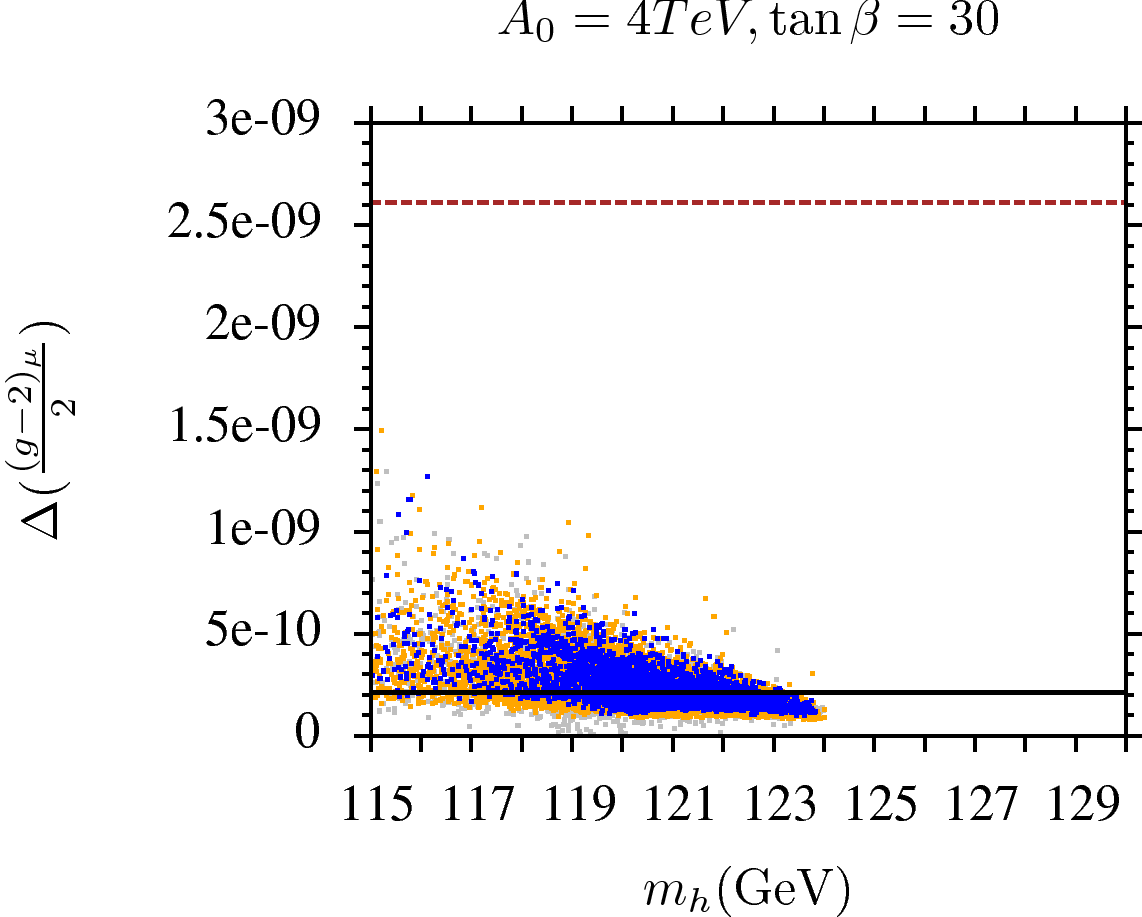}
}
\subfigure{
\includegraphics[width=7cm]{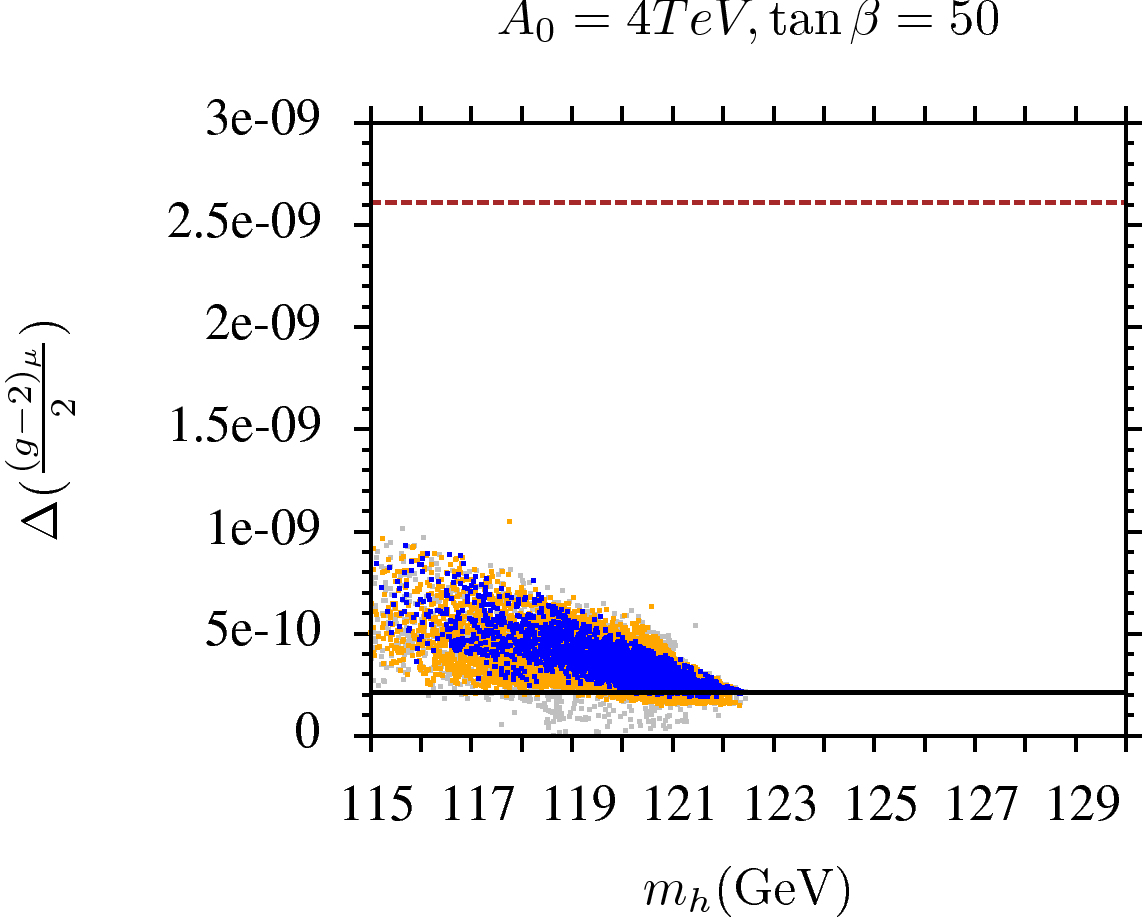}
}
\caption{
Color coding same as in Fig.~\ref{An3TeVg2mh}, 
 for $A_0=4$ TeV. 
}
\label{Ap4TeVg2mh}
\end{figure}

In Fig.~\ref{An3TeVg2mh} 
 we show our results in the ($m_h$, $\Delta a_\mu$)-plane 
 for $A_0 =-3$ TeV, $\tan\beta =10$, $30$ and $50$. 
 Gray points satisfy the requirements of REWSB and neutralino LSP. 
Orange points satisfy the mass bounds and B-physics bounds. 
Blue points are a subset of orange points and represent solutions 
 satisfying the WMAP9 bounds at 5$\sigma$ 
 as well as $m_{\tilde g}$, $m_{\tilde d_R} \geq 1$ TeV. 
The horizontal dashed red and solid black lines represent  
the central value of $\Delta a_{\mu}$ and the lower bar of the 3$\sigma$ variation from the central value respectively. 
For $\tan \beta =30$ and $50$ and $A_{0}=-3$ TeV, only a very limited parameter space   
 can simultaneously satisfy the Higgs boson mass bound 
 and the muon $g-2$ data, 
 while for $\tan \beta=10$, none of the points  can do this. 
Similar plots with $A_0 =-4$, $-5$ and $-6$ TeV, 
 are shown in Figs.~\ref{An4TeVg2mh}-\ref{An6TeVg2mh}. 
Analogous to Fig.~\ref{An3TeVg2mh}, 
 we find parameter regions for $\tan \beta=30$ and $50$ which are consistent with 
 the Higgs boson mass and the muon $g-2$ bounds. 
 The results for $A_0 =0$ and $4$ TeV are shown 
 in Figs.~\ref{A0TeVg2mh} and \ref{Ap4TeVg2mh}. 
For positive $A_0$ values, no satisfactory solution is found.

\begin{figure}
\centering
\subfiguretopcaptrue

\subfigure{
\includegraphics[width=7cm]{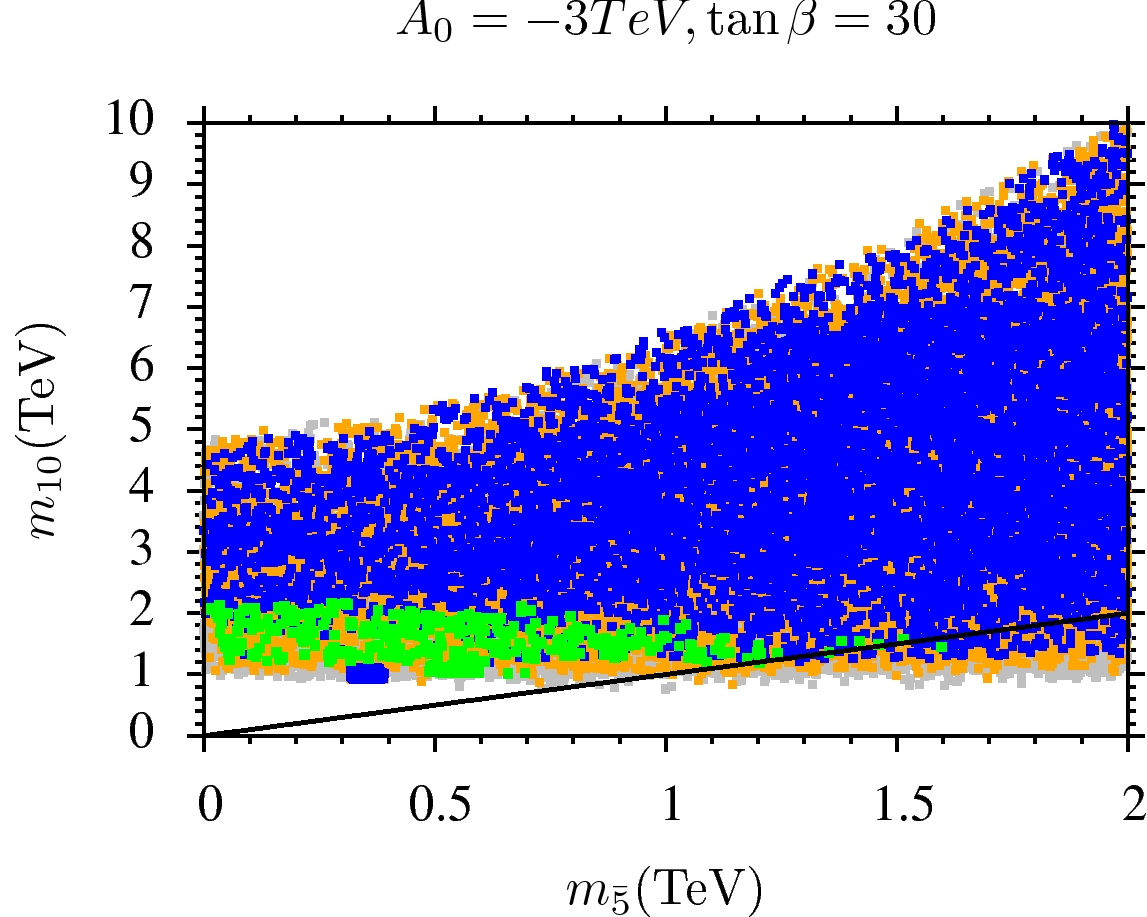}
}
\subfigure{
\includegraphics[width=7cm]{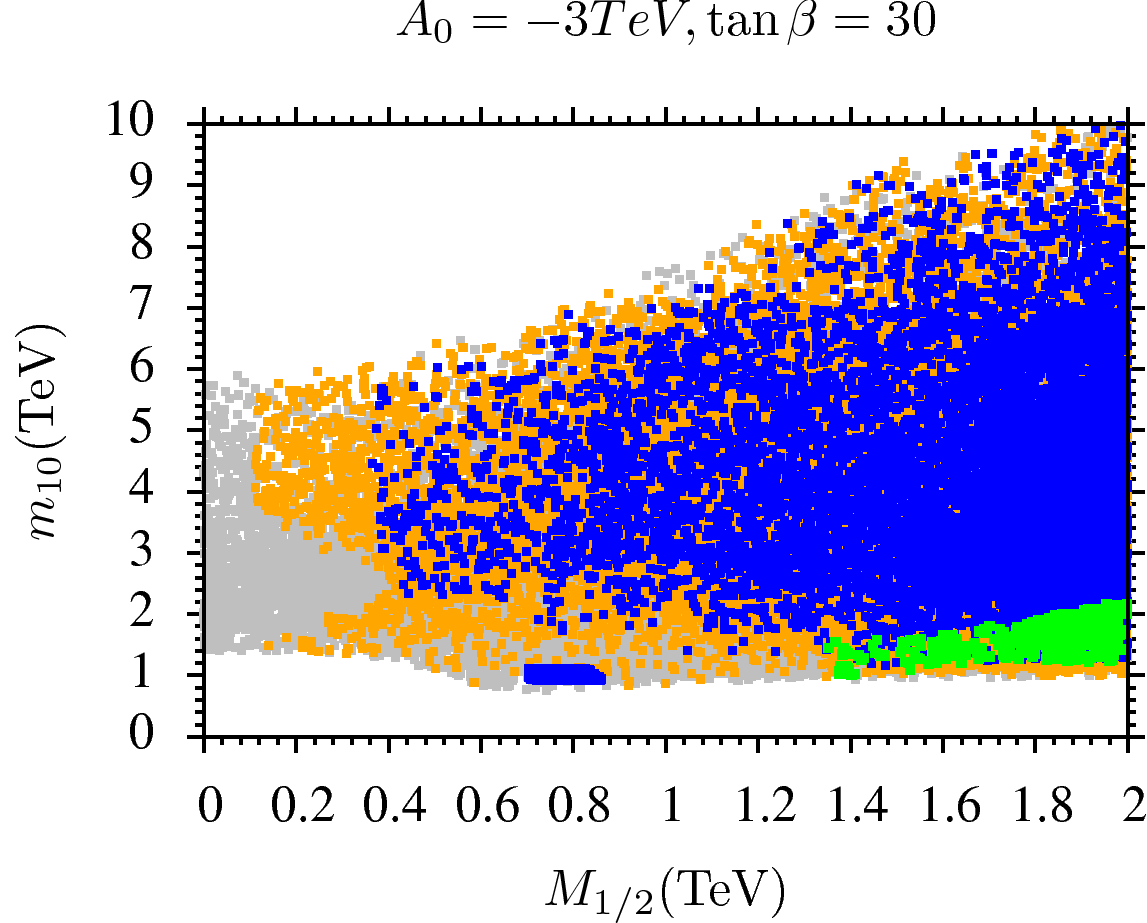}
}\\
\subfigure{
\includegraphics[width=7cm]{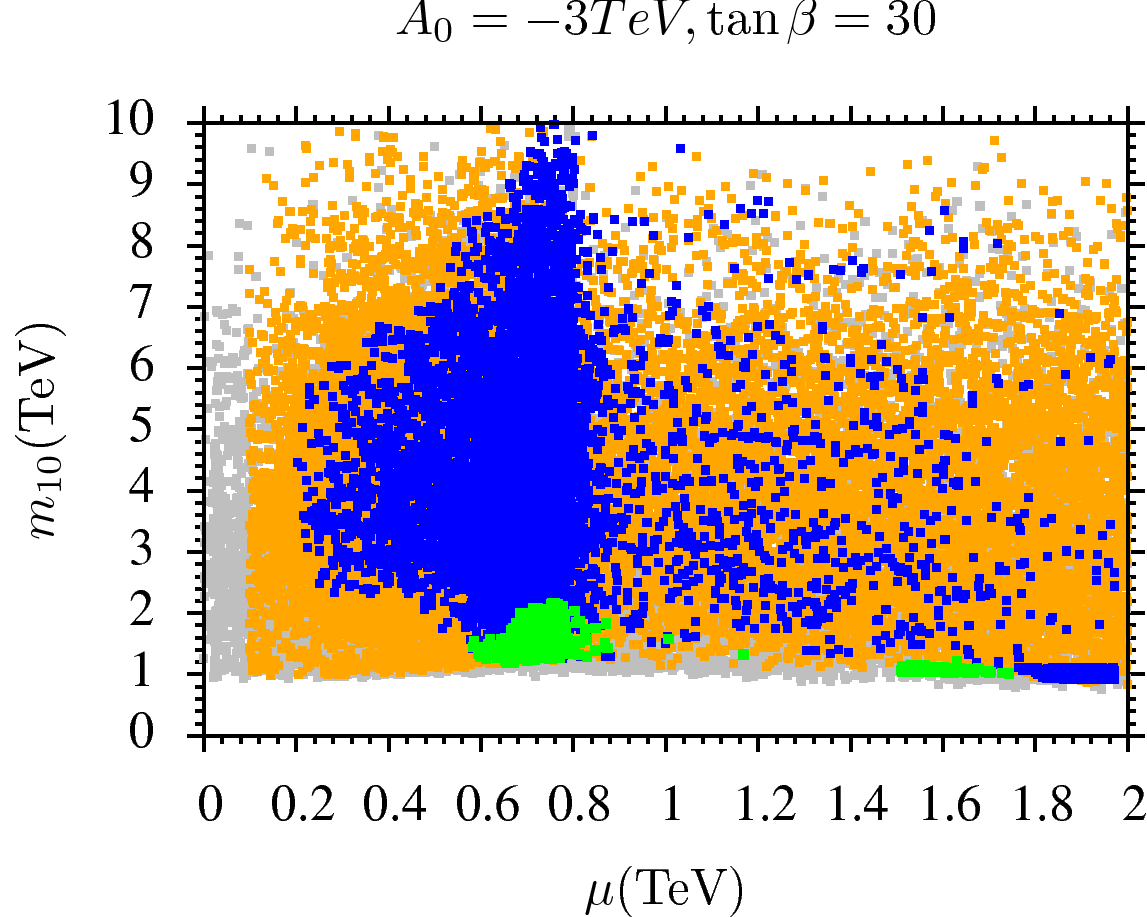}
}
\subfigure{
\includegraphics[width=7cm]{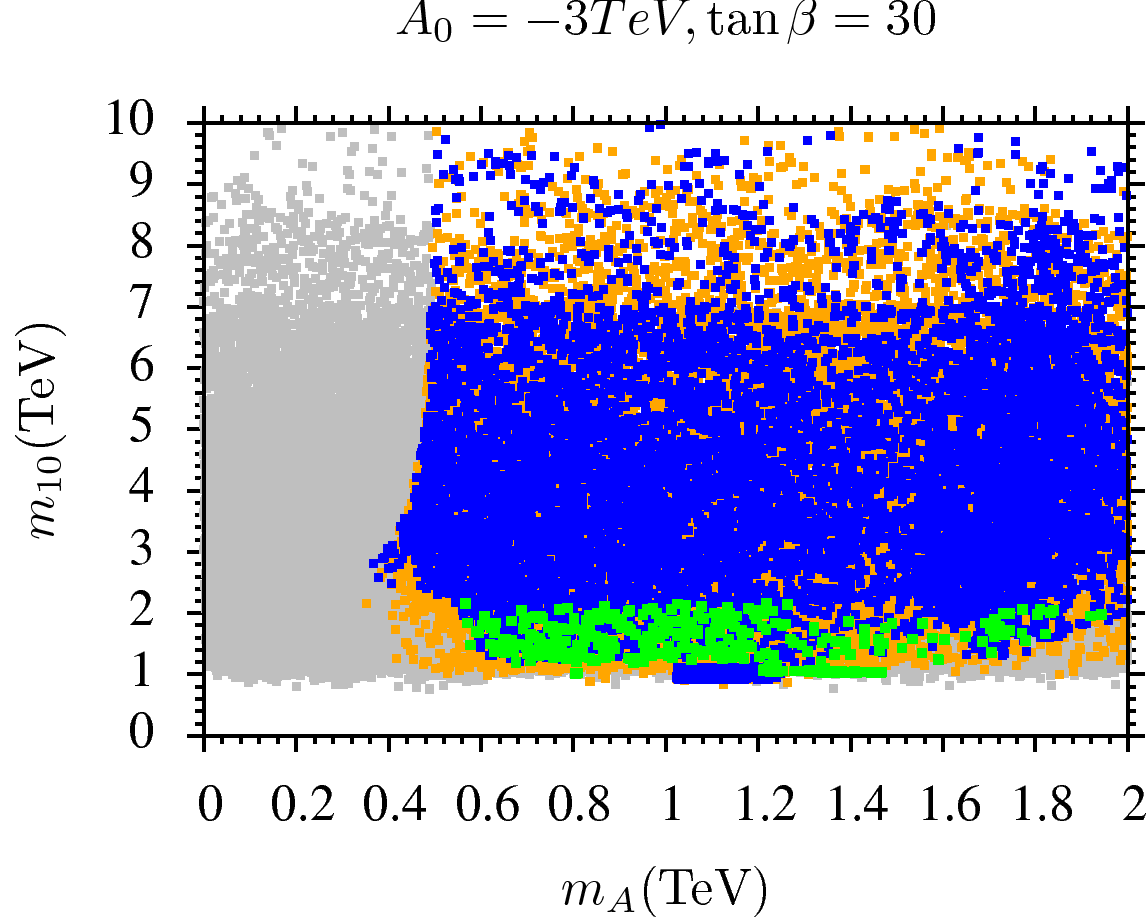}
}
\caption{
Plots in ($m_{10}$, $m_{\bar 5}$)-plane, 
 ($m_{10}$, $M_{1/2}$)-plane, 
 ($m_{\bar 5}$, $M_{1/2}$)-plane  
  and ($m_A$, $\mu$)-plane. 
Gray points satisfy the requirements of REWSB and neutralino LSP. 
Orange points satisfy mass bounds and B-physics bounds. 
Blue points are subset of the orange points and 
 represent solutions satisfying the WMAP9 (5$\sigma$) bounds, 
 $124 \; {\rm GeV} \leq m_h \leq 127$ GeV, 
 $m_{\tilde g}$, $m_{\tilde d_R} \gtrsim 1$ TeV. 
Green points are subset of the blue points and satisfy 
 the 3$\sigma$ bounds on $\Delta a_{\mu}$. 
}
\label{An3TeVtanb30funda}
\end{figure}

\begin{figure}
\centering
\subfiguretopcaptrue

\subfigure{
\includegraphics[width=7cm]{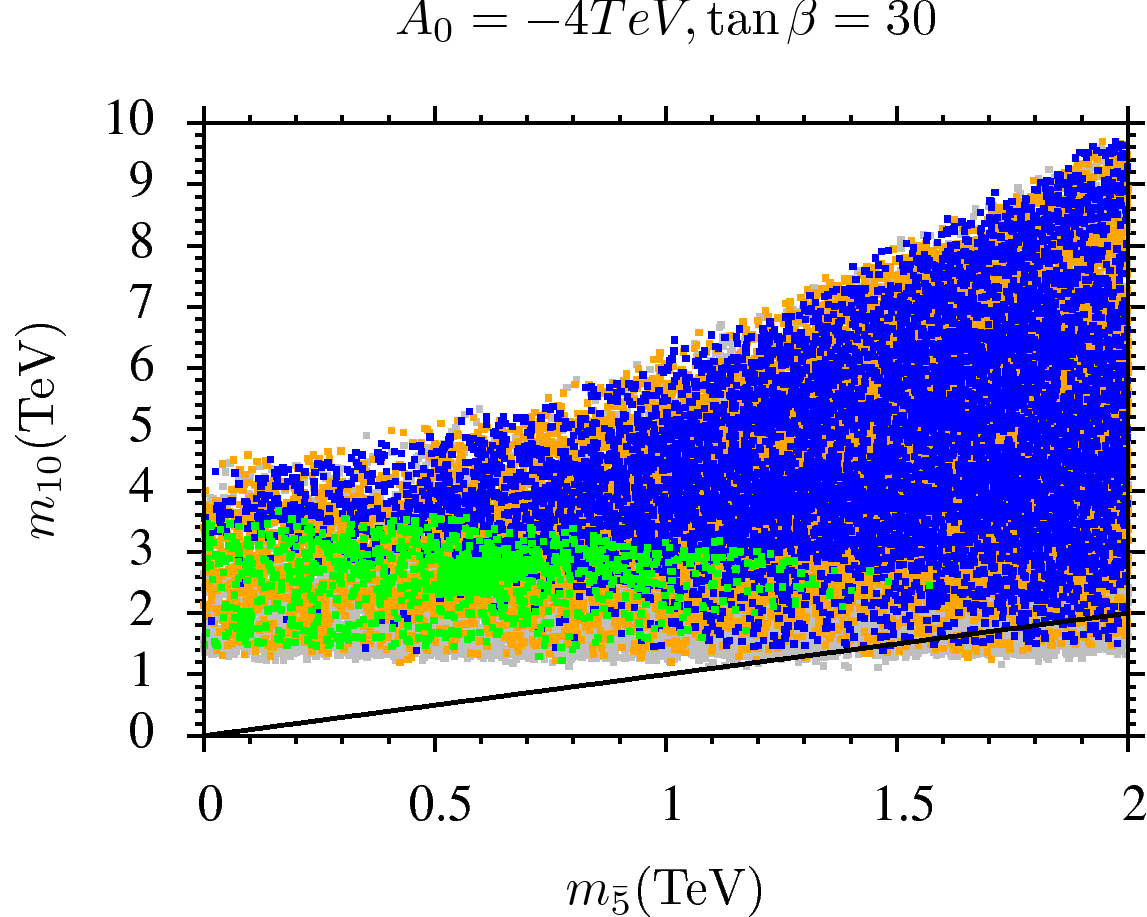}
}
\subfigure{
\includegraphics[width=7cm]{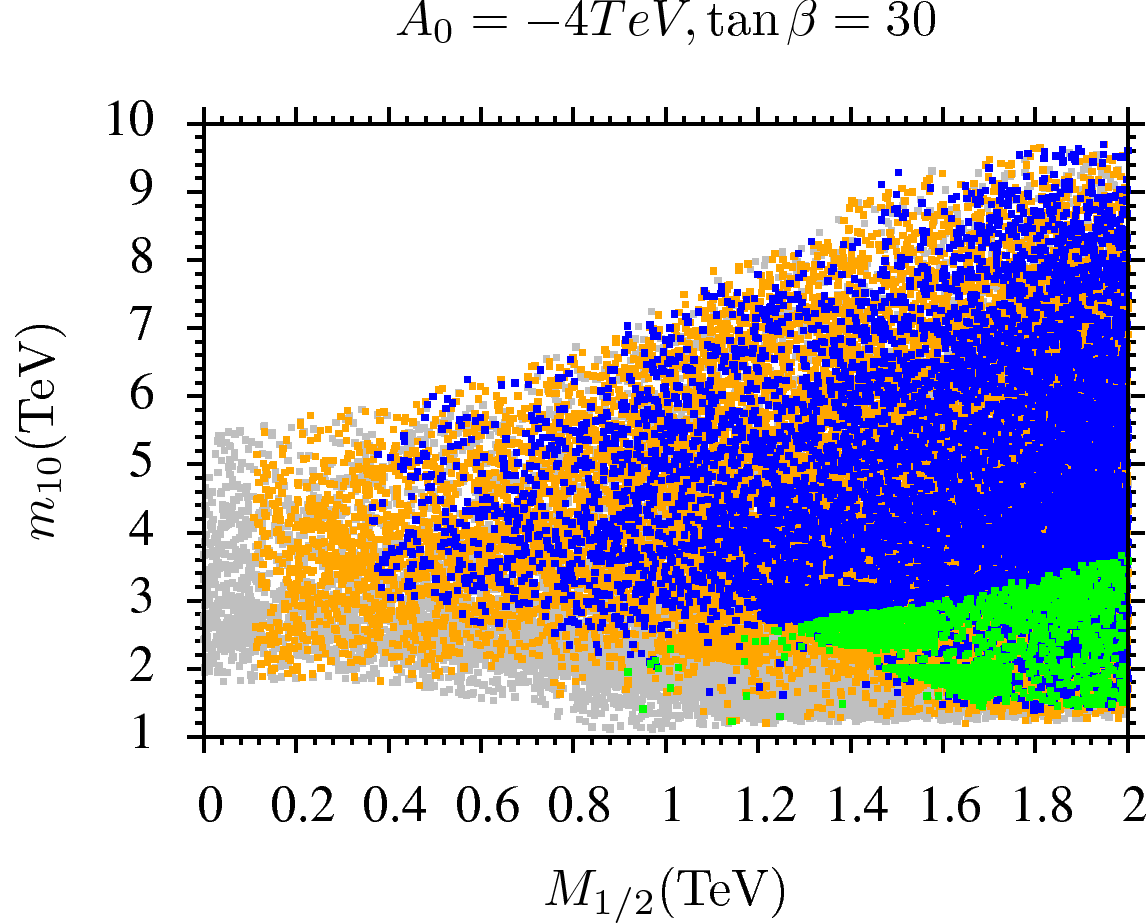}
}\\
\subfigure{
\includegraphics[width=7cm]{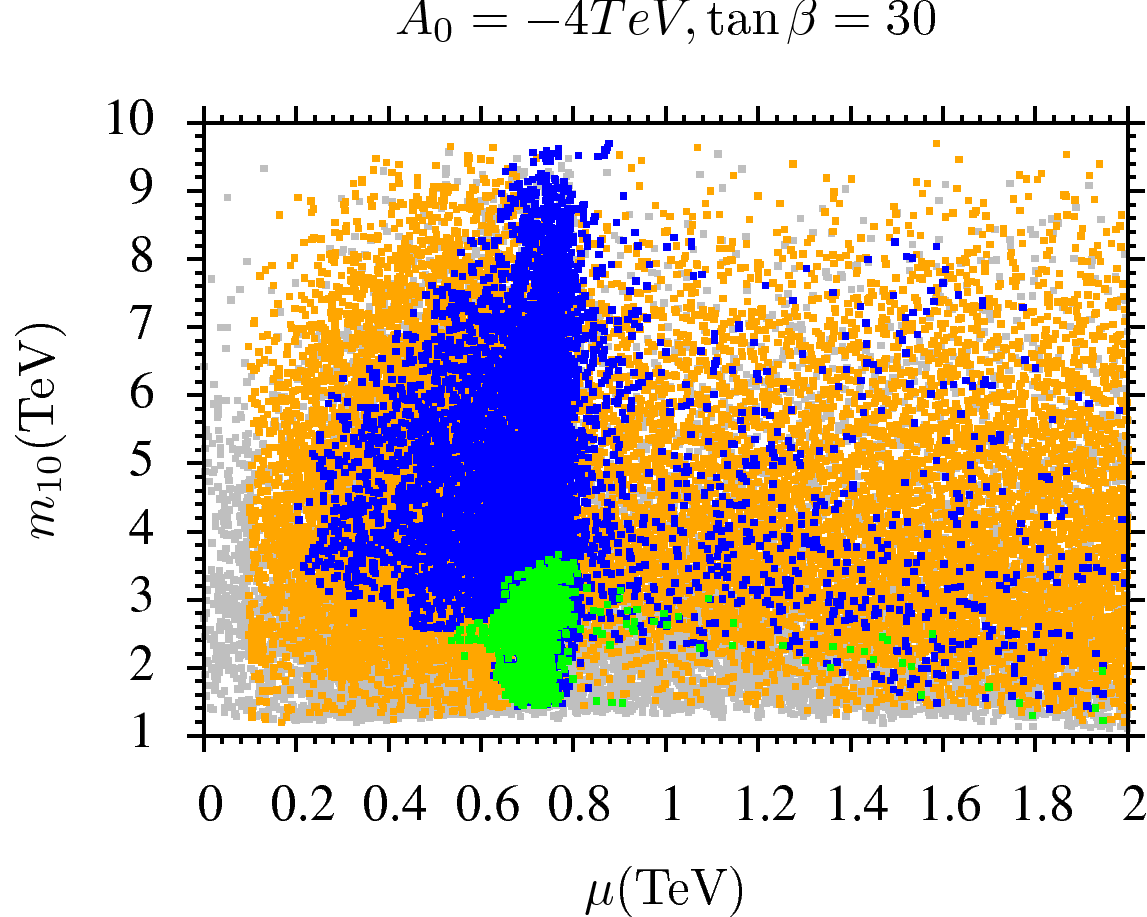}
}
\subfigure{
\includegraphics[width=7cm]{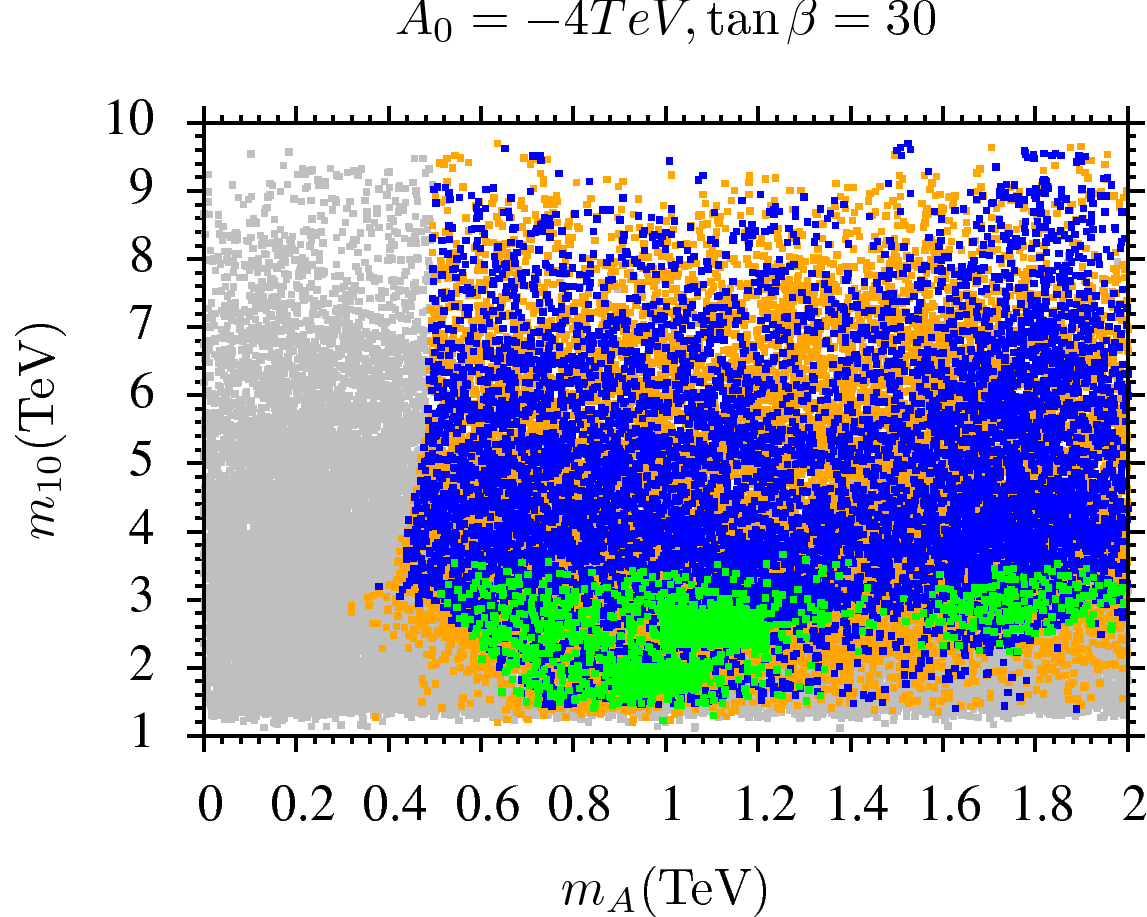}
}
\caption{
Color coding same as in Fig.~\ref{An3TeVtanb30funda}, 
 with $A_0=-4$ TeV. 
}
\label{An4TeVtanb30funda}
\end{figure}

\begin{figure}
\centering
\subfiguretopcaptrue

\subfigure{
\includegraphics[width=7cm]{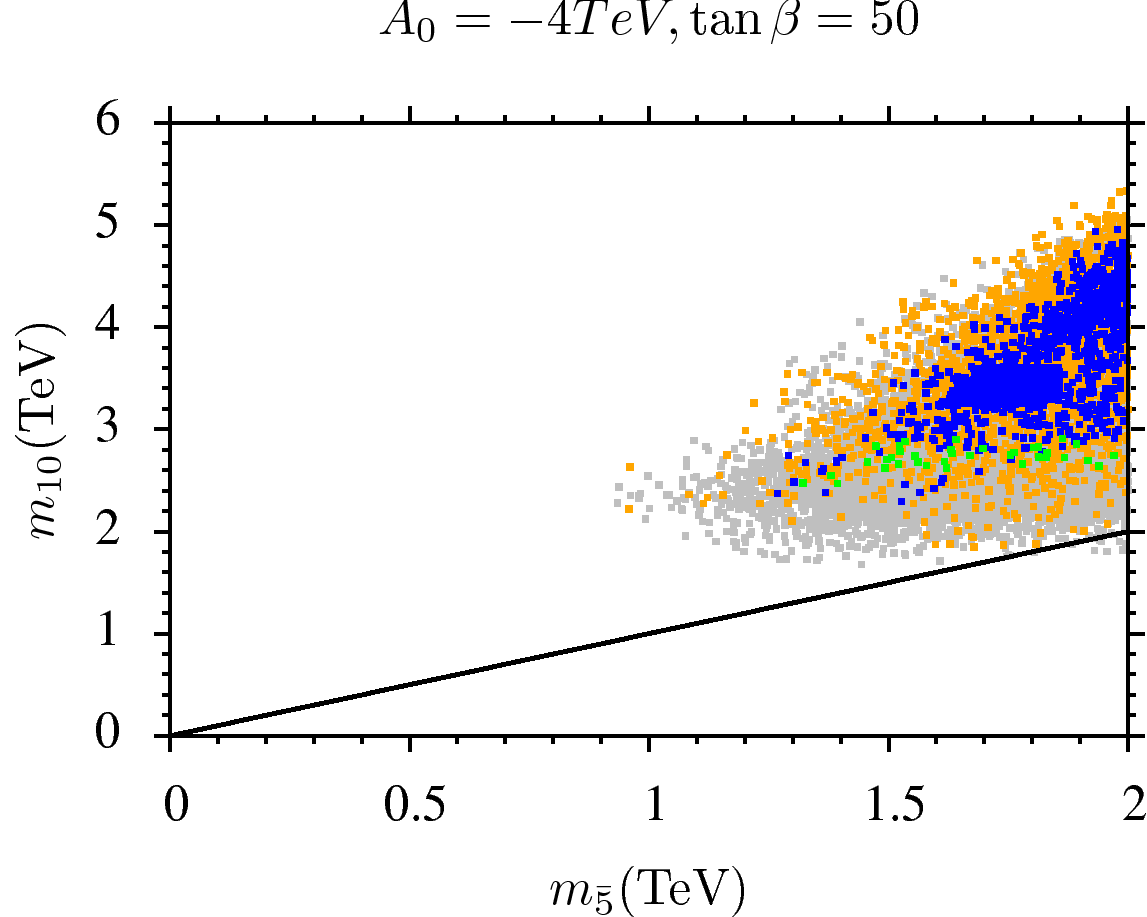}
}
\subfigure{
\includegraphics[width=7cm]{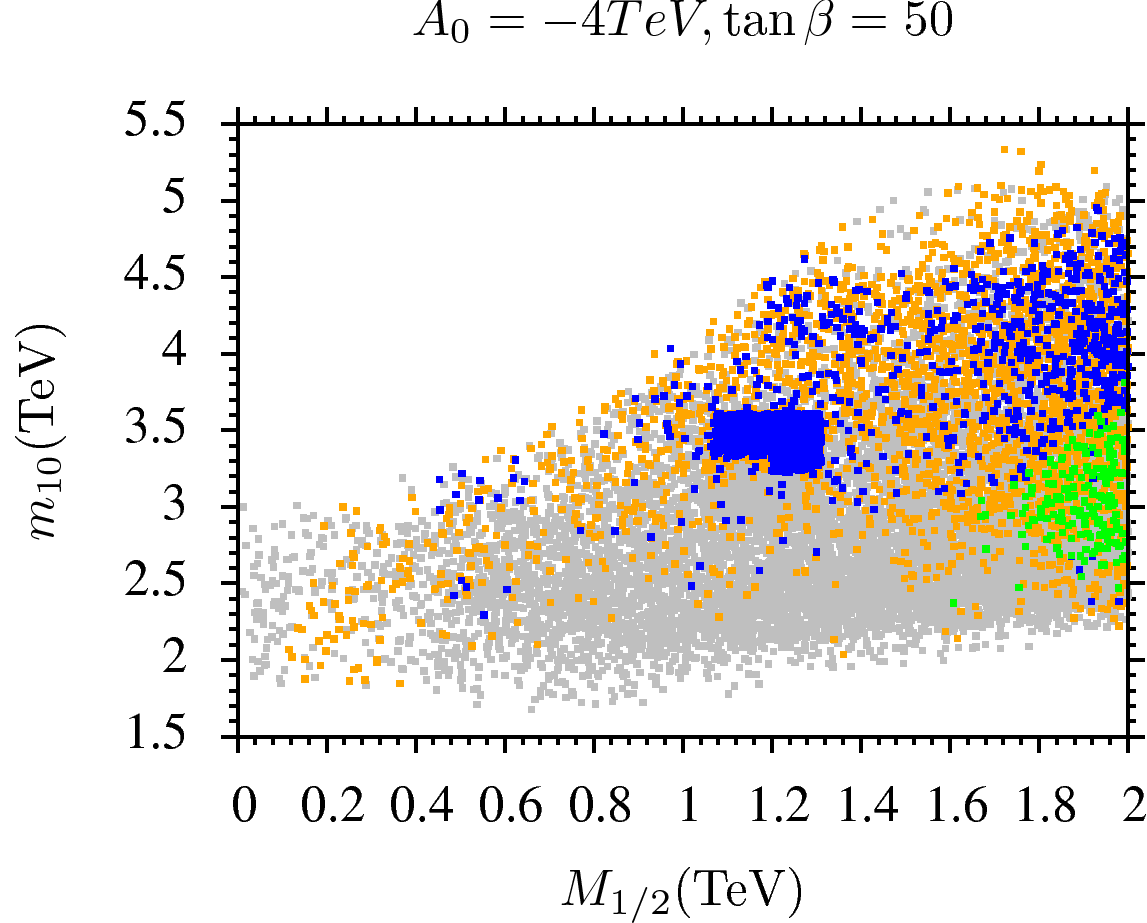}
}\\
\subfigure{
\includegraphics[width=7cm]{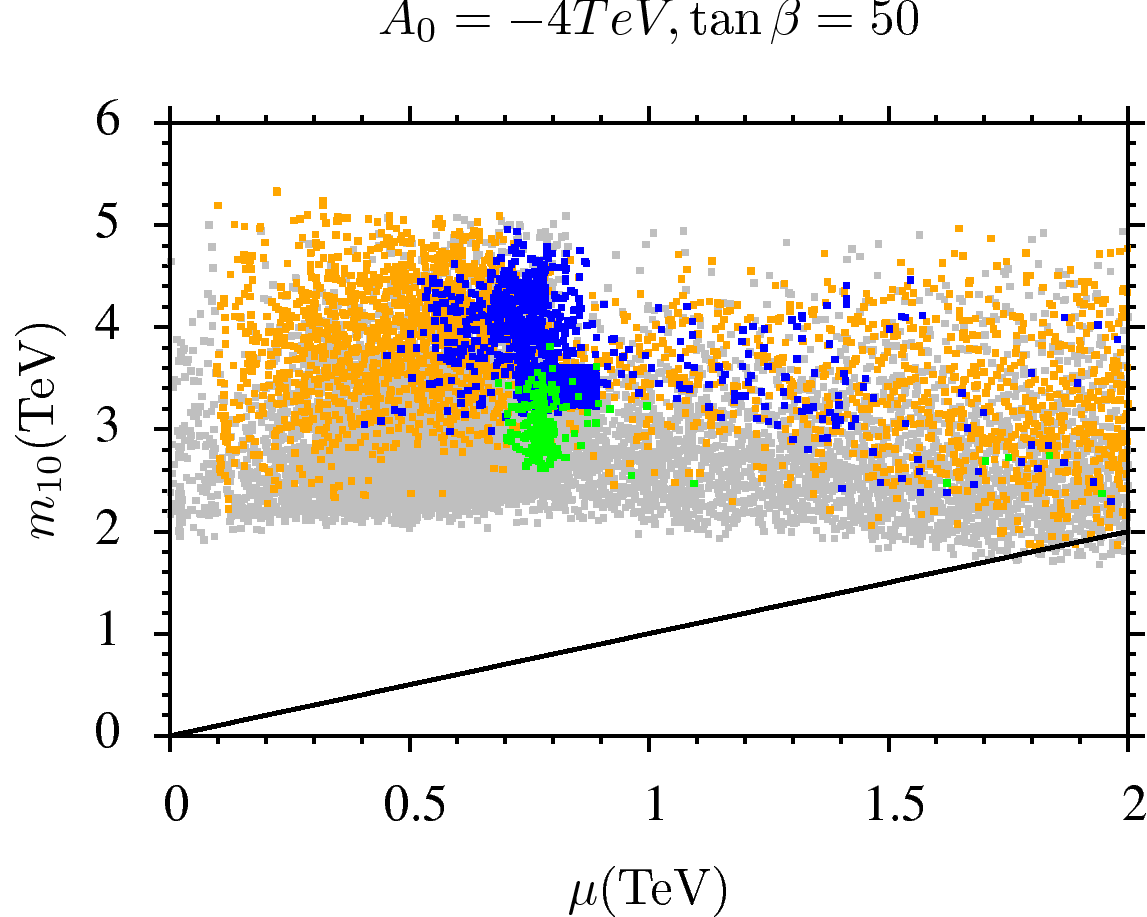}
}
\subfigure{
\includegraphics[width=7cm]{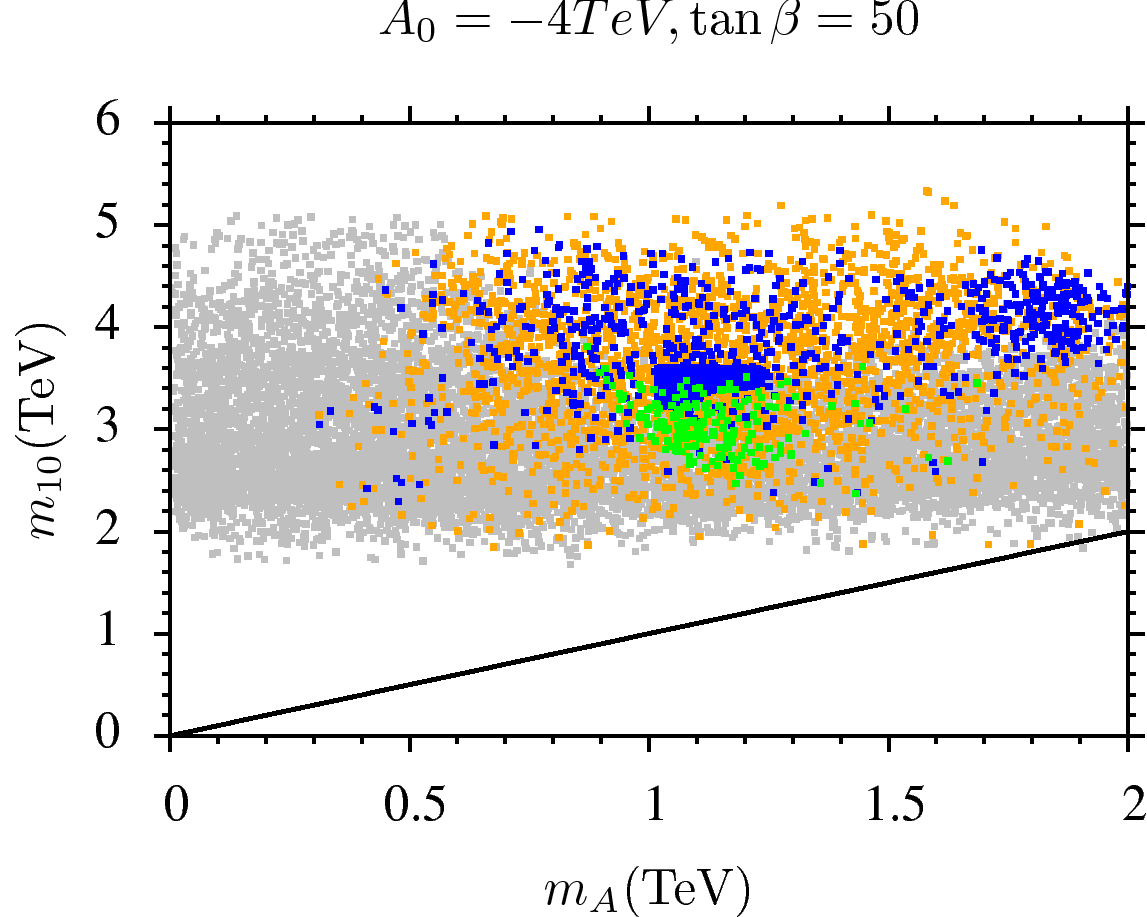}
}
\caption{
Color coding same as in Fig.~\ref{An3TeVtanb30funda},
 with $A_0=-4$ TeV.
}
\label{An4TeVtanb50funda}
\end{figure}

\begin{figure}
\centering
\subfiguretopcaptrue

\subfigure{
\includegraphics[width=7cm]{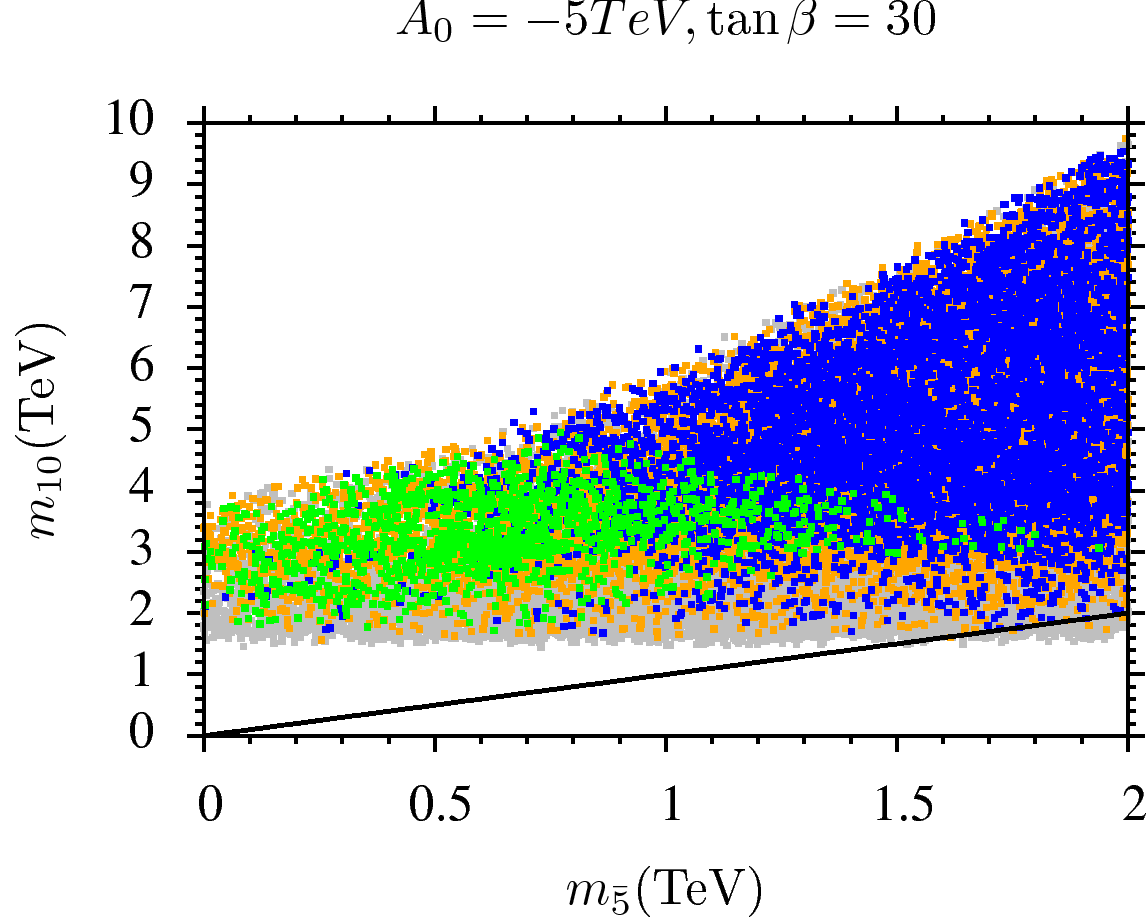}
}
\subfigure{
\includegraphics[width=7cm]{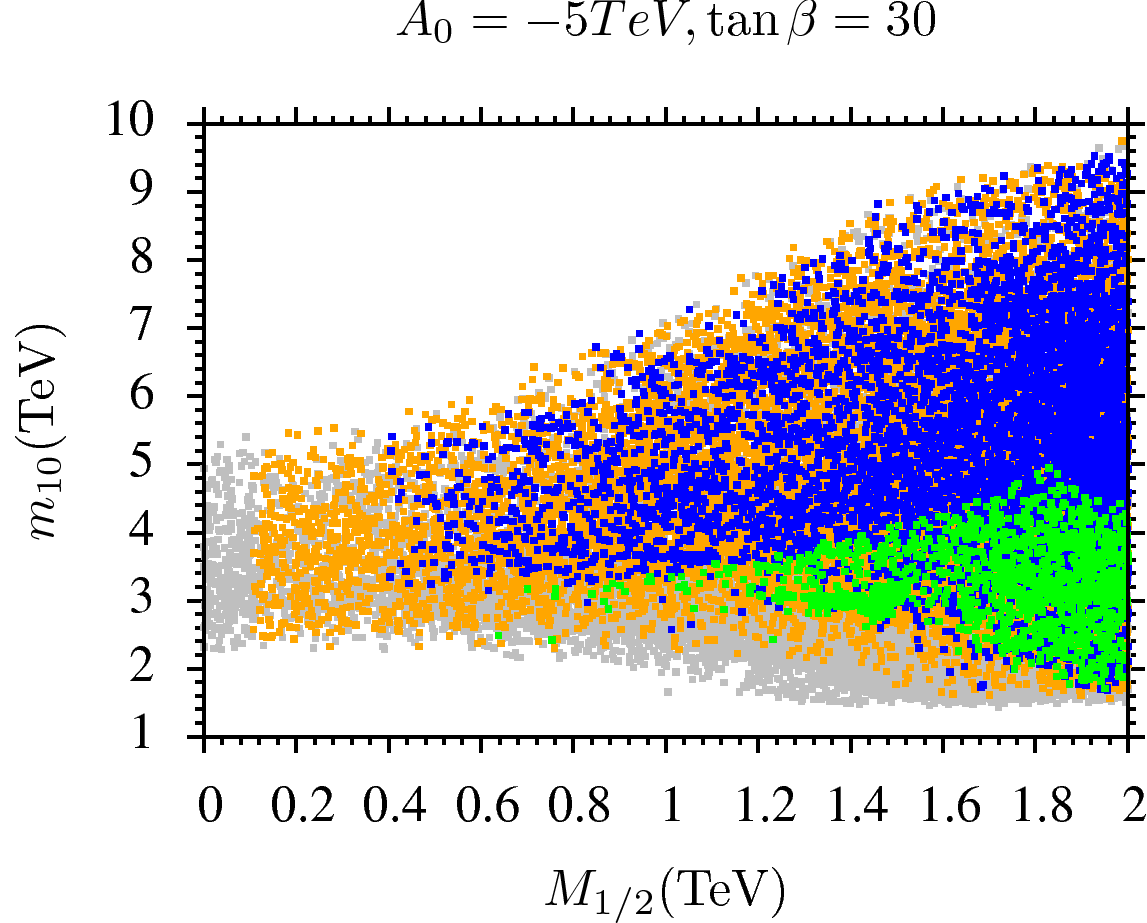}
}\\
\subfigure{
\includegraphics[width=7cm]{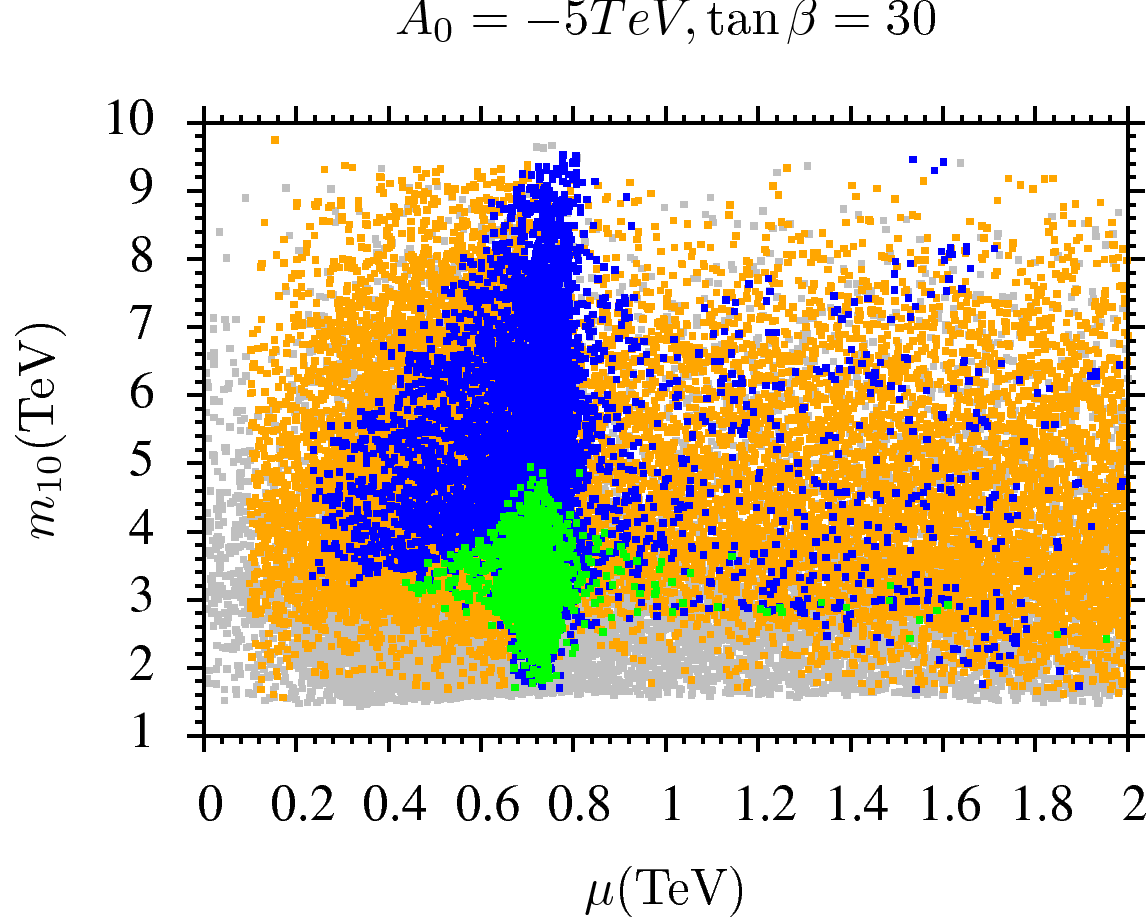}
}
\subfigure{
\includegraphics[width=7cm]{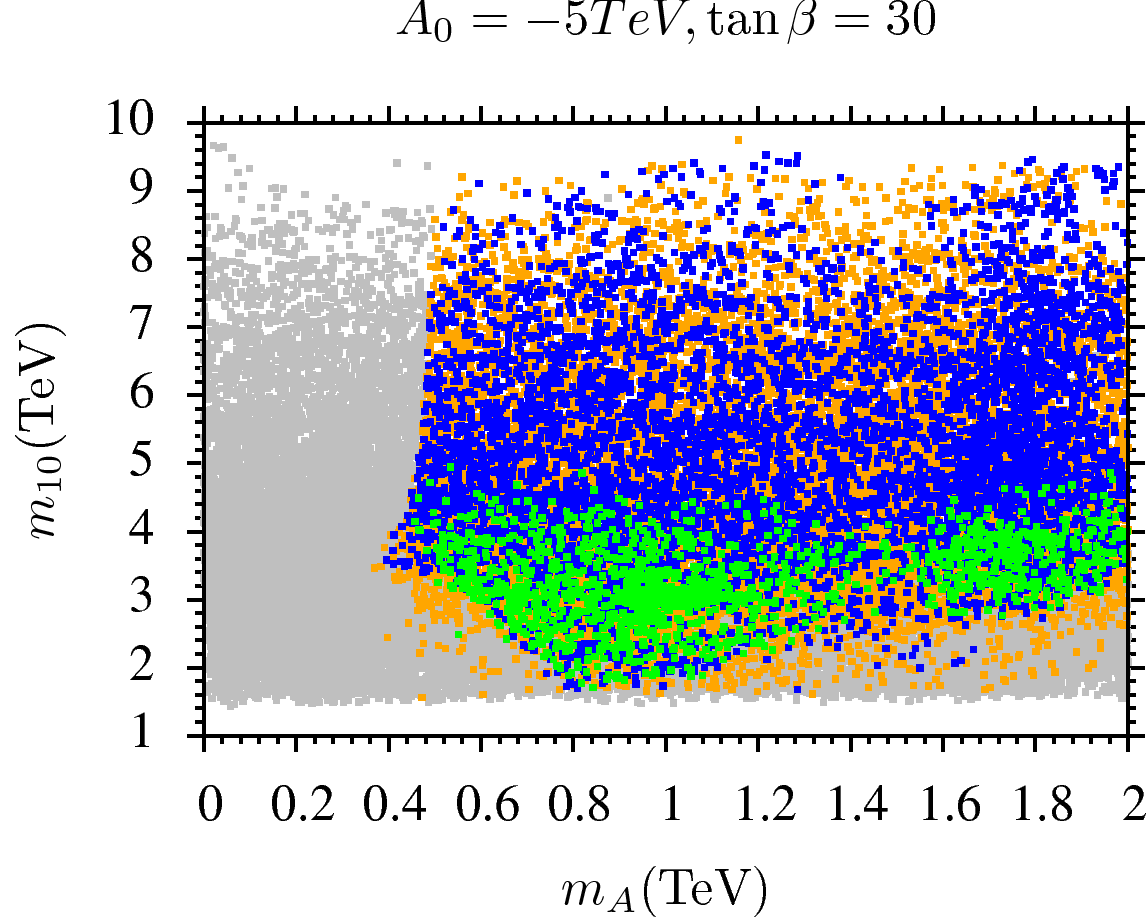}
}
\caption{
Color coding same as in Fig.~\ref{An3TeVtanb30funda}, 
for $A_0=-5$ TeV. 
}
\label{An5TeVtanb30funda}
\end{figure}

\begin{figure}
\centering
\subfiguretopcaptrue

\subfigure{
\includegraphics[width=7cm]{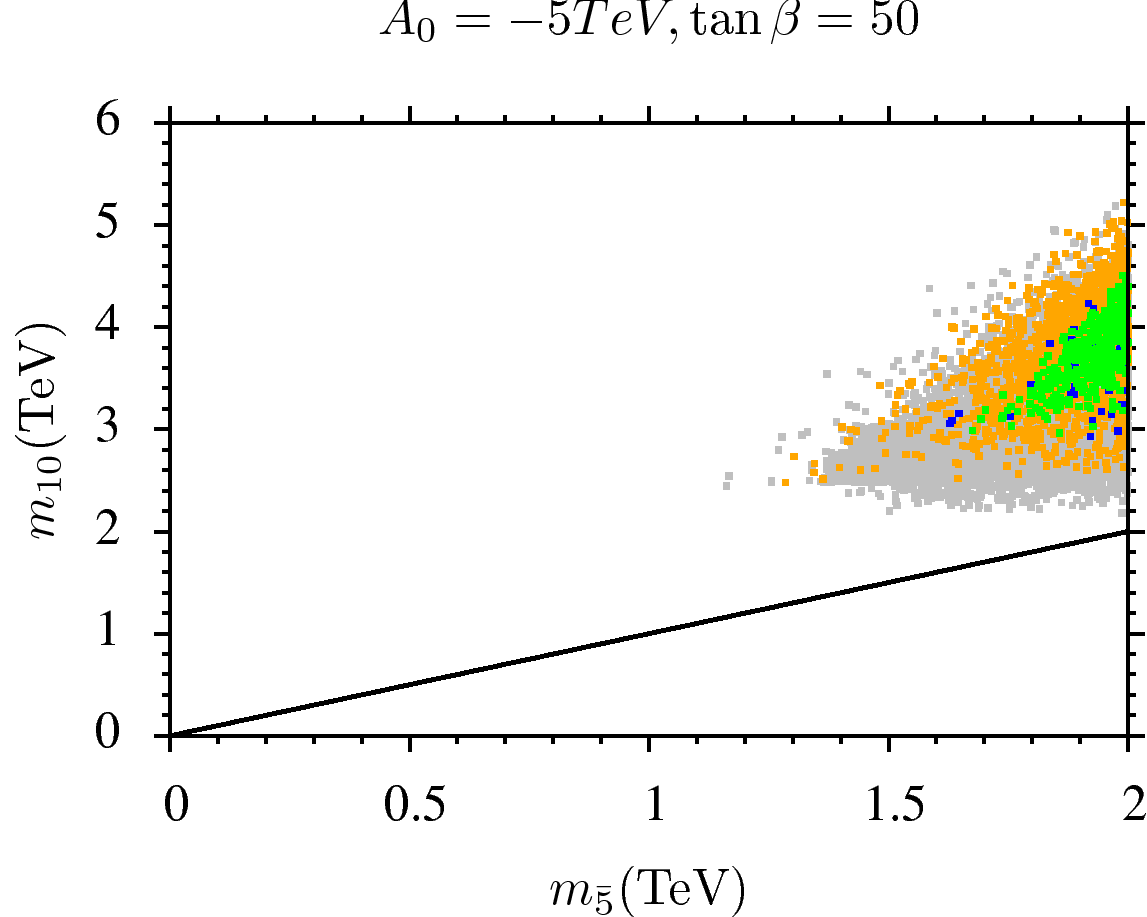}
}
\subfigure{
\includegraphics[width=7cm]{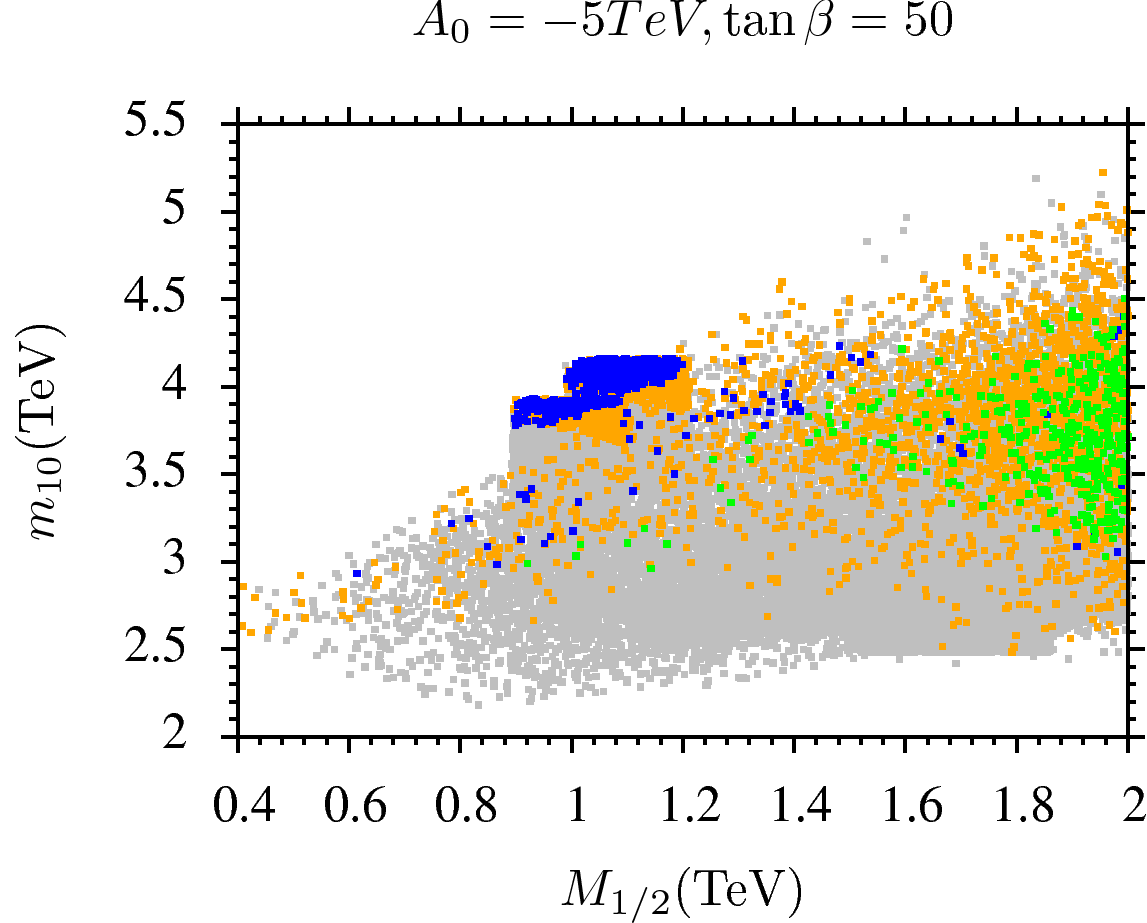}
}\\
\subfigure{
\includegraphics[width=7cm]{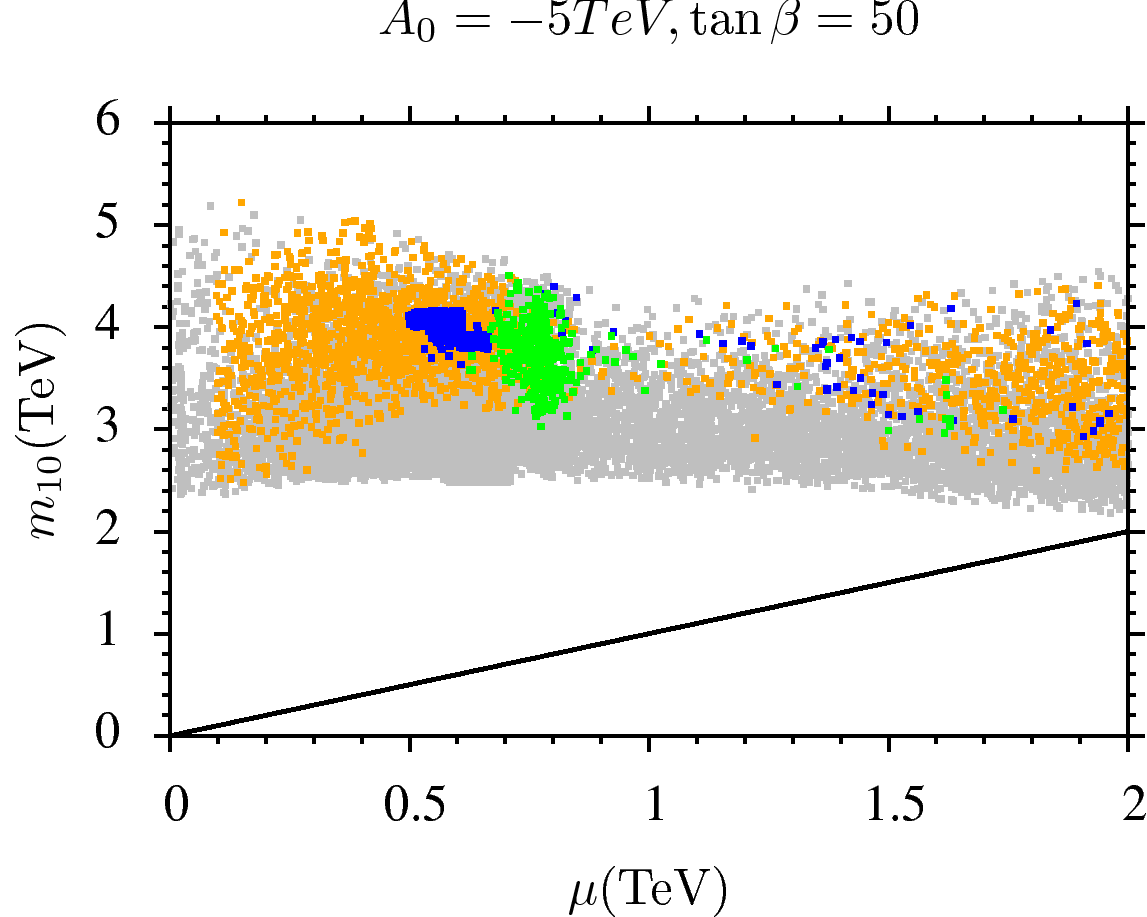}
}
\subfigure{
\includegraphics[width=7cm]{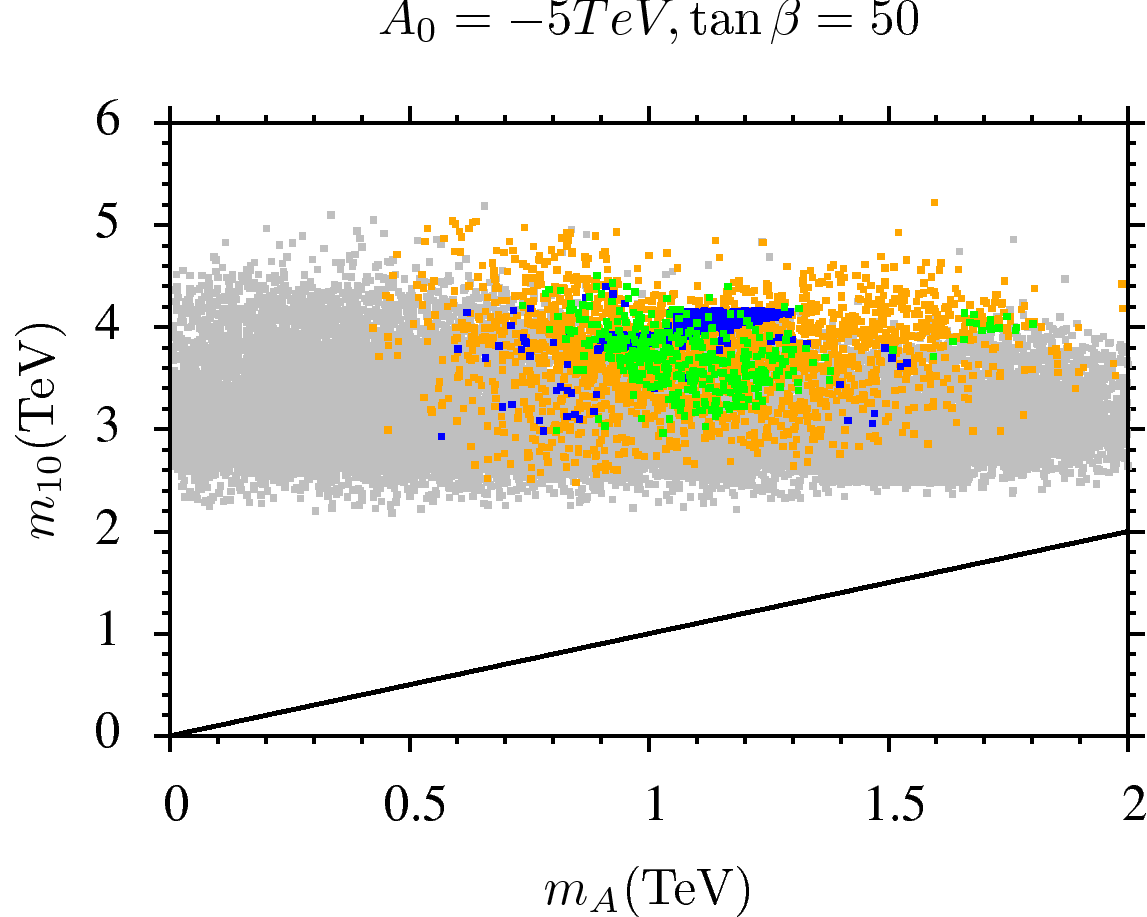}
}
\caption{
Color coding same as in Fig.~\ref{An3TeVtanb30funda},
 for $A_0=-5$ TeV.
}
\label{An5TeVtanb50funda}
\end{figure}

\begin{figure}
\centering
\subfiguretopcaptrue

\subfigure{
\includegraphics[width=7cm]{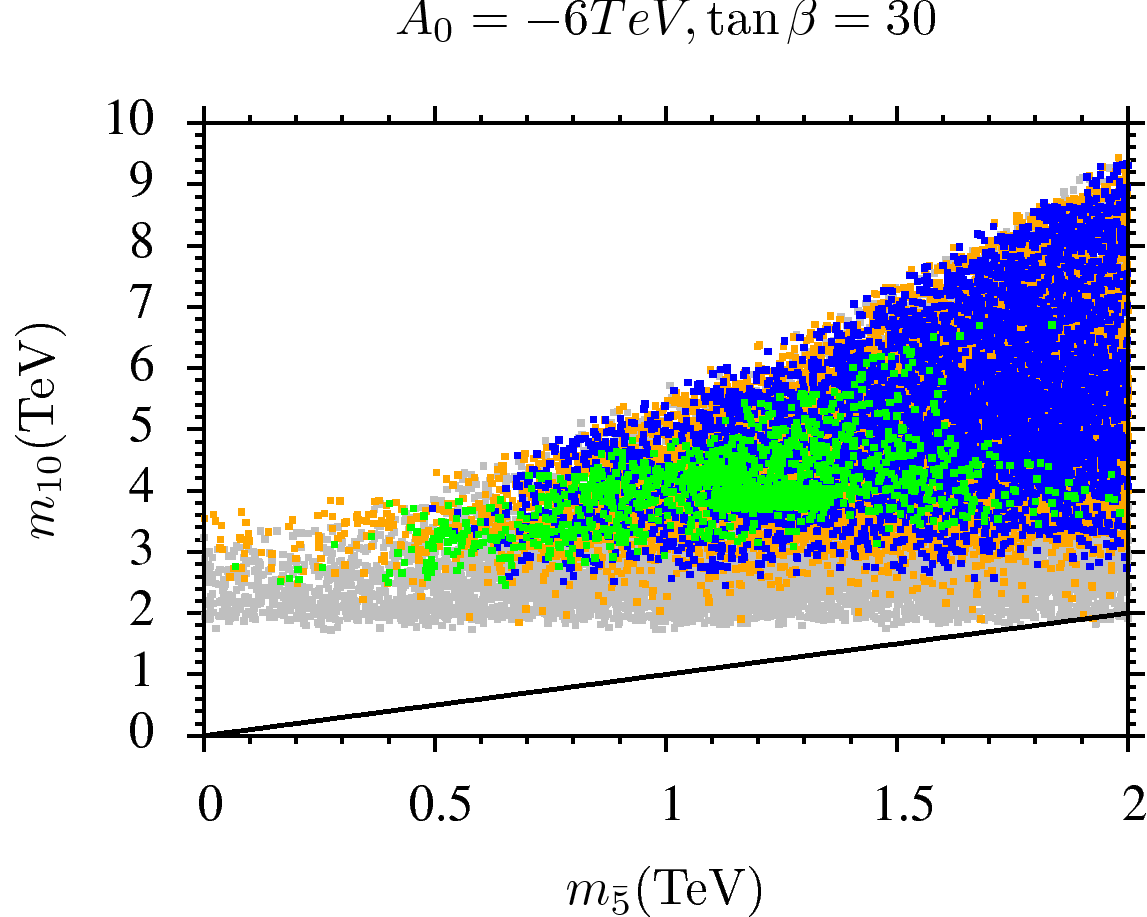}
}
\subfigure{
\includegraphics[width=7cm]{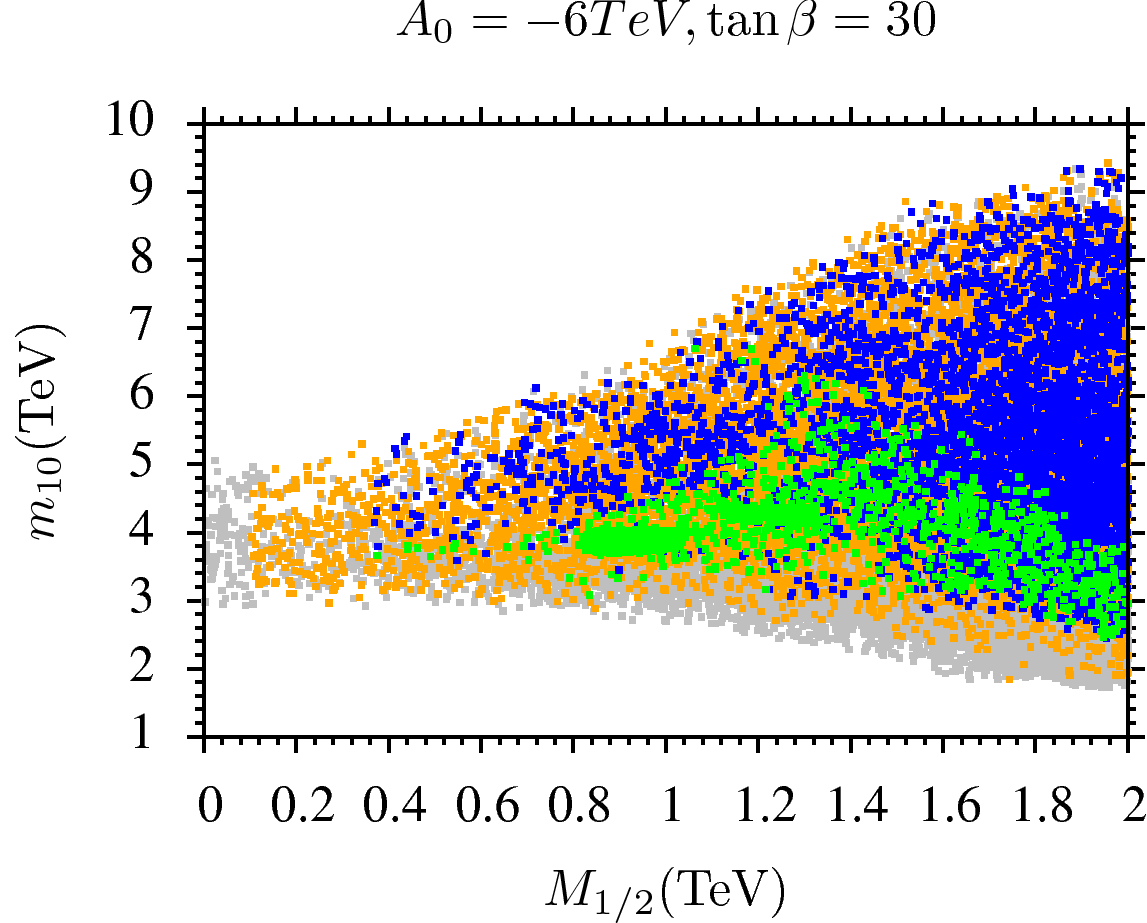}
}\\
\subfigure{
\includegraphics[width=7cm]{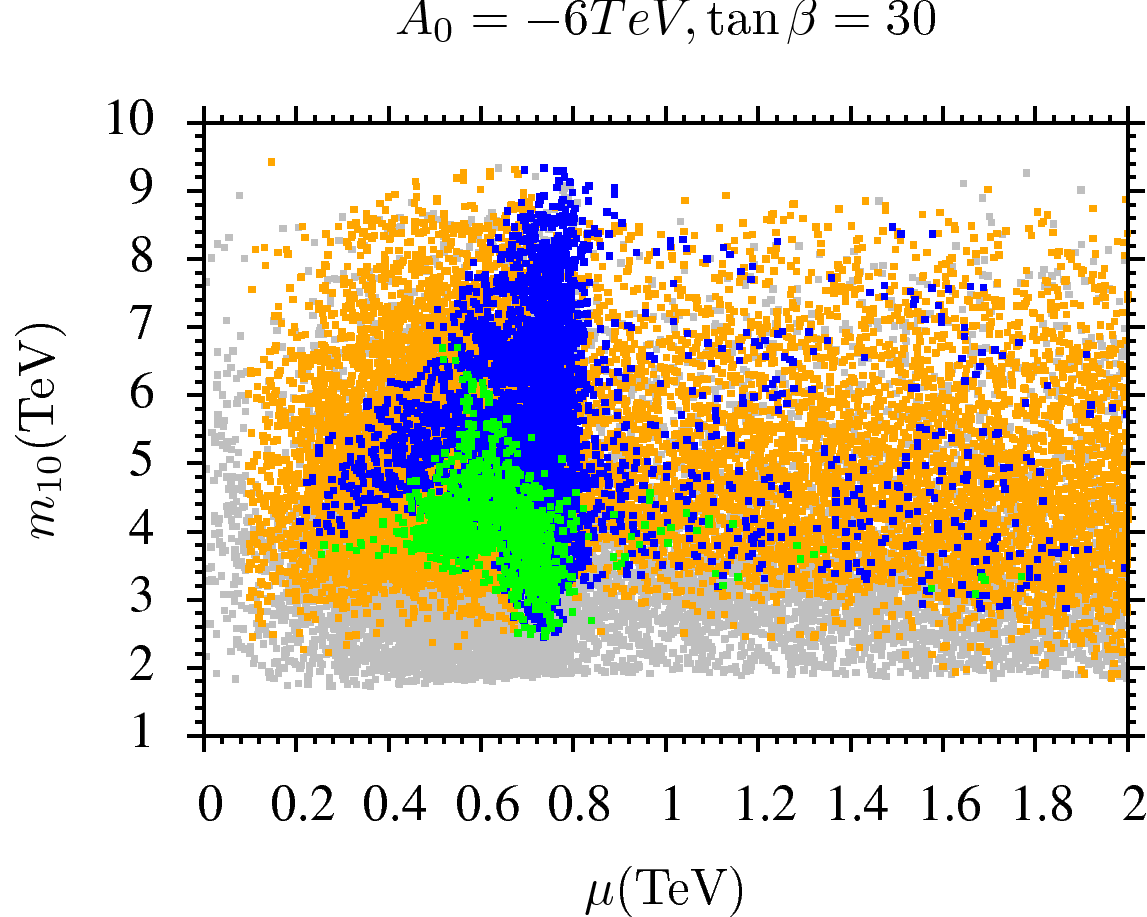}
}
\subfigure{
\includegraphics[width=7cm]{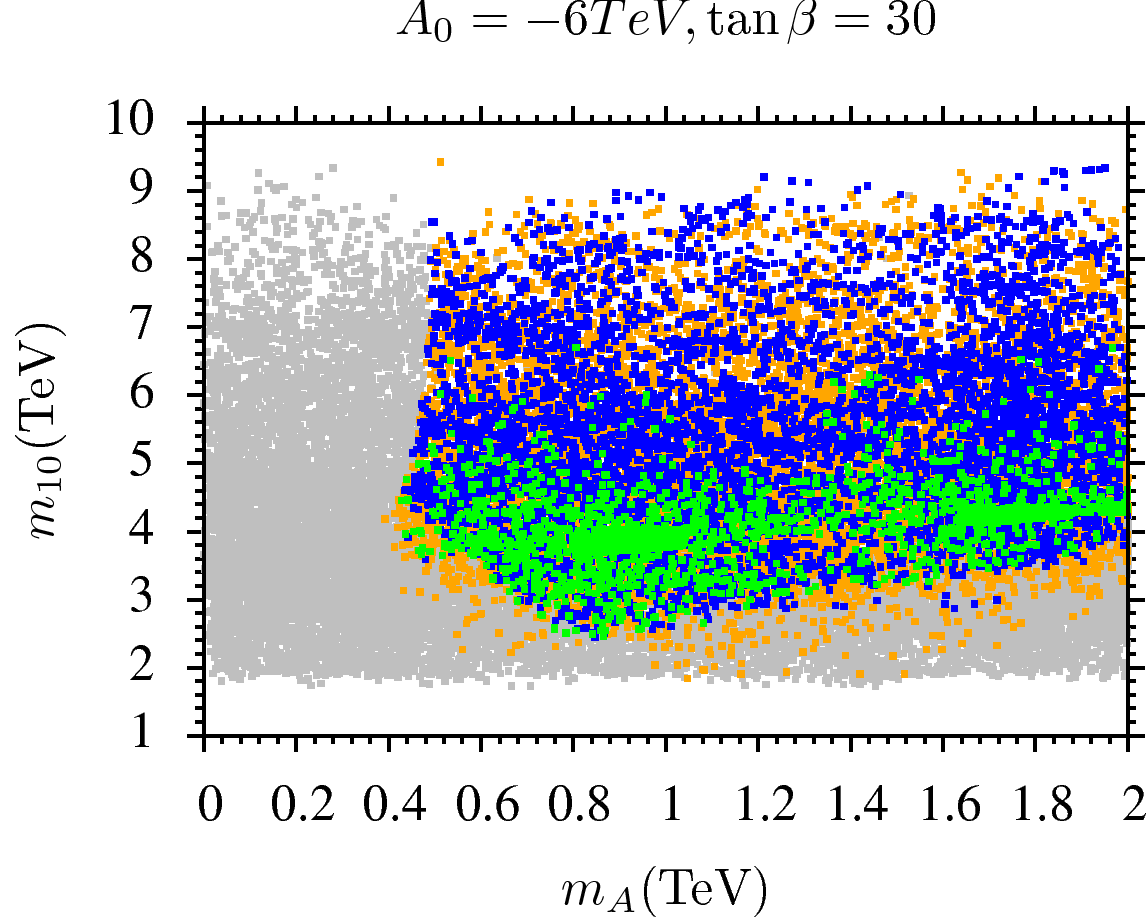}
}
\caption{
Color coding same as in Fig.~\ref{An3TeVtanb30funda}, 
 for $A_0=-6$ TeV. 
}
\label{An6TeVtanb30funda}
\end{figure}

In Fig.~\ref{An3TeVtanb30funda}, we present our results 
 in the fundamental parameter space 
 ($m_{\bar 5}$, $m_{10}$, $M_{1/2}$, $\mu$ and $m_A$) 
 for $A_0=-3$ TeV and $\tan\beta=30$. 
Gray points satisfy the requirements of REWSB and neutralino LSP. 
Orange points satisfy, in addition, the mass bounds and B-physics bounds. 
Blue points are a subset of the orange points and represent 
 solutions satisfying the WMAP9 bounds at 5$\sigma$,  
 $124\,{\rm GeV} \leq m_h \leq 127\,{\rm GeV}$, 
 and $m_{\tilde g}$, $m_{\tilde d_R} \geq$ 1TeV. 
Green points are a subset of the blue points 
 and  satisfy the 3$\sigma$ bound on $\Delta a_{\mu}$. 
In the top-left panel, 
 the black solid line represents the relation $m_{\bar 5}=m_{10}$. 
We see that most of the green points appear 
 in the region $m_{\bar 5} < m_{10}$. 
In other words, it is quite difficult to stay within 
 the 3$\sigma$ bound of $\Delta a_{\mu}$ 
 in the context of CMSSM/NUHM2. 
The green points appear for 
 $m_{\bar 5} \lesssim 1.6$ TeV,
 $1 \; {\rm TeV} \lesssim m_{10} \lesssim 2$ TeV, 
 $1.3 \; {\rm TeV} \lesssim M_{1/2}$, 
 $0.6 \; {\rm TeV} \lesssim \mu \lesssim 0.9$ TeV 
 and $0.8 \; {\rm TeV} \lesssim m_A$. 
The results for different values of $A_0=-4$, $-5$, $-6$ TeV 
 are depicted in Figs.~\ref{An4TeVtanb30funda}-\ref{An6TeVtanb30funda}. 
We see that our results are more or less similar 
 to those in Fig.~\ref{An3TeVtanb30funda}. 
However, note that for $A_0=-4$, $-5$, $-6$ TeV, 
 the condition $m_{\bar 5} < m_{10}$ is necessary 
 to have green points. 
In all figures, the Higgsino mass parameter is always found 
 to be relatively small, namely $\mu = 0.6-0.8$ TeV 
 for most of the green points.

\begin{figure}
\centering
\subfiguretopcaptrue

\subfigure{
\includegraphics[width=7cm]{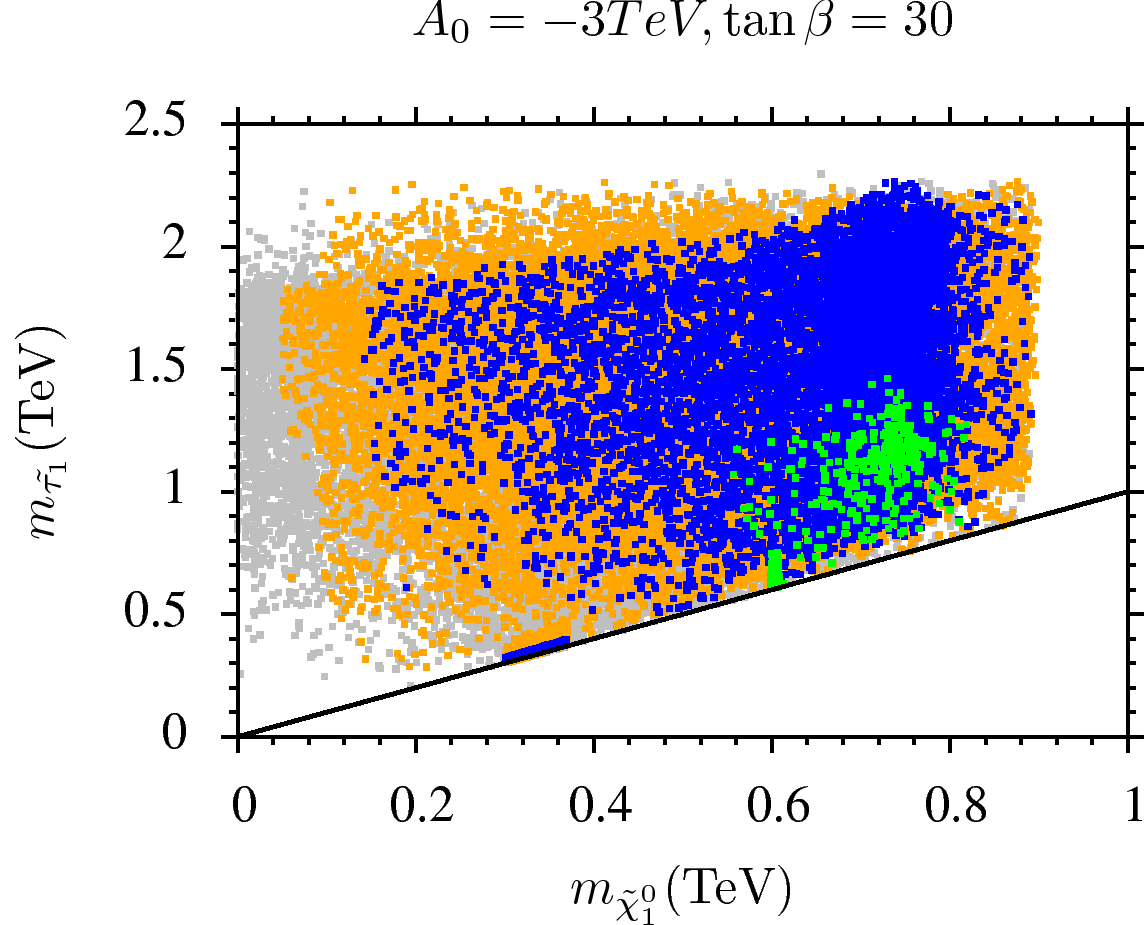}
}
\subfigure{
\includegraphics[width=7cm]{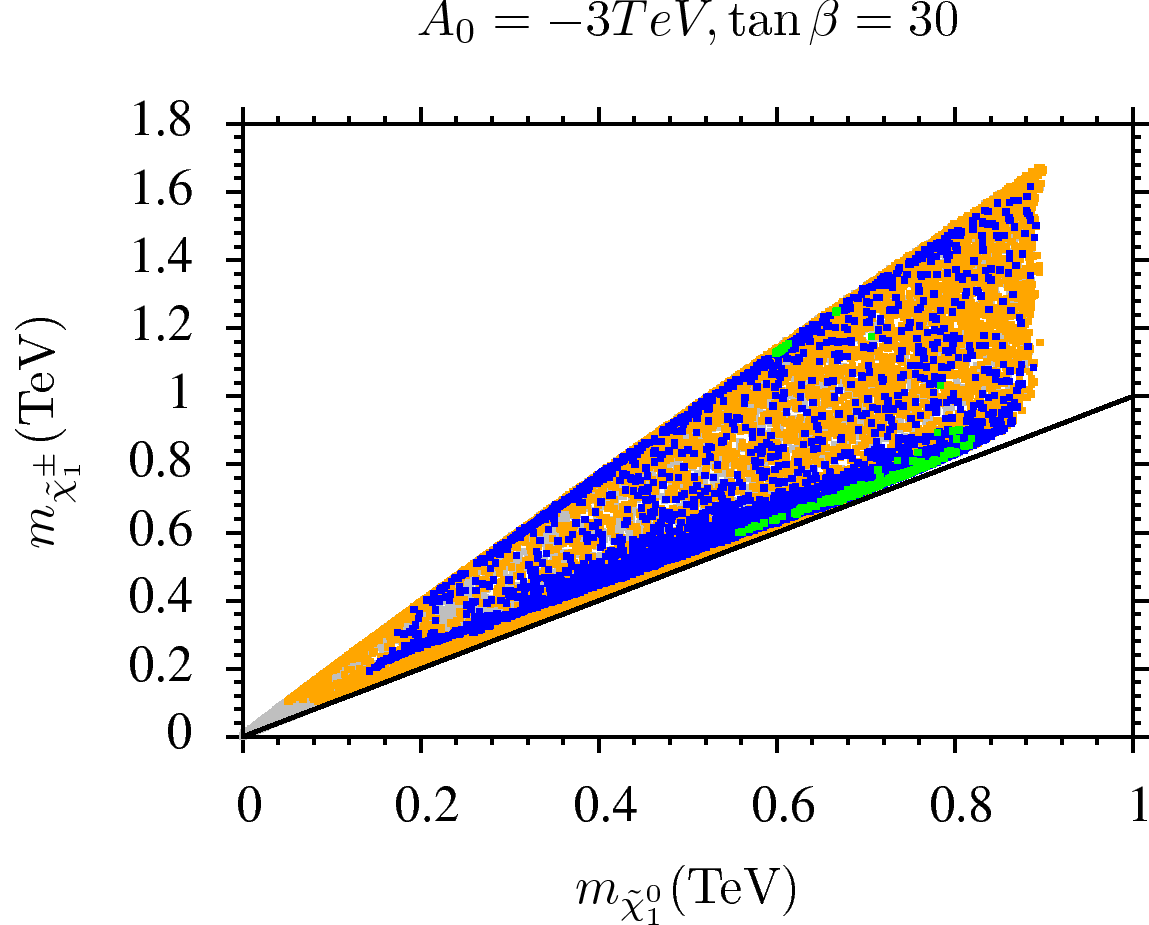}
}\\
\subfigure{
\includegraphics[width=7cm]{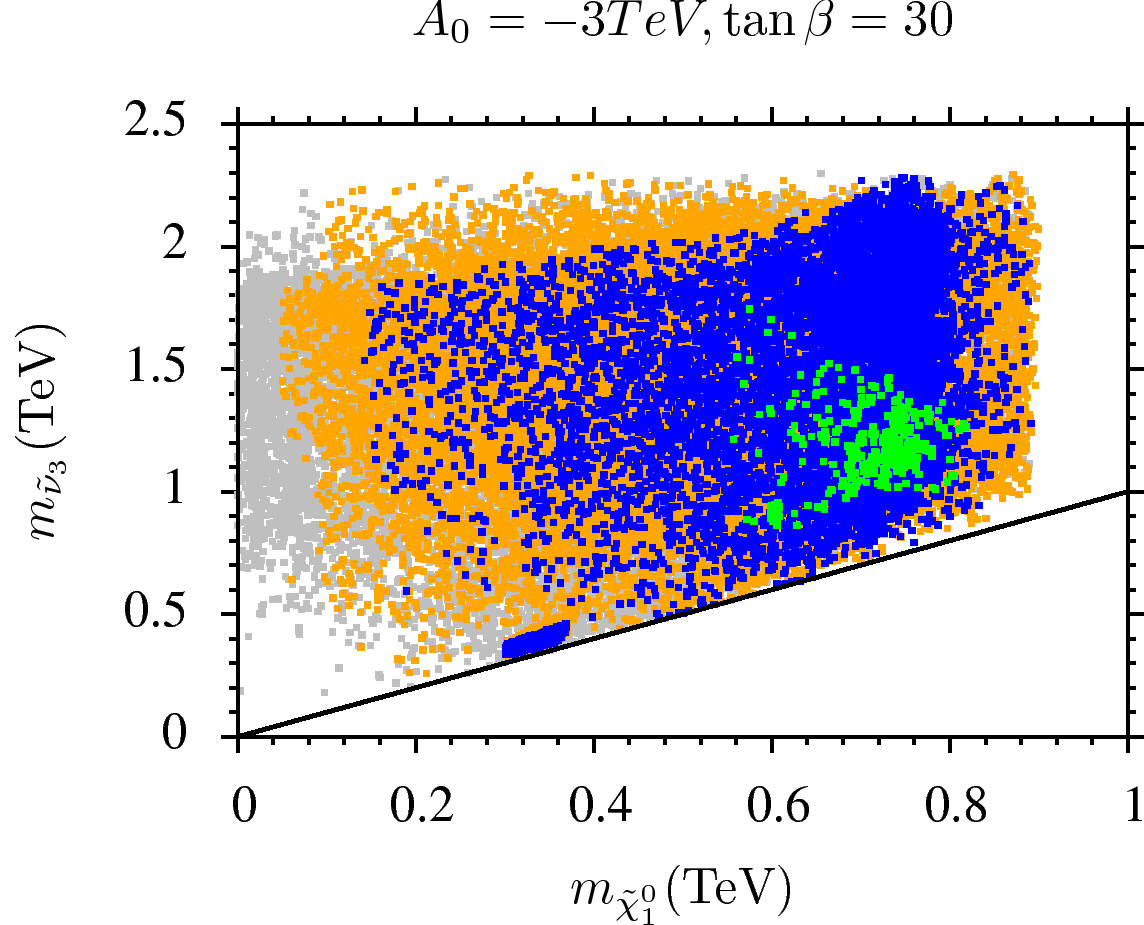}
}
\subfigure{
\includegraphics[width=7cm]{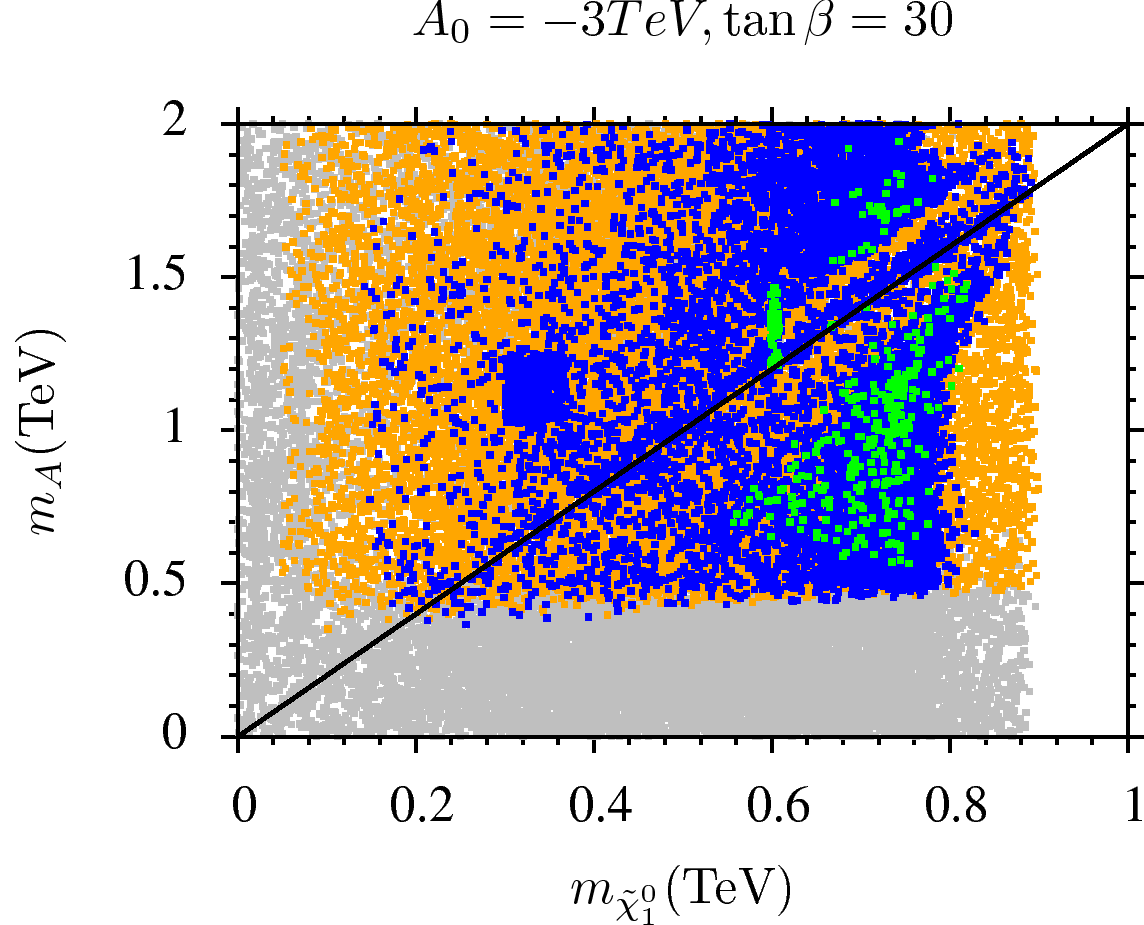}
}
\caption{
Results for $A_0=-3$ TeV and $\tan \beta=30$.
Color coding same as in Fig.~\ref{An3TeVtanb30funda}.
}
\label{An3TeVtanb30spectrum}
\end{figure}

In Fig.~\ref{An3TeVtanb30spectrum} 
 we present our results for $A_0=-3$ TeV and $\tan \beta =30$ 
 in terms of sparticle masses: LSP neutralino $m_{\tilde {\chi}^0_1}$,  
 lighter stau $m_{\tilde \tau_1}$, 
 lighter chargino $m_{\tilde {\chi}^{\pm}_1}$,
 lighter sneutrino $m_{\tilde \nu_3}$, and 
 CP-odd Higgs boson $m_{A}$. 
Here the color coding is the same as in Fig~\ref{An3TeVtanb30funda}. 
The solid lines represent the mass degeneracy of the LSP neutralino 
 with the lighter stau, lighter chargino and lighter sneutrino. 
In the bottom-right panel, the solid line corresponds 
 to the relation $2 m_{\tilde {\chi}^0_1}=m_A$. 
The region $0.6 \; {\rm TeV} \lesssim \mu \lesssim 0.9$ TeV 
 corresponds to $m_{\tilde {\chi}^0_1} = 0.55-0.8$ TeV, 
 and we can see from the top-right panel 
 that the neutralino LSP is closely-degenerate in mass 
 with the lighter chargino for most of the green points. 
A more careful investigation of this region 
reveals that the neutralino LSP is a bino-Higgsino admixture.

\begin{figure}
\centering
\subfiguretopcaptrue

\subfigure{
\includegraphics[width=7cm]{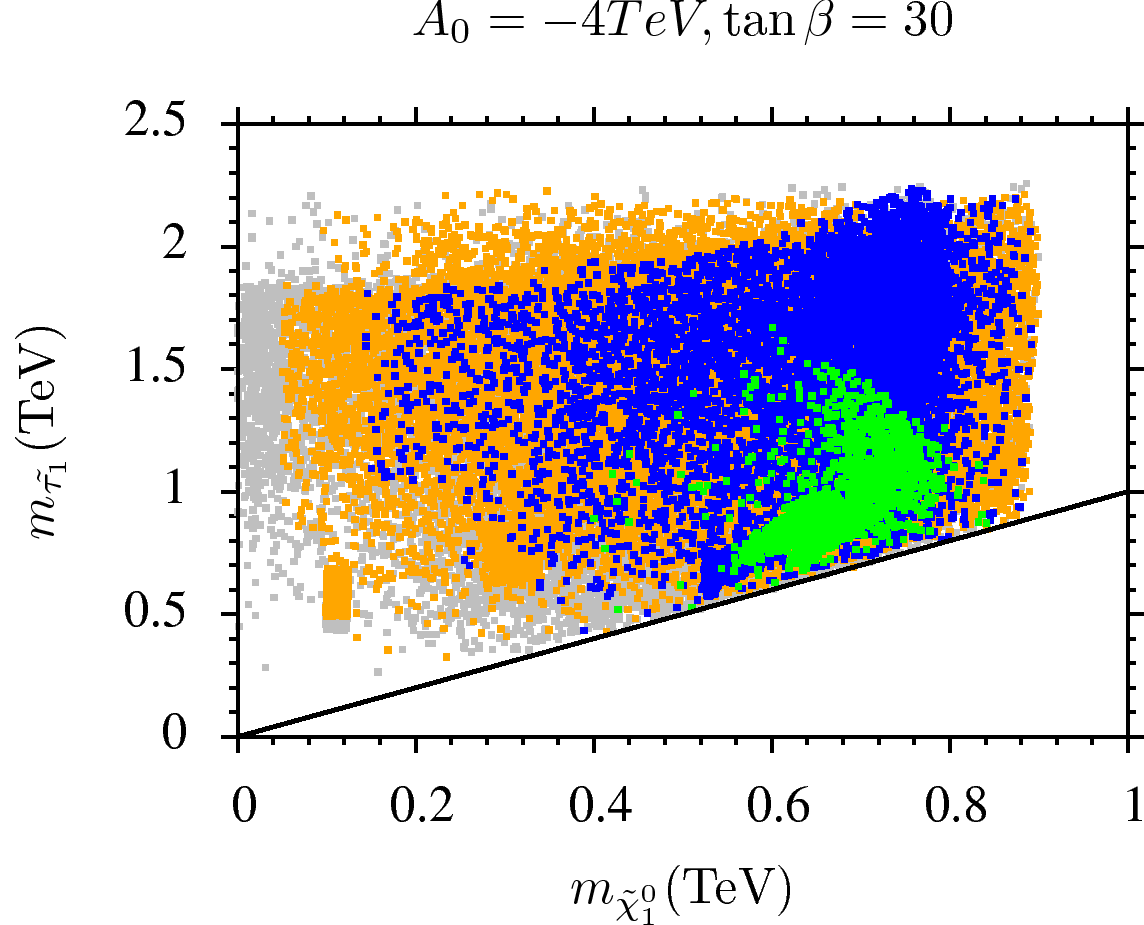}
}
\subfigure{
\includegraphics[width=7cm]{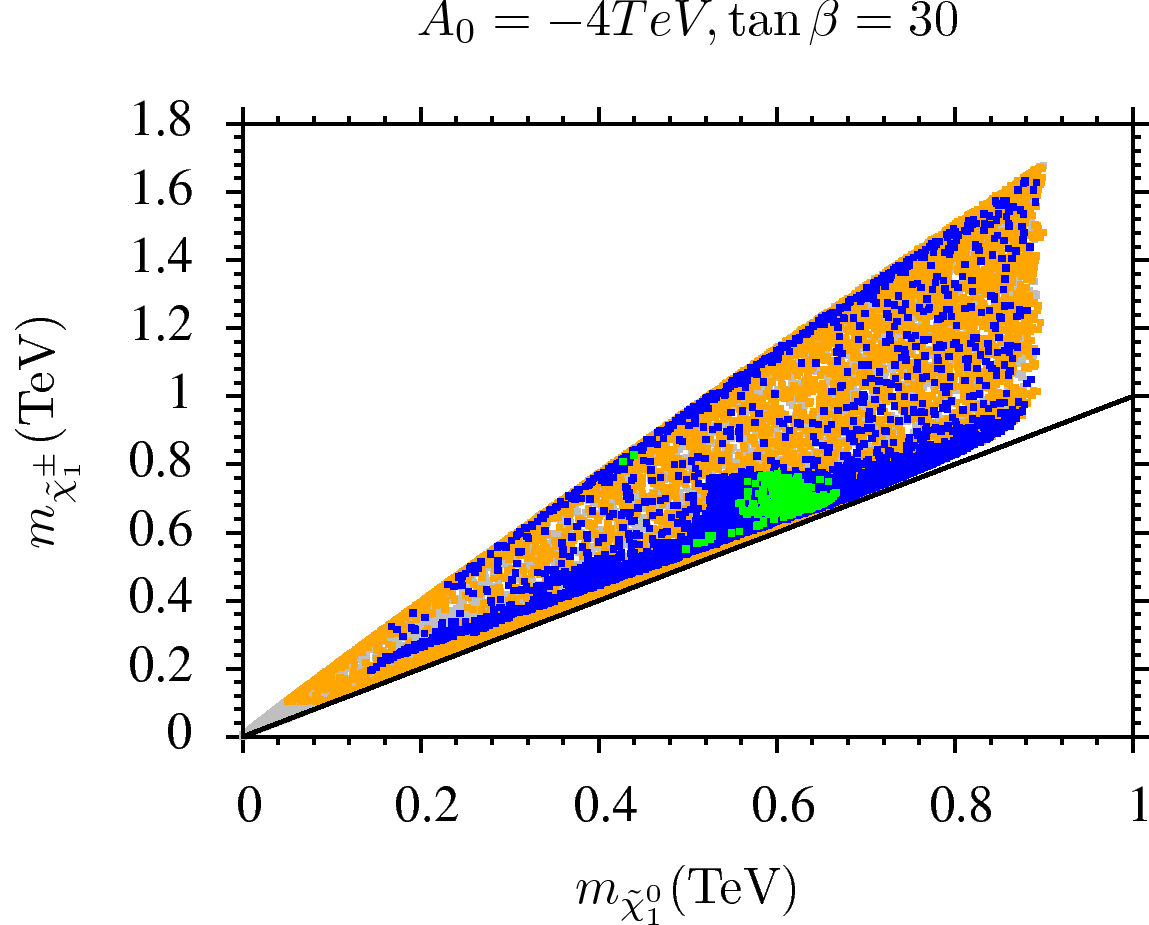}
}\\
\subfigure{
\includegraphics[width=7cm]{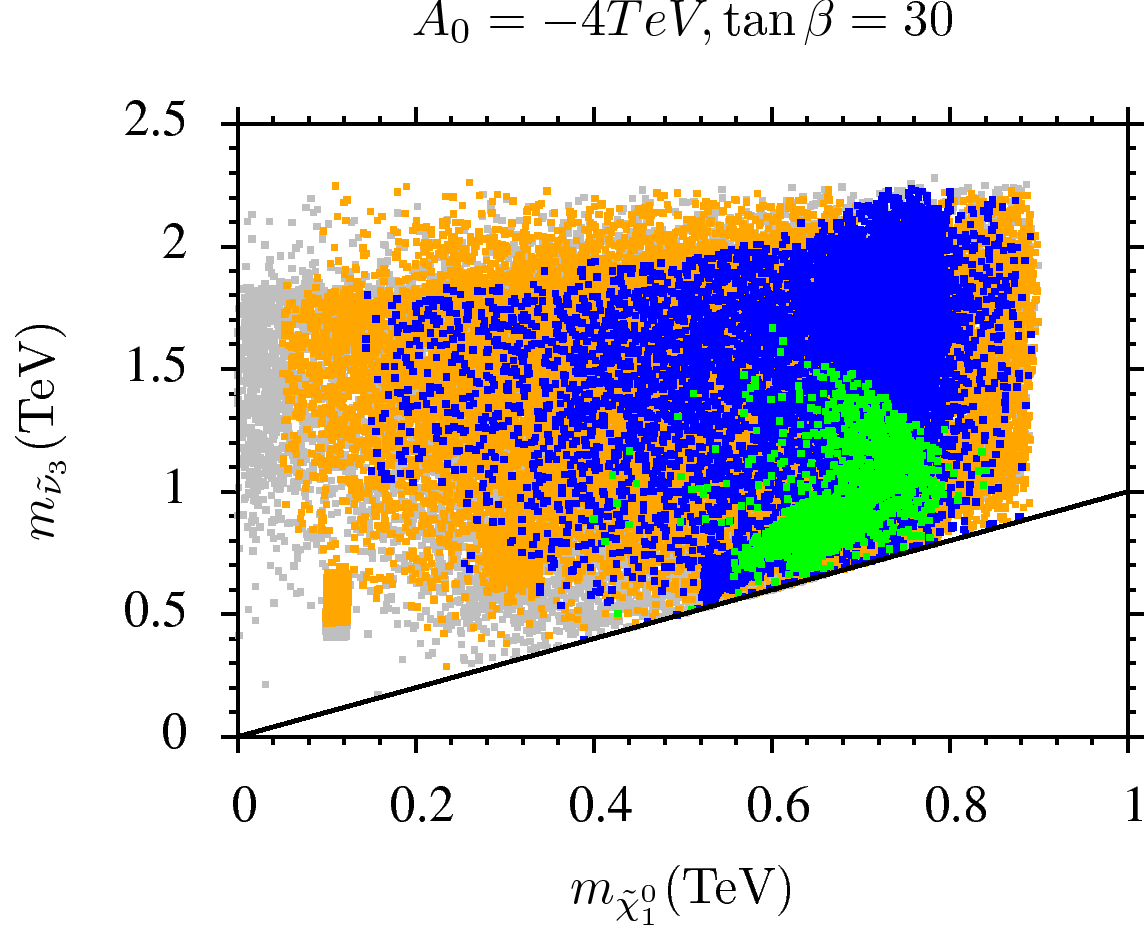}
}
\subfigure{
\includegraphics[width=7cm]{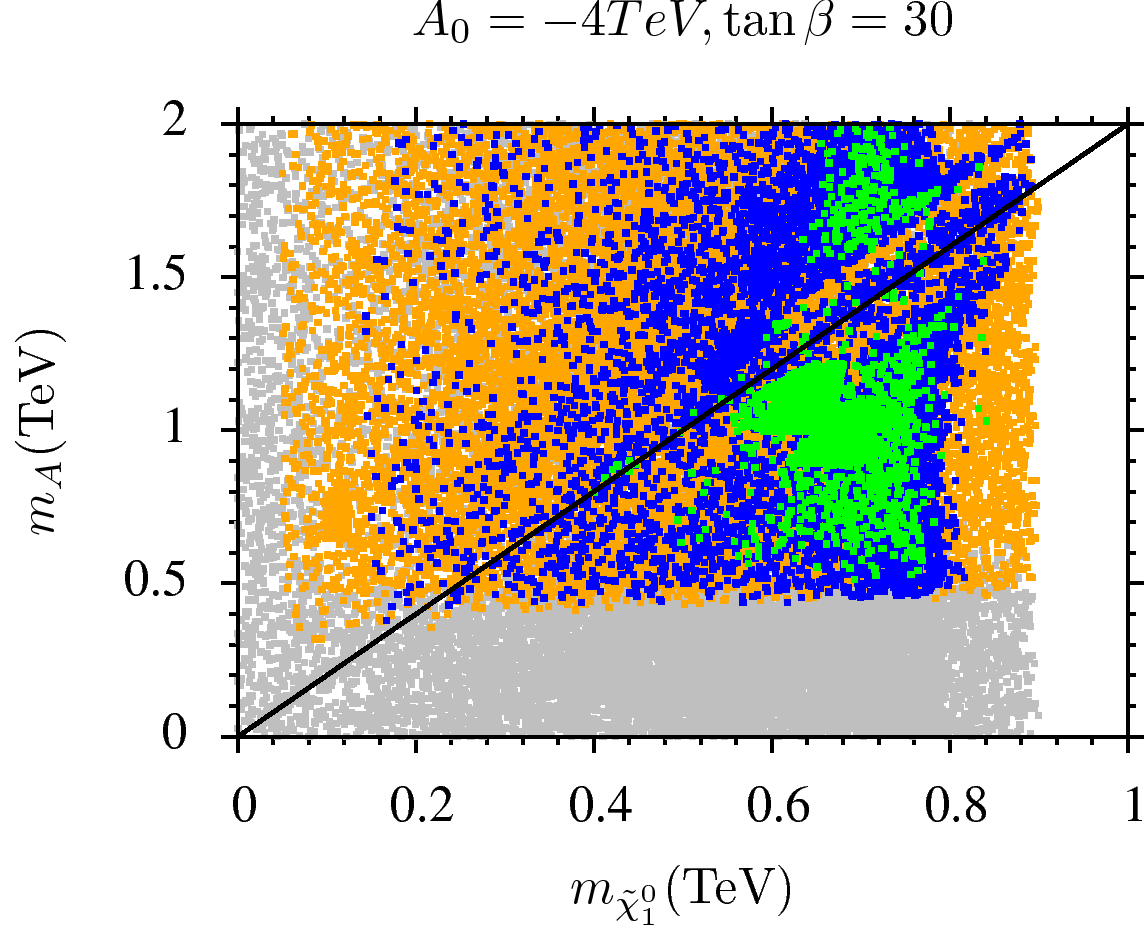}
}
\caption{
Color coding same as in Fig.~\ref{An3TeVtanb30spectrum},  
for $A_0=-4$ TeV. 
}
\label{An4TeVtanb30spectrum}
\end{figure}

\begin{figure}
\centering
\subfiguretopcaptrue

\subfigure{
\includegraphics[width=7cm]{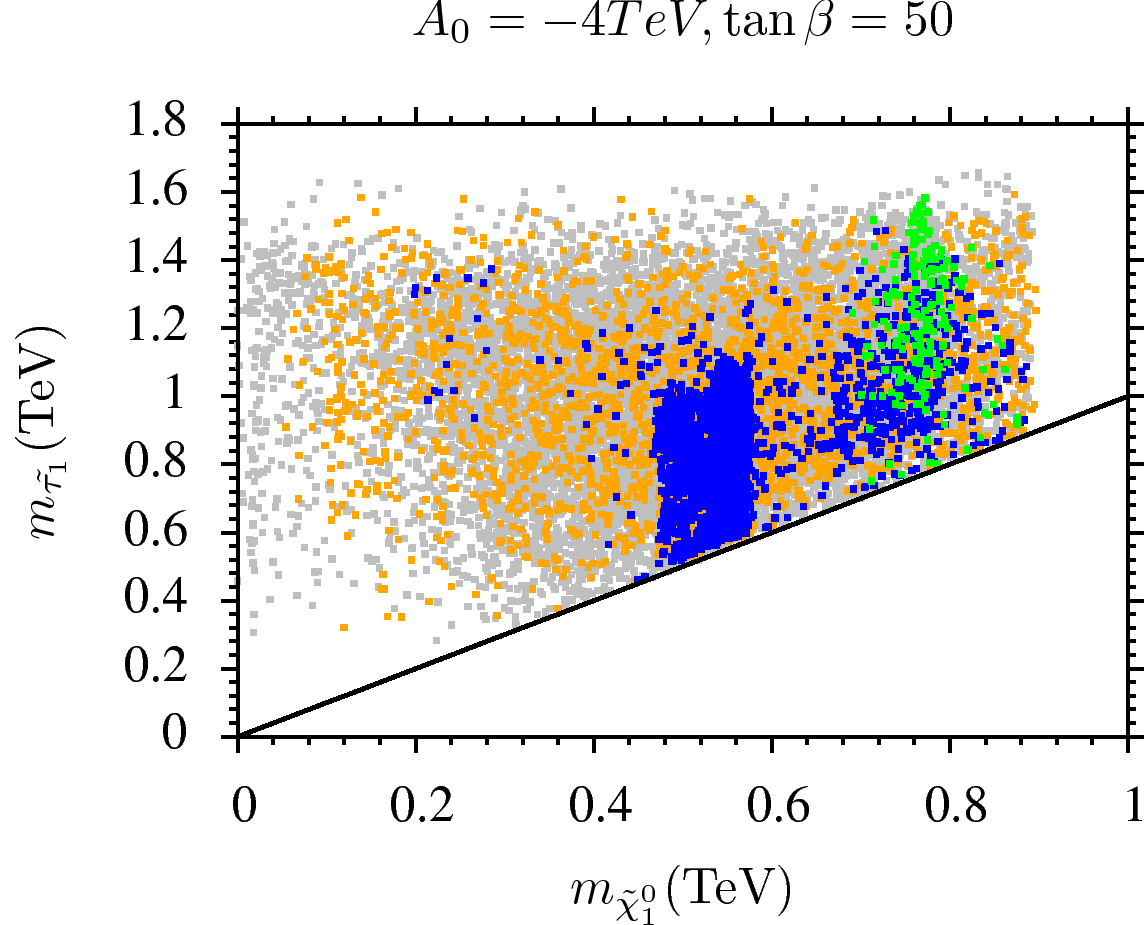}
}
\subfigure{
\includegraphics[width=7cm]{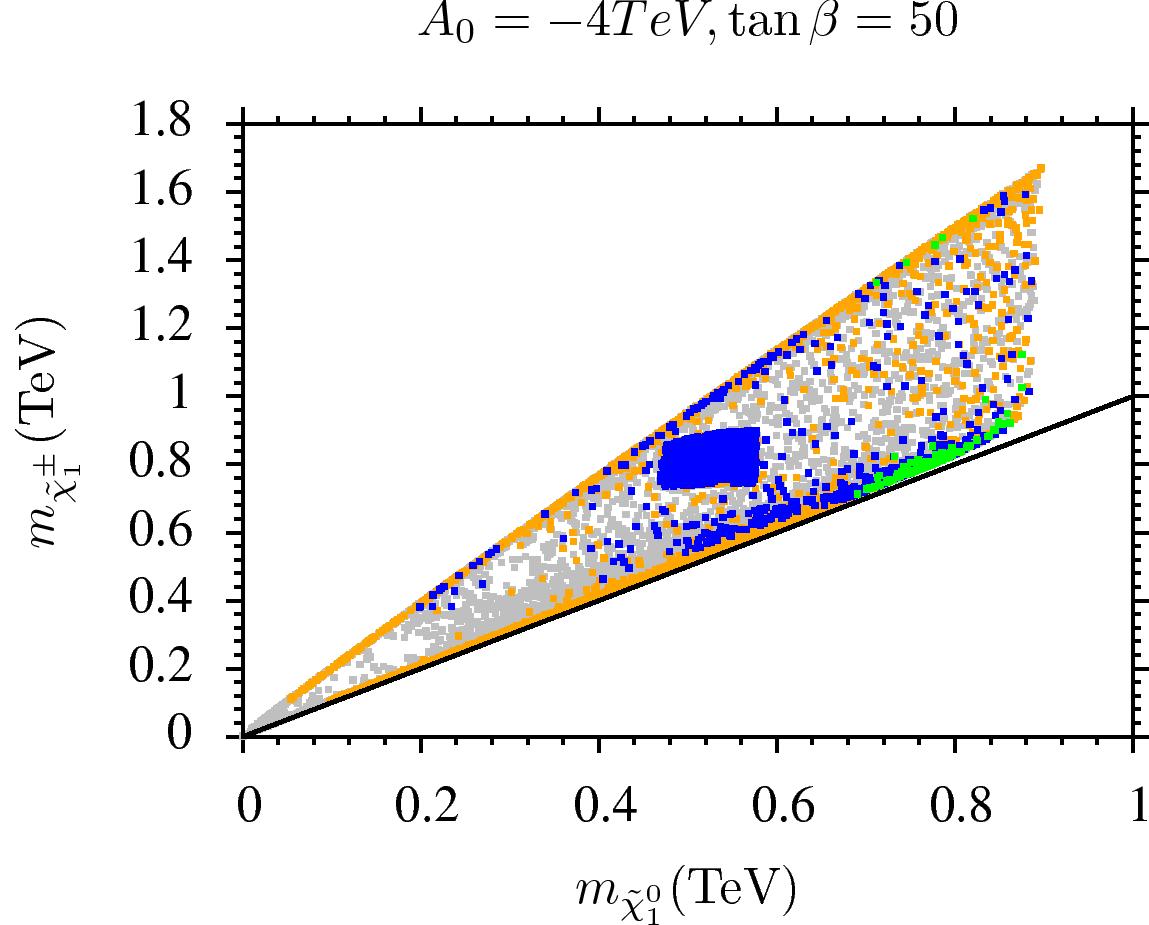}
}\\
\subfigure{
\includegraphics[width=7cm]{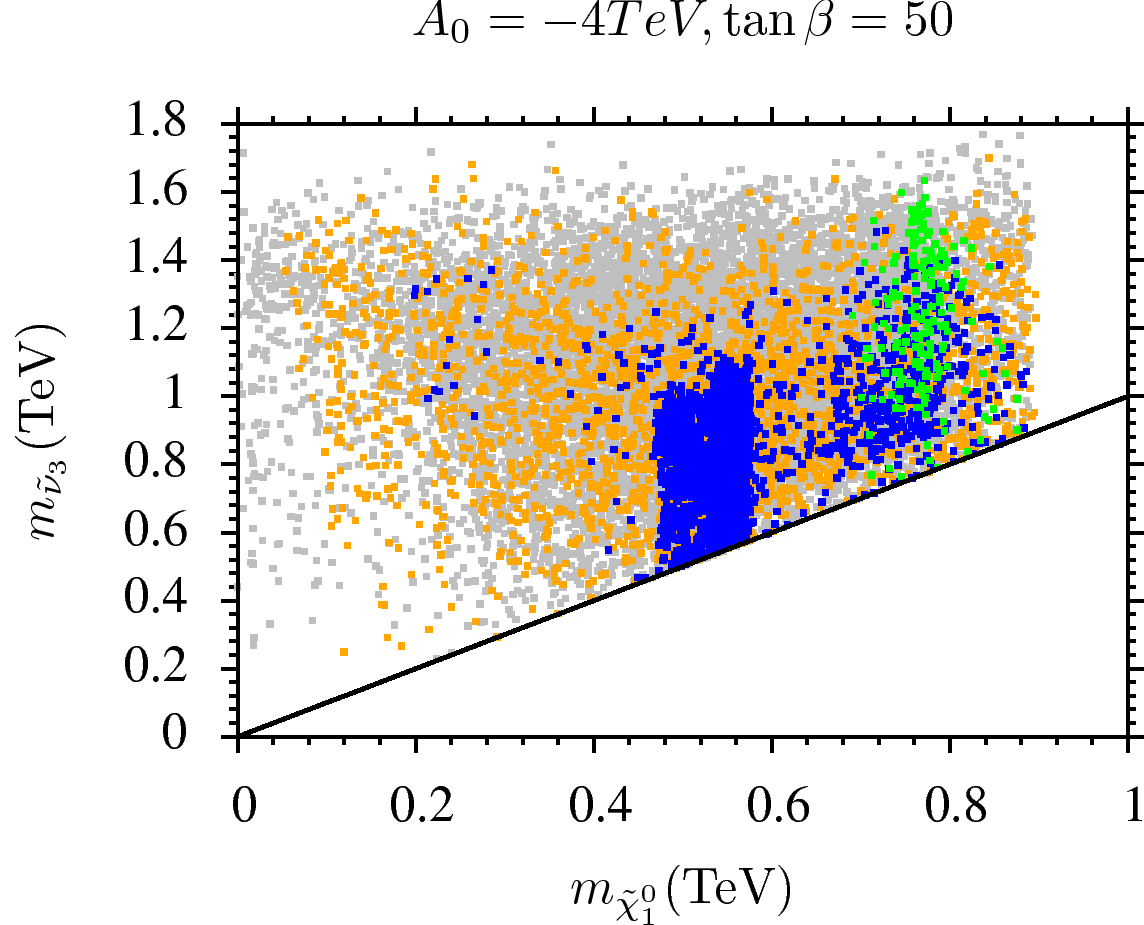}
}
\subfigure{
\includegraphics[width=7cm]{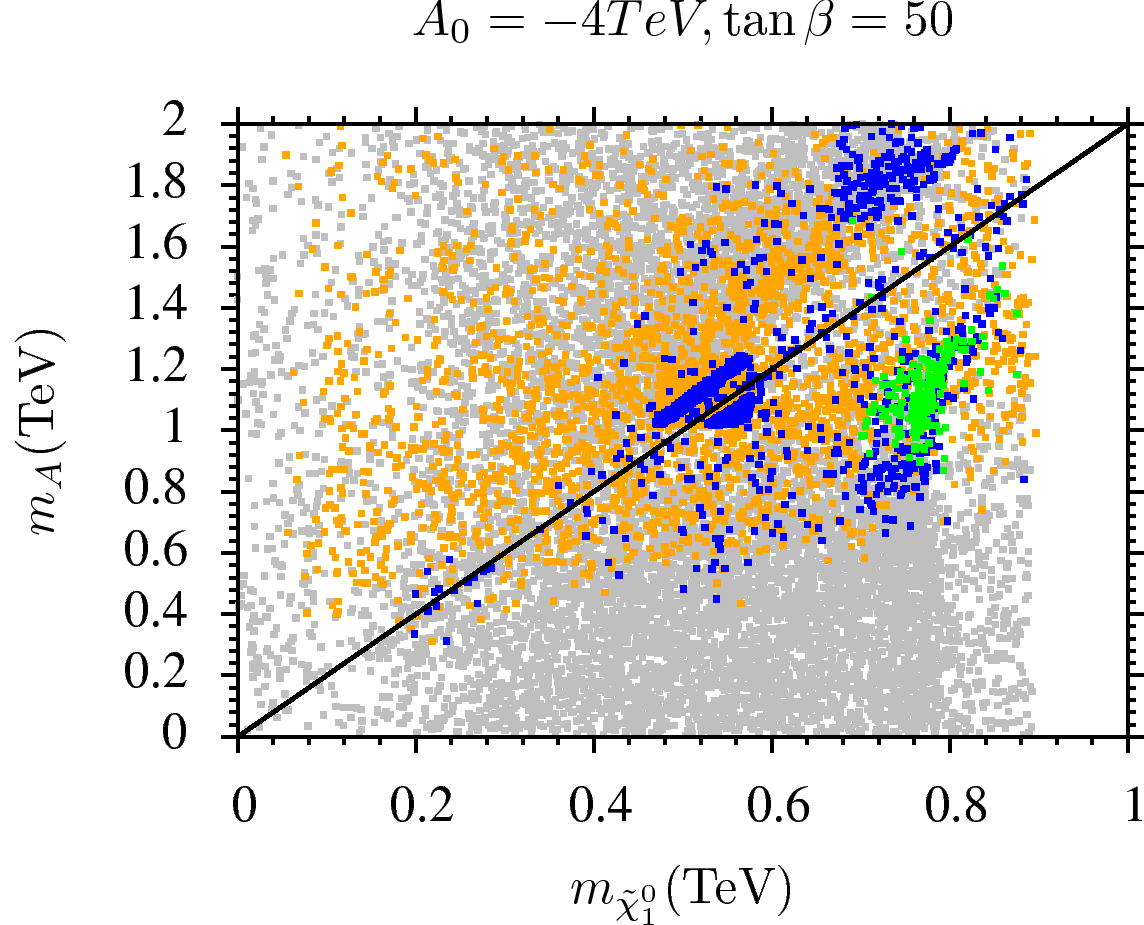}
}
\caption{
Color coding same as in Fig.~\ref{An3TeVtanb30spectrum},
for $A_0=-4$ TeV.
}
\label{An5TeVtanb50spectrum}
\end{figure}

\begin{figure}
\centering
\subfiguretopcaptrue

\subfigure{
\includegraphics[width=7cm]{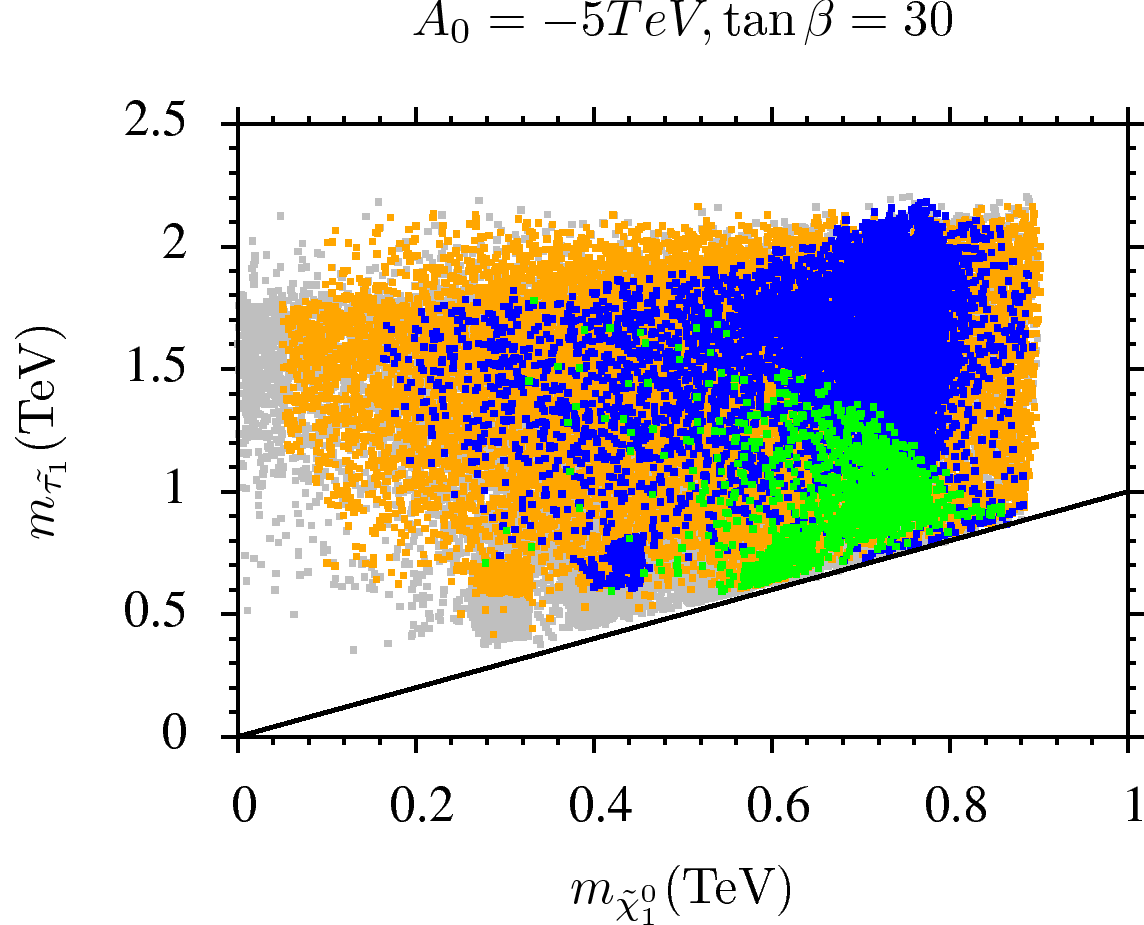}
}
\subfigure{
\includegraphics[width=7cm]{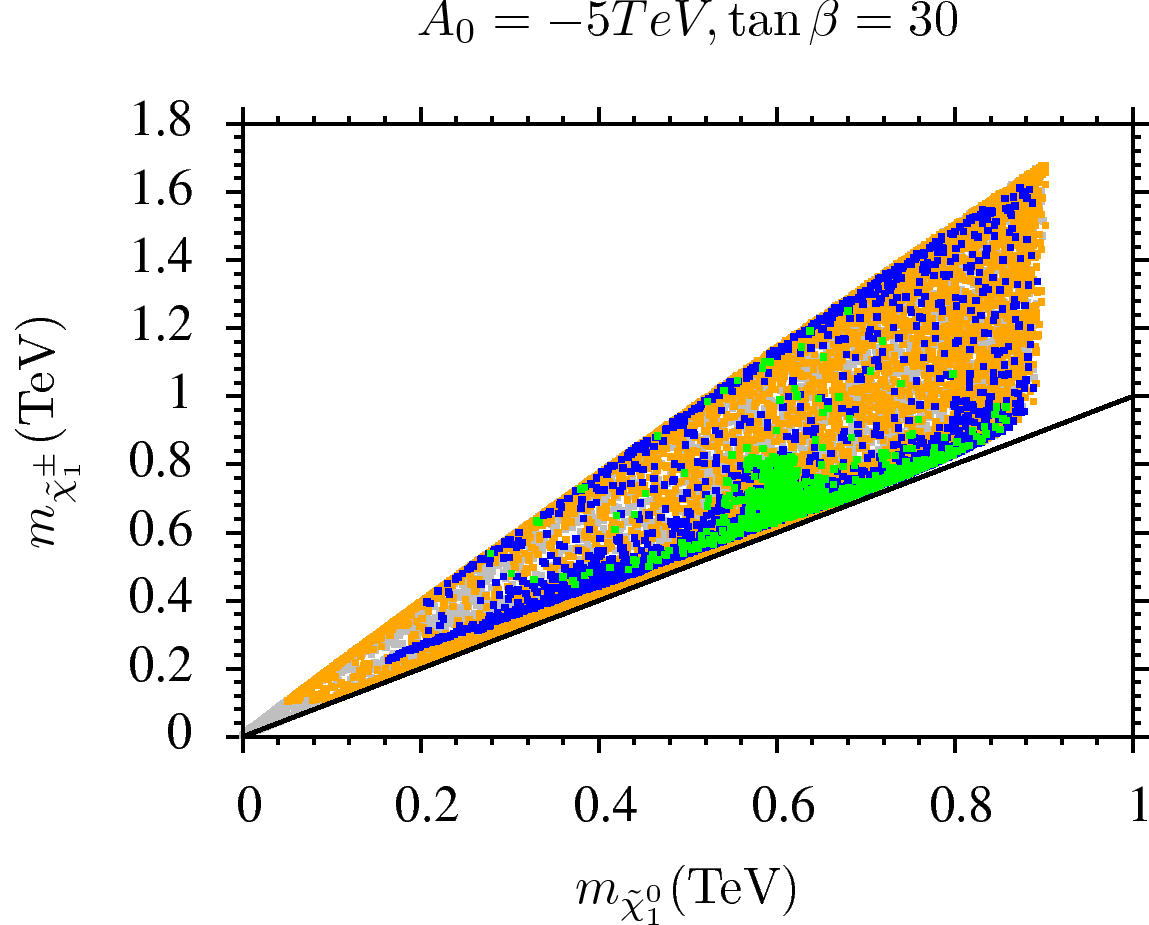}
}\\
\subfigure{
\includegraphics[width=7cm]{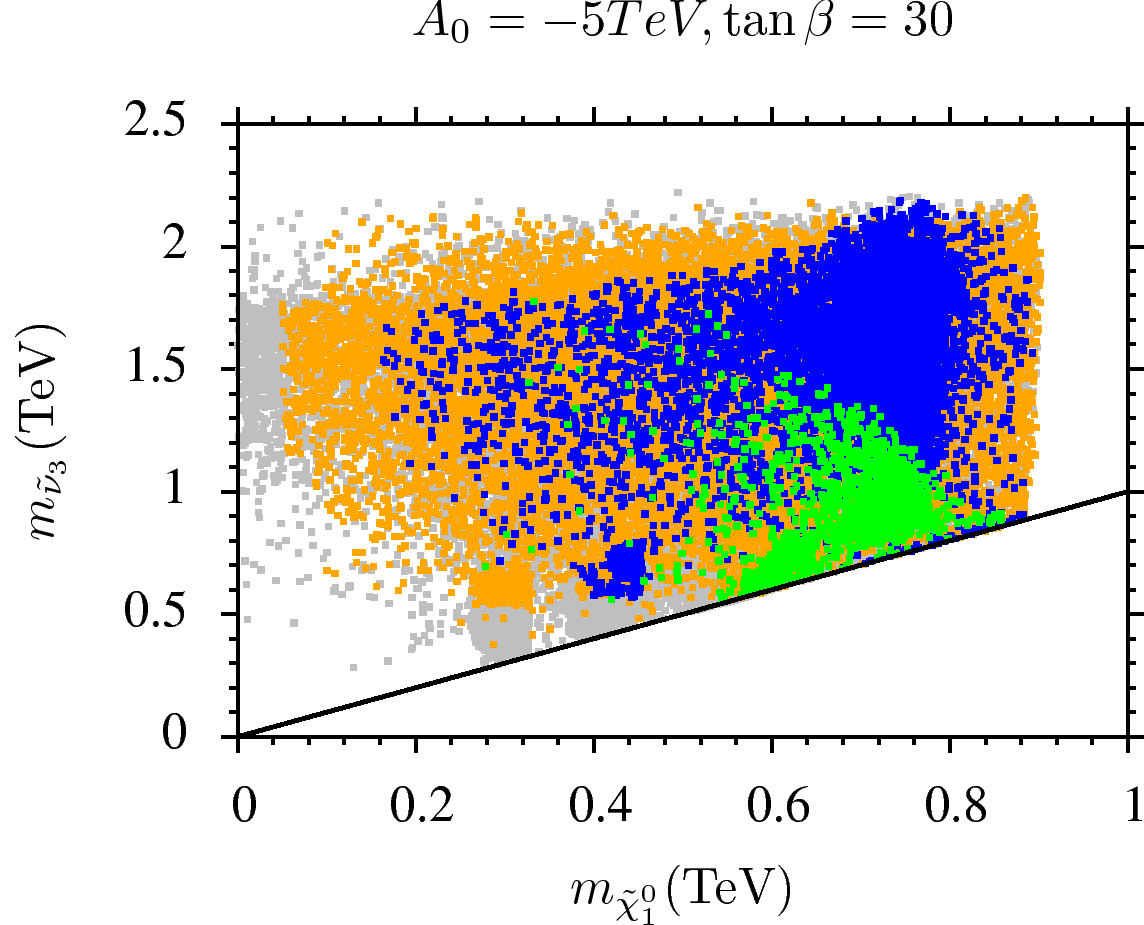}
}
\subfigure{
\includegraphics[width=7cm]{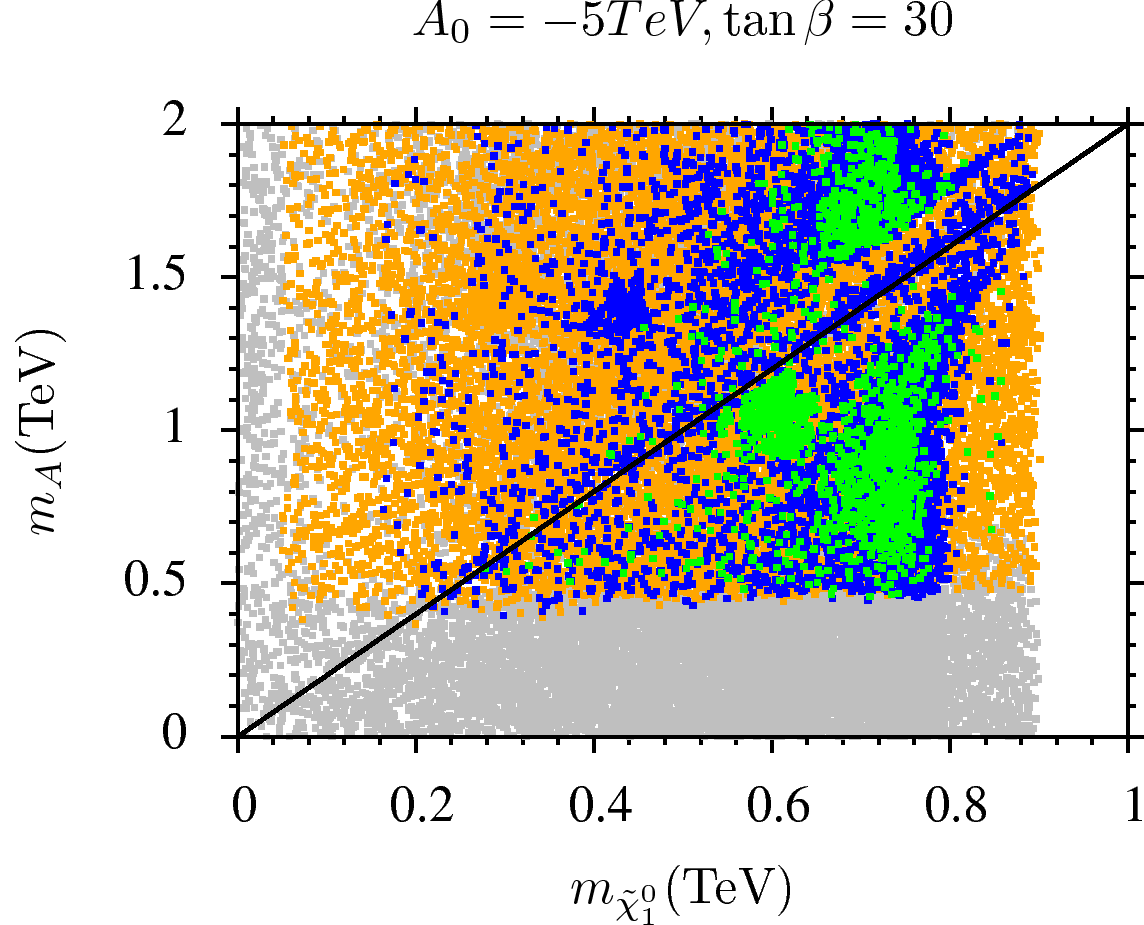}
}
\caption{
Color coding same as in Fig.~\ref{An3TeVtanb30spectrum},  
for $A_0=-5$ TeV. 
}
\label{An5TeVtan30spectrum}
\end{figure}

\begin{figure}
\centering
\subfiguretopcaptrue

\subfigure{
\includegraphics[width=7cm]{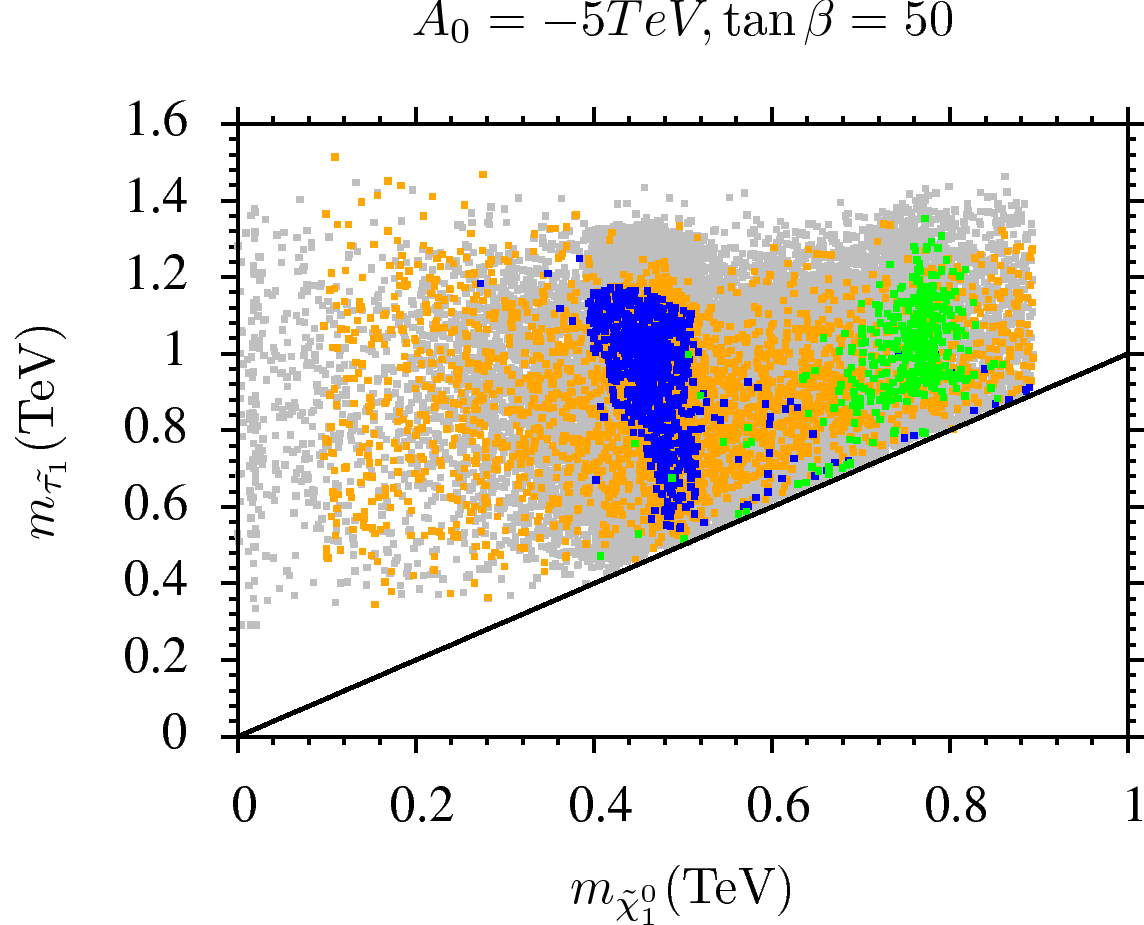}
}
\subfigure{
\includegraphics[width=7cm]{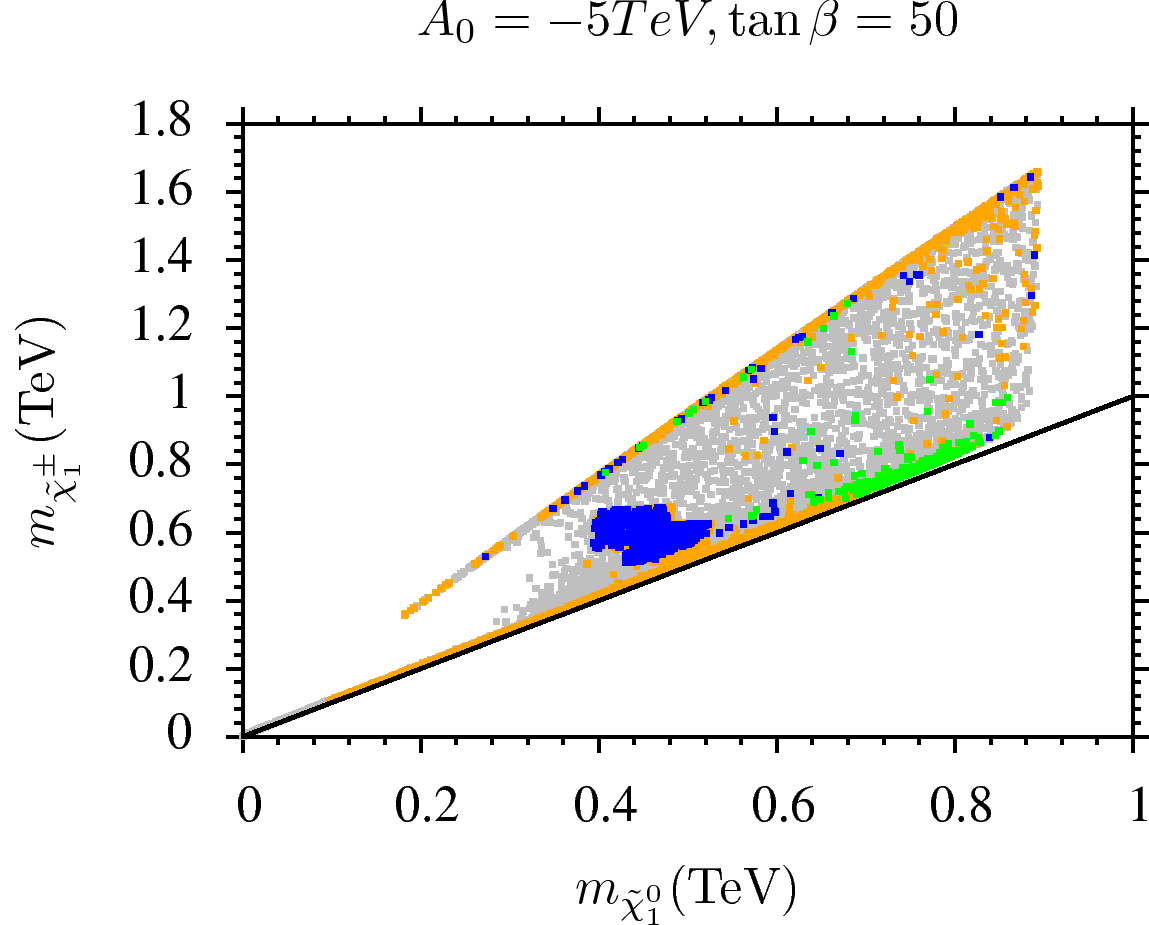}
}\\
\subfigure{
\includegraphics[width=7cm]{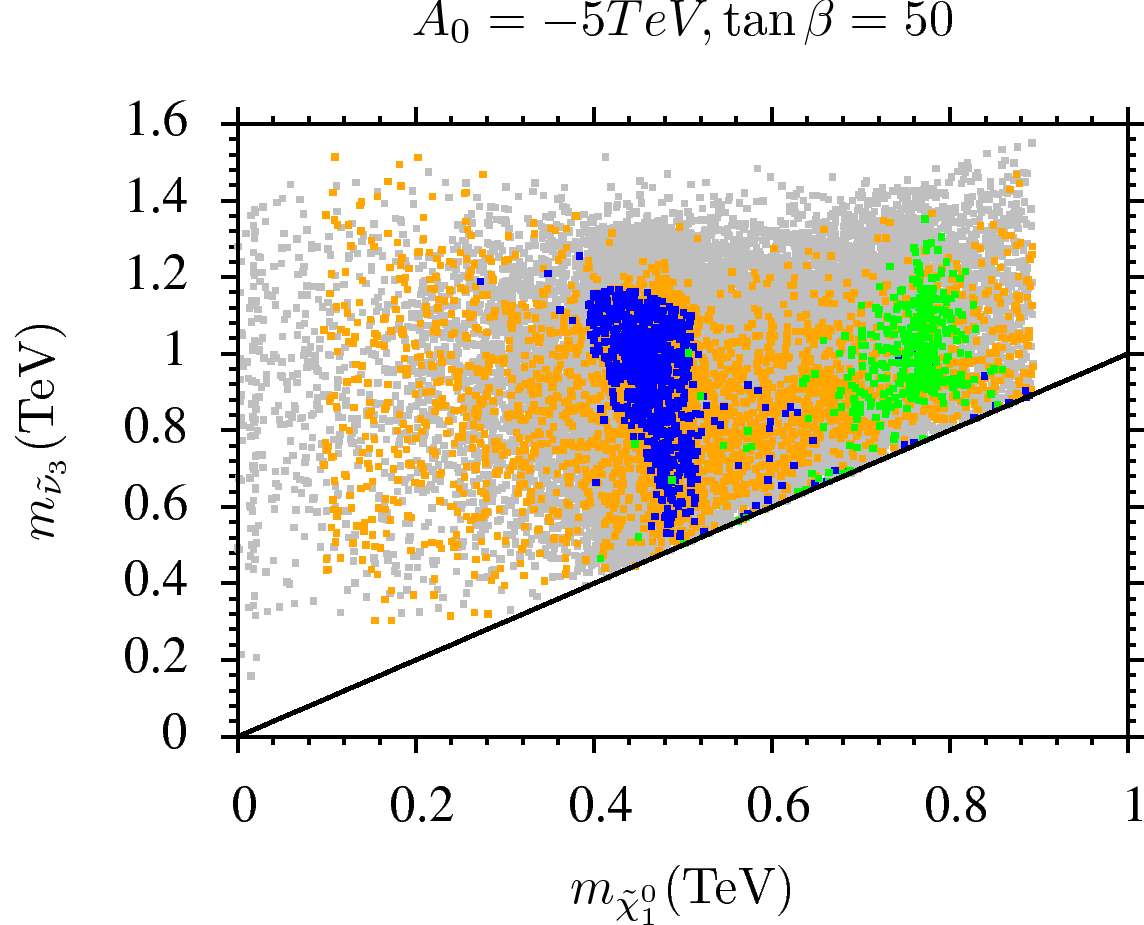}
}
\subfigure{
\includegraphics[width=7cm]{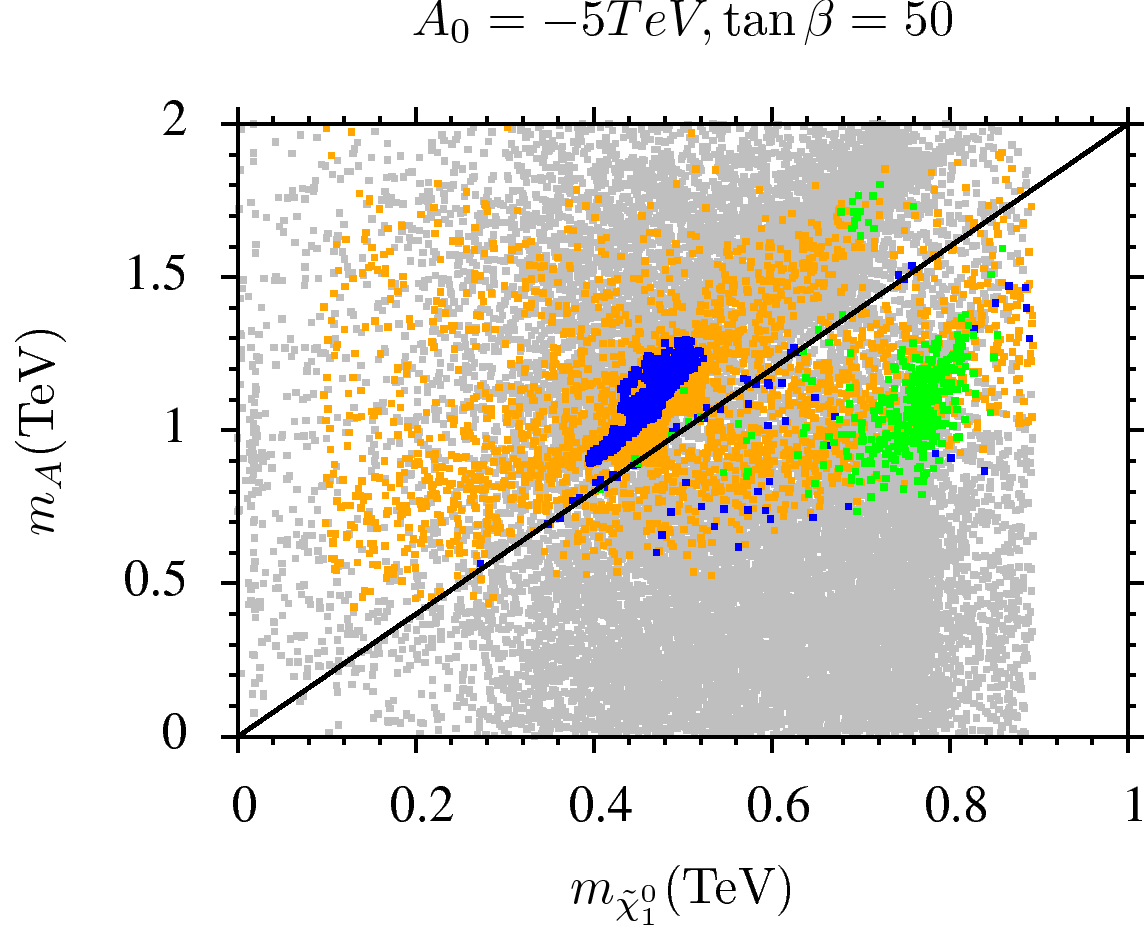}
}
\caption{
Color coding same as in Fig.~\ref{An3TeVtanb30spectrum},
for $A_0=-5$ TeV.
}
\label{An5TeVtan50spectrum}
\end{figure}

\begin{figure}
\centering
\subfiguretopcaptrue

\subfigure{
\includegraphics[width=7cm]{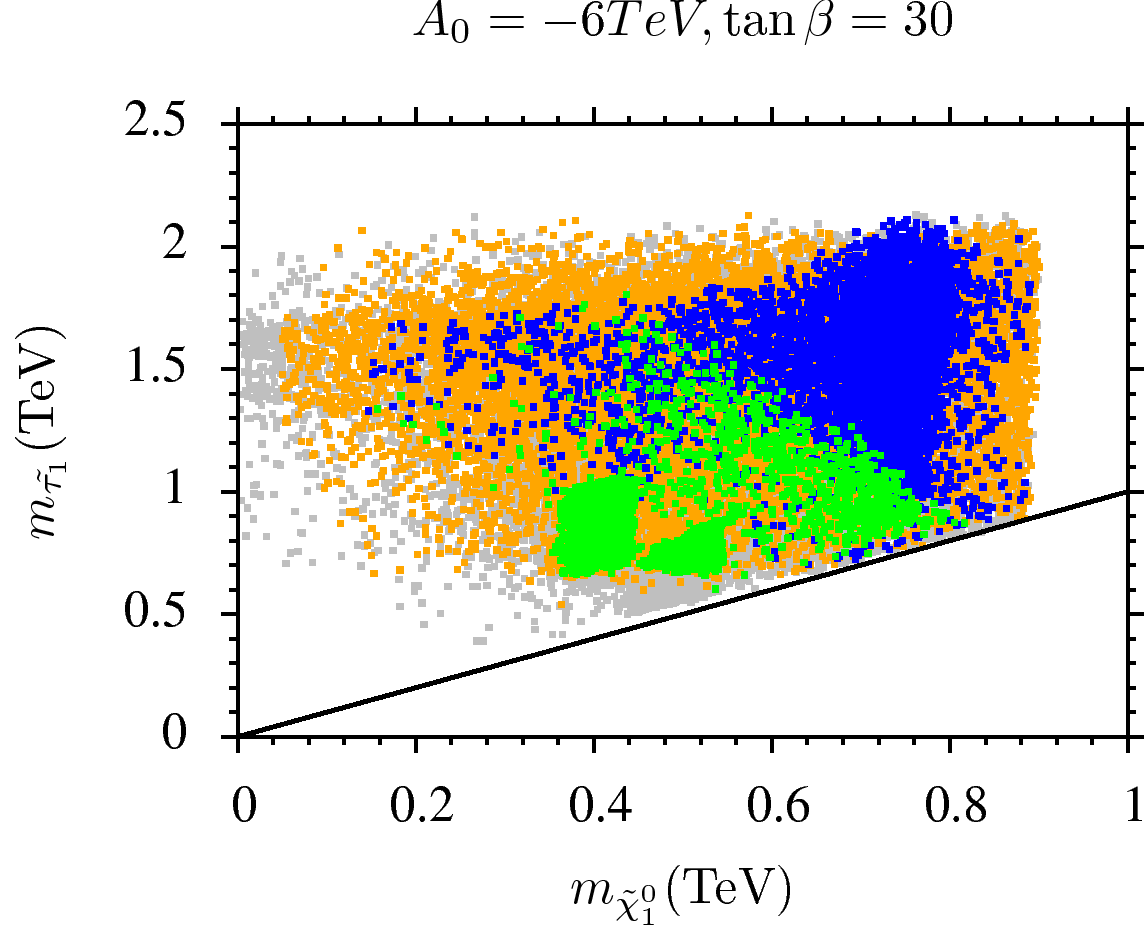}
}
\subfigure{
\includegraphics[width=7cm]{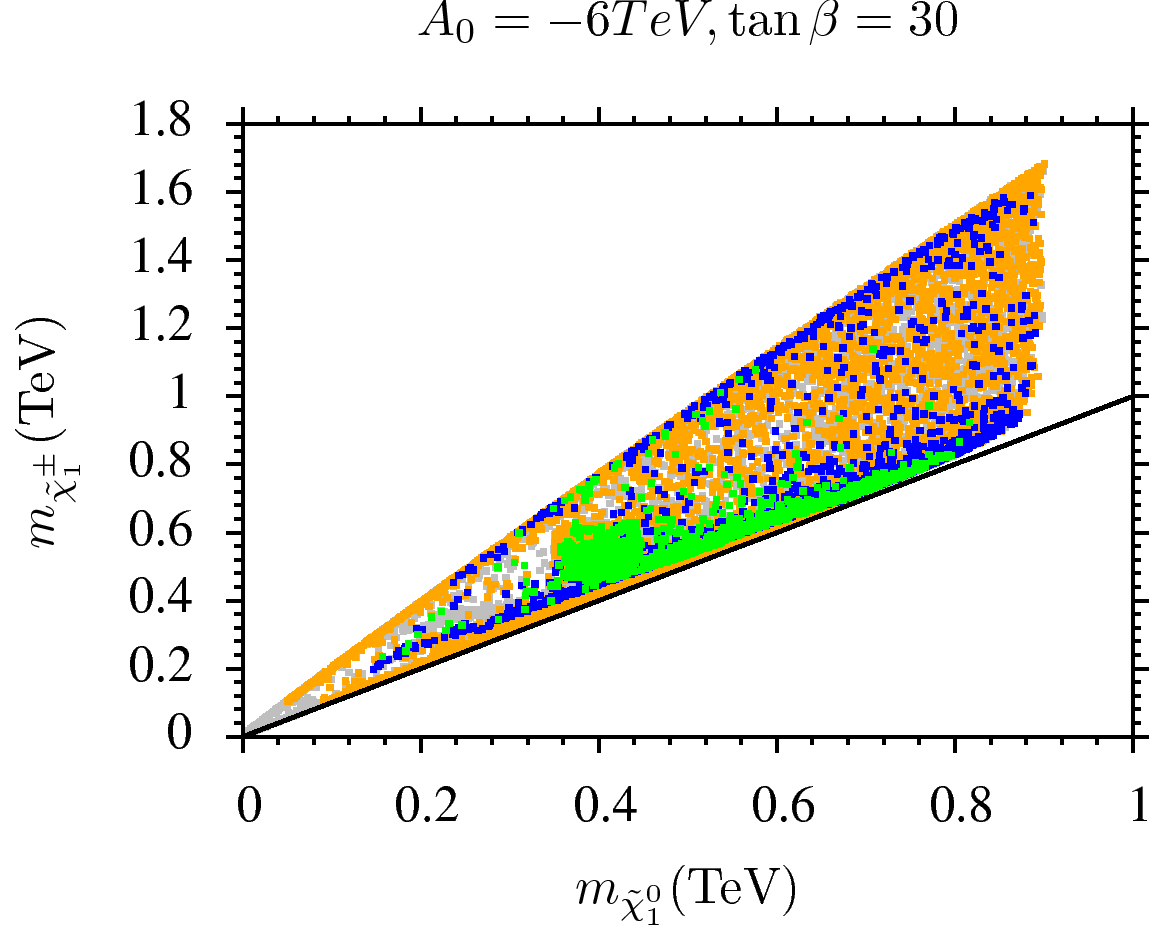}
}\\
\subfigure{
\includegraphics[width=7cm]{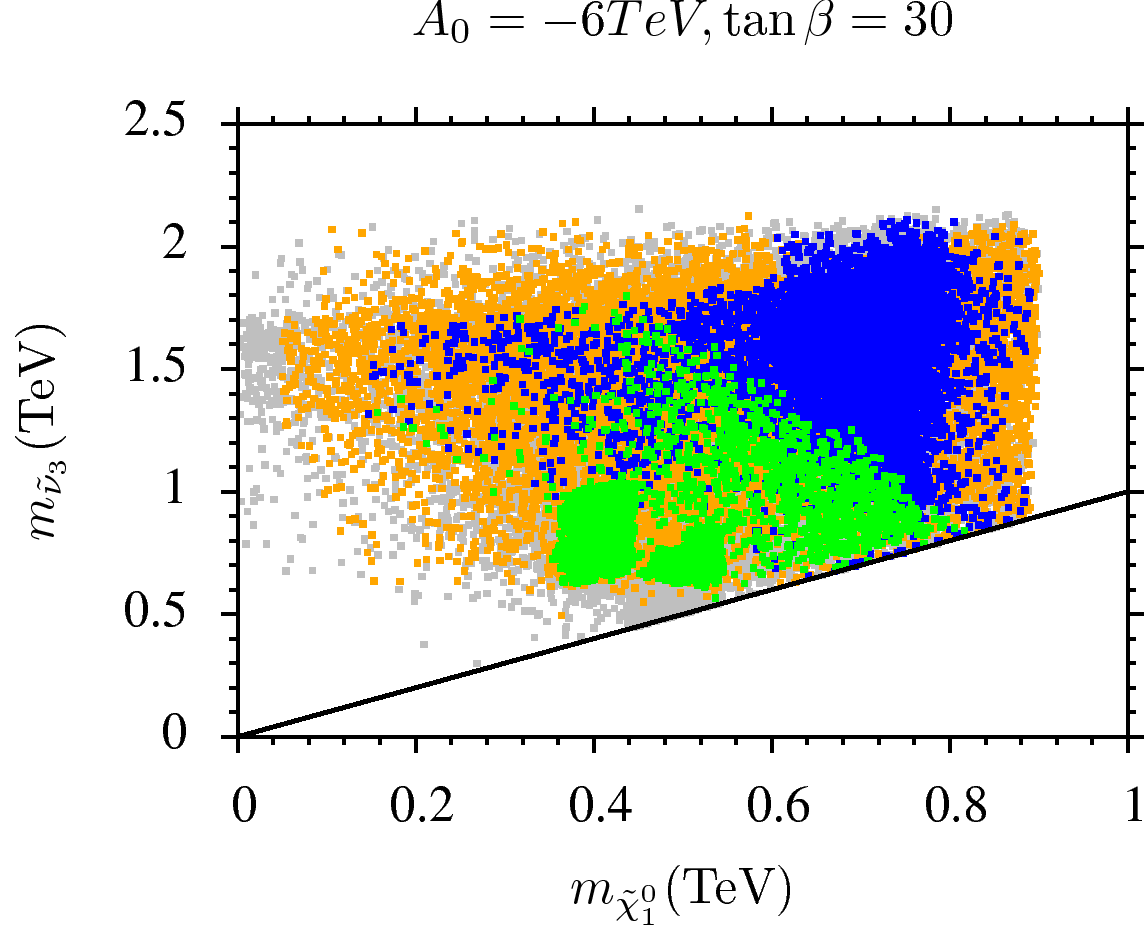}
}
\subfigure{
\includegraphics[width=7cm]{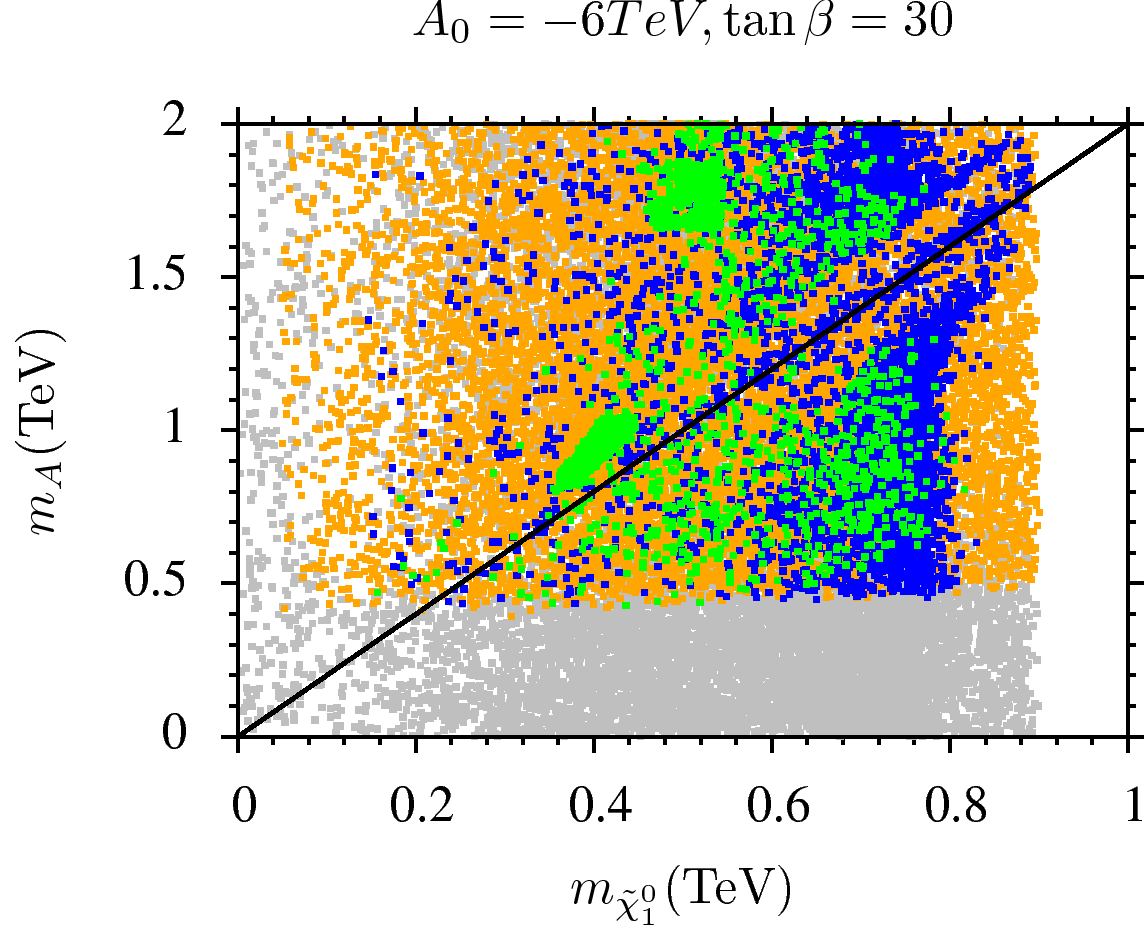}
}
\caption{
Color coding same as in Fig.~\ref{An3TeVtanb30spectrum},  
for $A_0=-6$ TeV. 
}
\label{An6TeVtanb30spectrum}
\end{figure}

The results for $A_0=-4$ TeV are depicted in Fig.~\ref{An4TeVtanb30spectrum}
 with the same color coding as in Fig.~\ref{An3TeVtanb30funda}. 
Here we note from the top-right panel that the neutralino LSP 
 is nearly-degenerate in mass with the lighter chargino for most of 
 the green points. 
In the top-left and bottom-left panels, we see that the stau and sneutrino 
 are nearly-degenerate with the neutralino LSP. 
Further investigation shows that 
 for most of these points, the sneutrino is closer in mass to the 
 neutralino LSP than the stau, 
 and the correct neutralino LSP relic density 
 is realized through the sneutrino-neutralino 
 coannihilation process 
 (the sneutrino-neutralino coannihilation region). 
However, we also found the stau-neutralino 
 coannihilation region for a limited set of points.
In the bottom-right panel, 
 some of the green points lie close to the solid line 
 corresponding to the relation $2 m_{\tilde {\chi}^0_1}=m_A$, 
 where the desired neutralino relic abundance is achieved 
 by efficient  annihilation processes through the A-resonance 
 (A-resonance region). 
The results for values of $A_0=-4$, $-5$ and $-6$ TeV 
 and $\tan \beta =30$ and $50$ are shown 
 in Figs.~\ref{An4TeVtanb30spectrum}-\ref{An6TeVtanb30spectrum}. 
We see that they are more or less similar 
 to those in Fig.~\ref{An3TeVtanb30spectrum}.

\begin{figure}
\centering
\subfiguretopcaptrue

\subfigure{
\includegraphics[width=7cm]{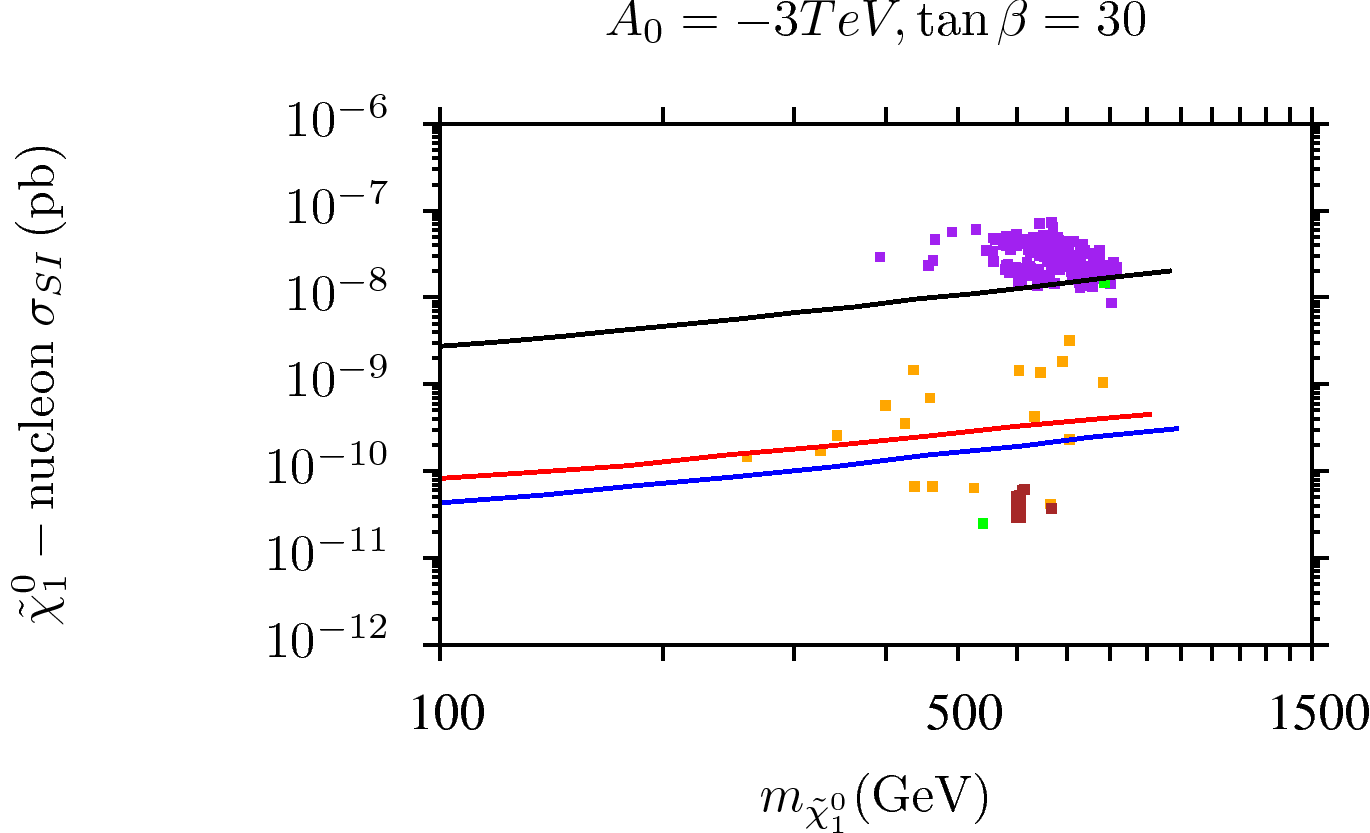}
}
\subfigure{
\includegraphics[width=7cm]{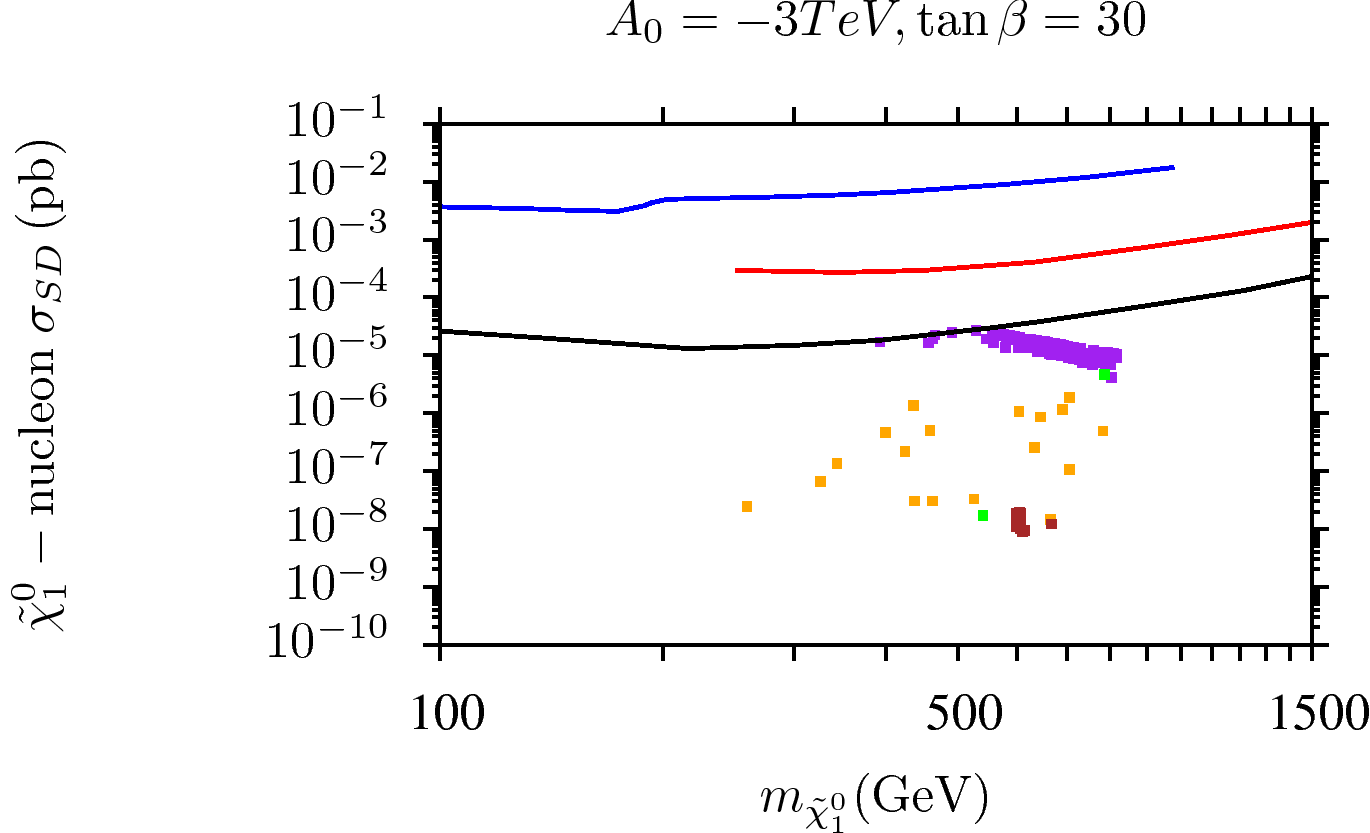}
}
\caption{
The SI and SD cross sections of neutralino elastic  
 scattering with nucleon, 
 along with the current and future bounds of 
 direct and indirect dark matter detection experiments. 
All points satisfy the requirements of REWSB, 
 neutralino LSP, mass bounds, B-physics bounds, 
 the  WMAP9 (5$\sigma$) bounds, Higgs boson mass and $\Delta a_{\mu}$ bound. 
Purple, green, orange and brown points respectively, represent the results 
 in the bino-Higgsino mixed dark matter region,
 the sneutrino-neutralino coannihilation region,
 the A-resonance and the stau-neutralino coannihilation region.
In the left panel, the black line represents the current  
 upper bound set by XENON 100, 
 while the blue (red) line represents future reach 
 of XENON 1T (SuperCDMS). 
In the right panel, the current upper bounds set  
 by Super-K (blue line) and IceCube DeepCore (red line)
 are shown. 
Future IceCube DeepCore bound is depicted by the black line. 
}
\label{An3TeVXection}
\end{figure}

For the green points which satisfy all constraints, 
 we also calculate the spin-independent (SI) and spin-dependent (SD)  
 elastic scattering cross sections of the dark matter neutralino  
 with a nucleon, and compare our results  with the current 
 and future bounds of direct dark matter detection 
 experiments. 
In Fig.~\ref{An3TeVXection},  
 we show the cross sections for the case 
 with $A_0=-3$ TeV and $\tan \beta=30$. 
The points all satisfy the requirements of REWSB,
 neutralino LSP, mass bounds, 
 B-physics bounds, the WMAP9 bound, 
 Higgs boson mass and $\Delta a_{\mu}$ bounds. 
The purple, green, orange and brown points represent the results, 
 respectively, in the bino-Higgsino mixed dark matter region, 
 the sneutrino-neutralino coannihilation region, 
 the A-resonance and the stau-neutralino coannihilation region.
In the left panel, the black line represents 
 the current upper bound set by XENON 100 ~\cite{Xenon100}, 
 while the blue (red) line represents the future reach 
 of XENON 1T~\cite{Xenon1T} (SuperCDMS~\cite{SuperCDMS}) experiment. 
In the right panel, 
 the current upper bounds from Super-K~\cite{SuperK} (blue line) 
 and the IceCube DeepCore ~\cite{IceCube} (red line), and the  
 future reach of IceCube DeepCore experiments (black line) are shown. 
We note that almost all of the purple points 
 (the bino-Higgsino mixed dark matter region) 
 have already been excluded by the current XENON 100 results.

\begin{figure}
\centering
\subfiguretopcaptrue

\subfigure{
\includegraphics[width=7cm]{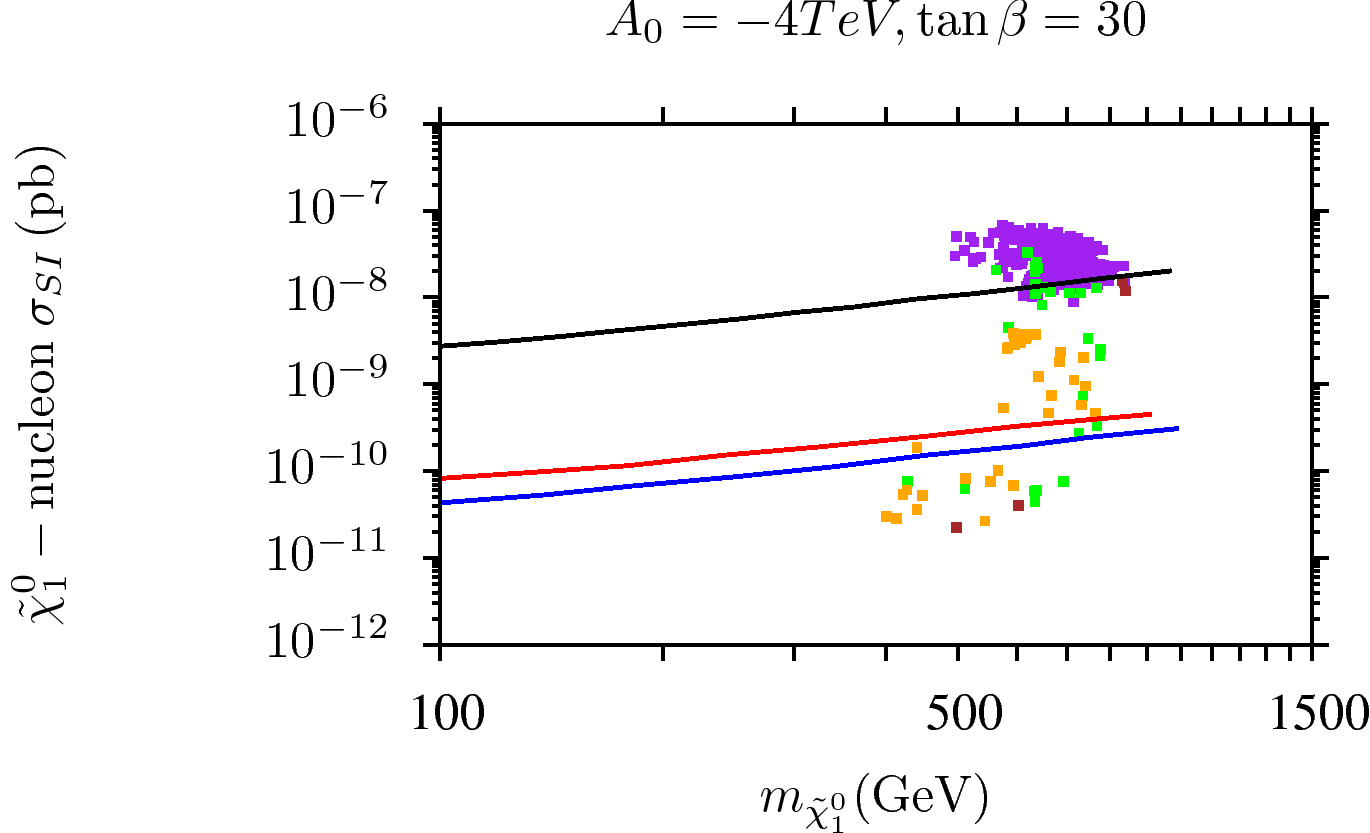}
}
\subfigure{
\includegraphics[width=7cm]{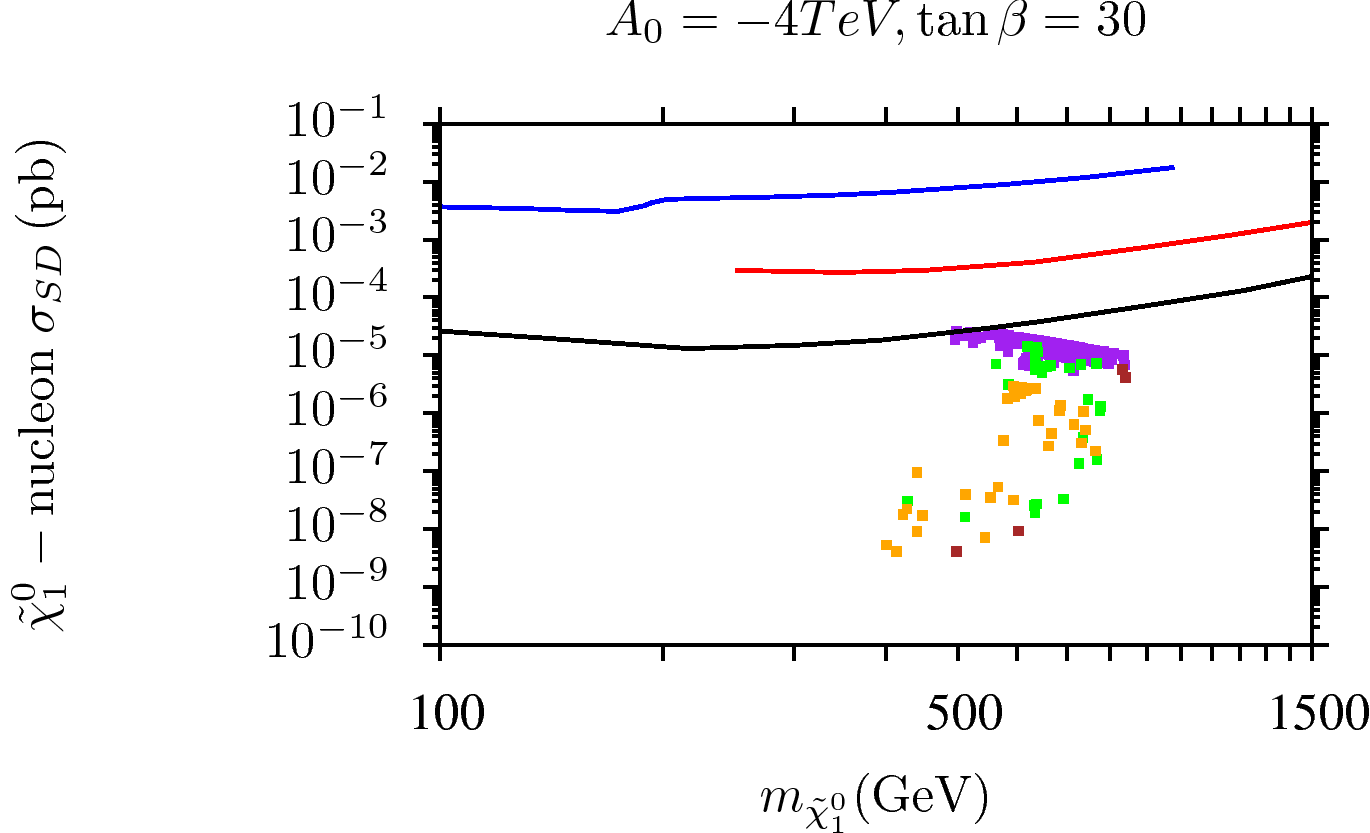}
}
\caption{
Color coding same as in Fig.~\ref{An3TeVXection},
for $A_0=-4$ TeV and $\tan\beta=30$.
}

\label{An4TeVXection}
\end{figure}

\begin{figure}
\centering
\subfiguretopcaptrue

\subfigure{
\includegraphics[width=7cm]{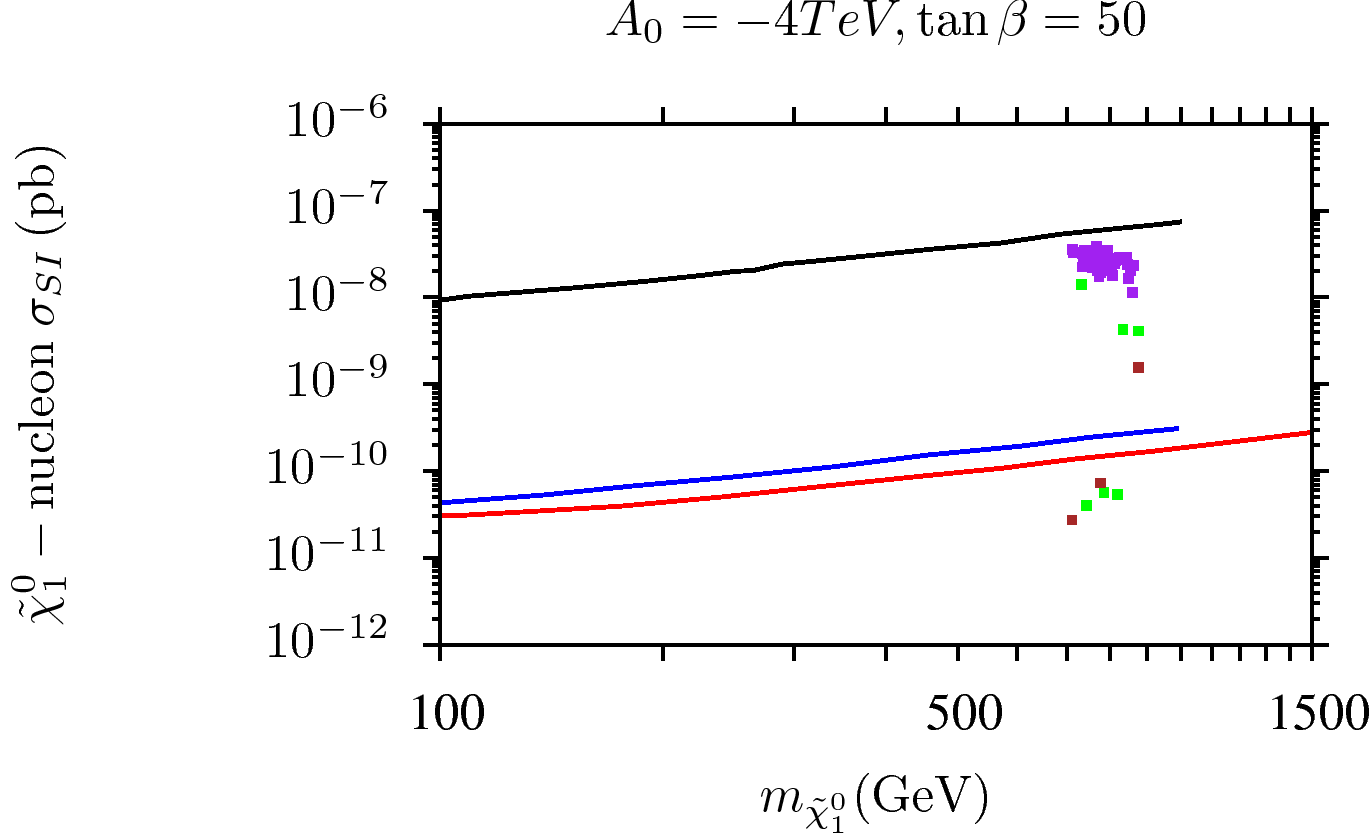}
}
\subfigure{
\includegraphics[width=7cm]{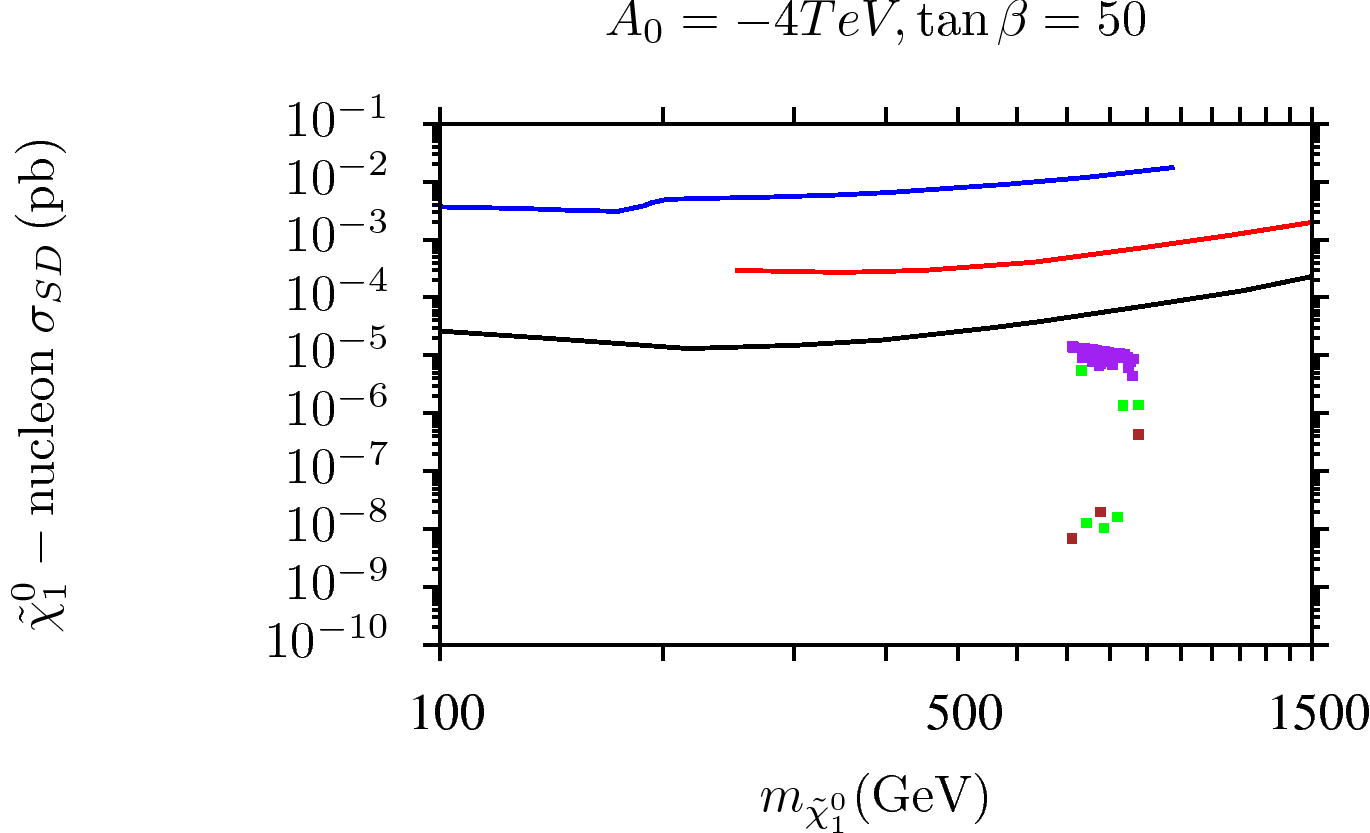}
}
\caption{
Color coding same as in Fig.~\ref{An3TeVXection},
for $A_0=-4$ TeV and $\tan\beta=50$.
}

\label{An4TeVtanb50Xection}
\end{figure}

\begin{figure}
\centering
\subfiguretopcaptrue

\subfigure{
\includegraphics[width=7cm]{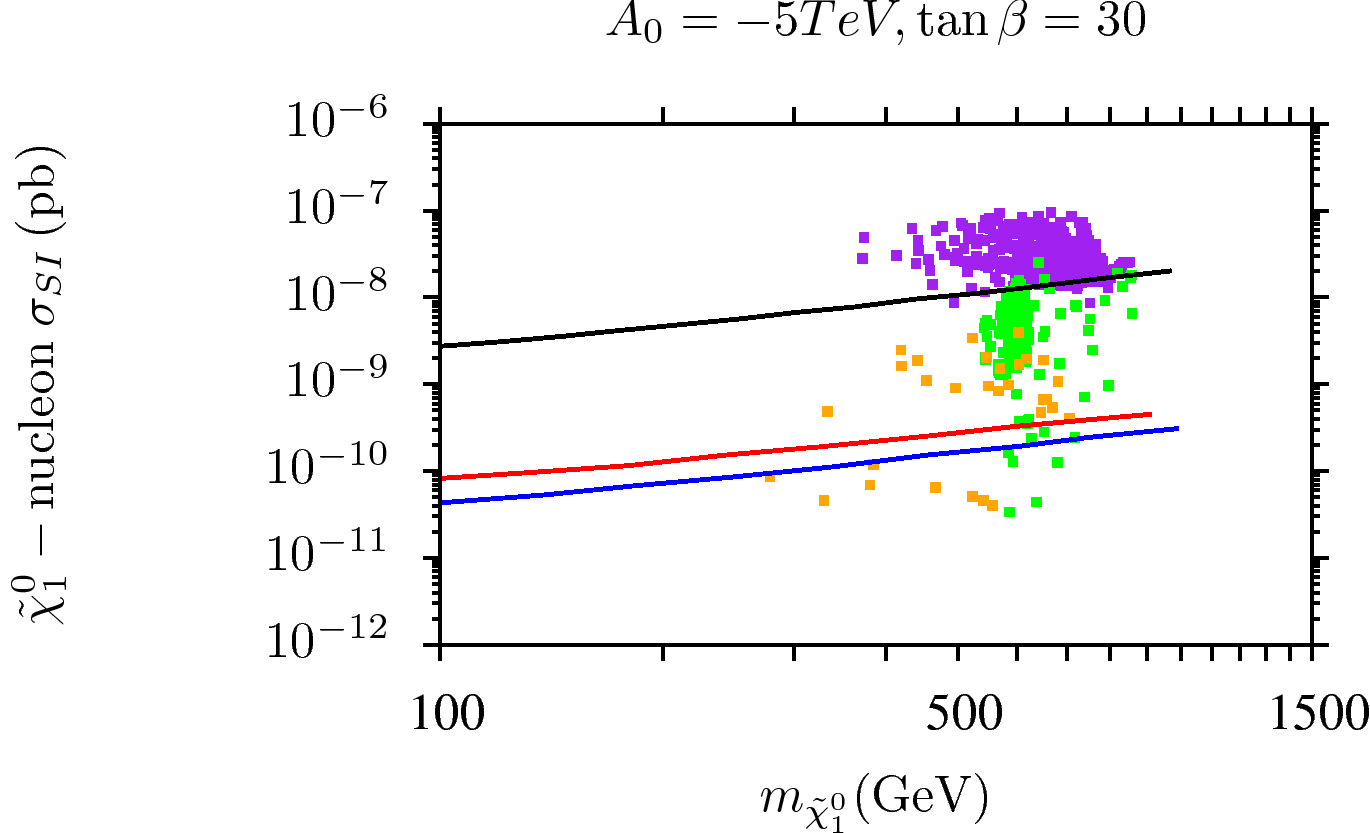}
}
\subfigure{
\includegraphics[width=7cm]{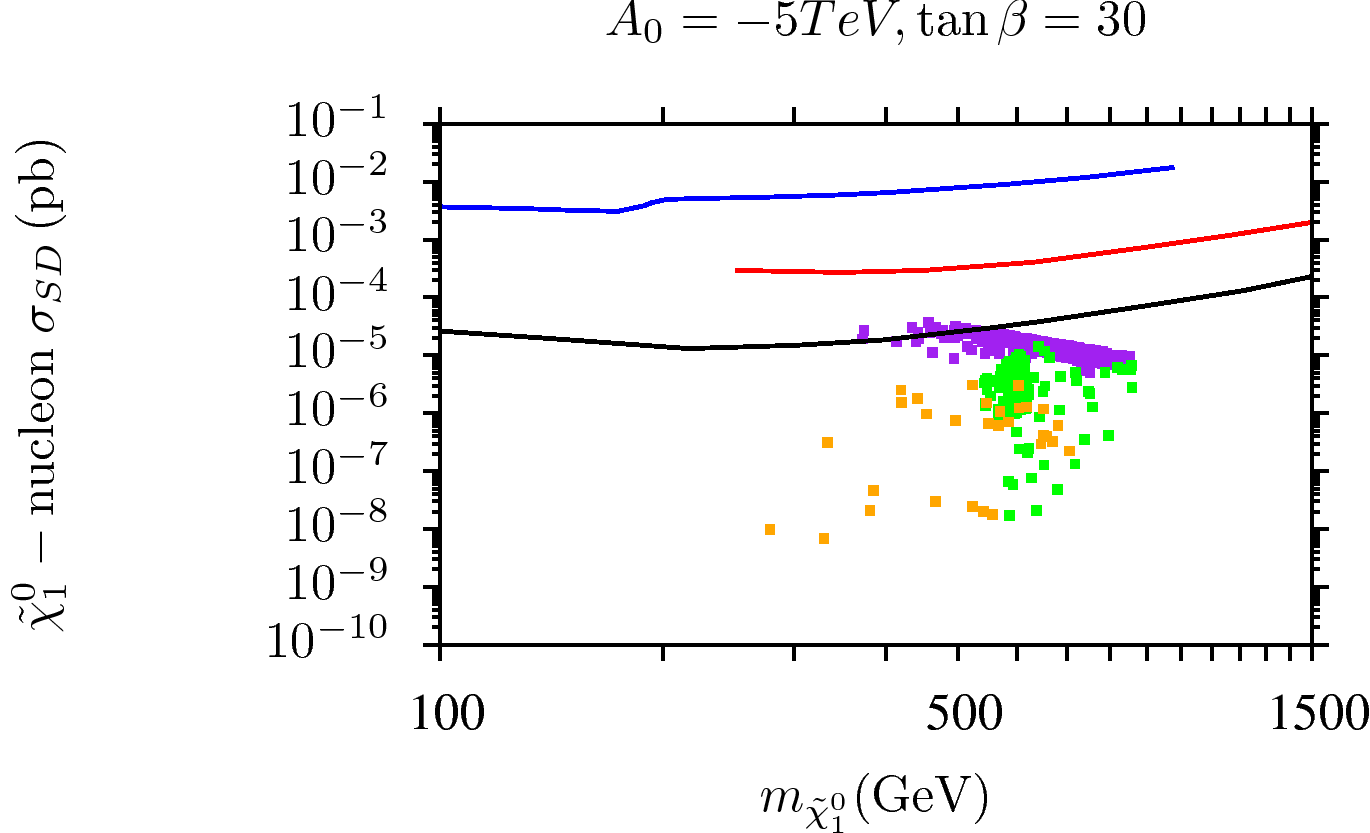}
}
\caption{
Color coding same as in Fig.~\ref{An3TeVXection},
for $A_0=-5$ TeV and $\tan\beta=30$.
}
\label{An5TeVXection}
\end{figure}

\begin{figure}
\centering
\subfiguretopcaptrue

\subfigure{
\includegraphics[width=7cm]{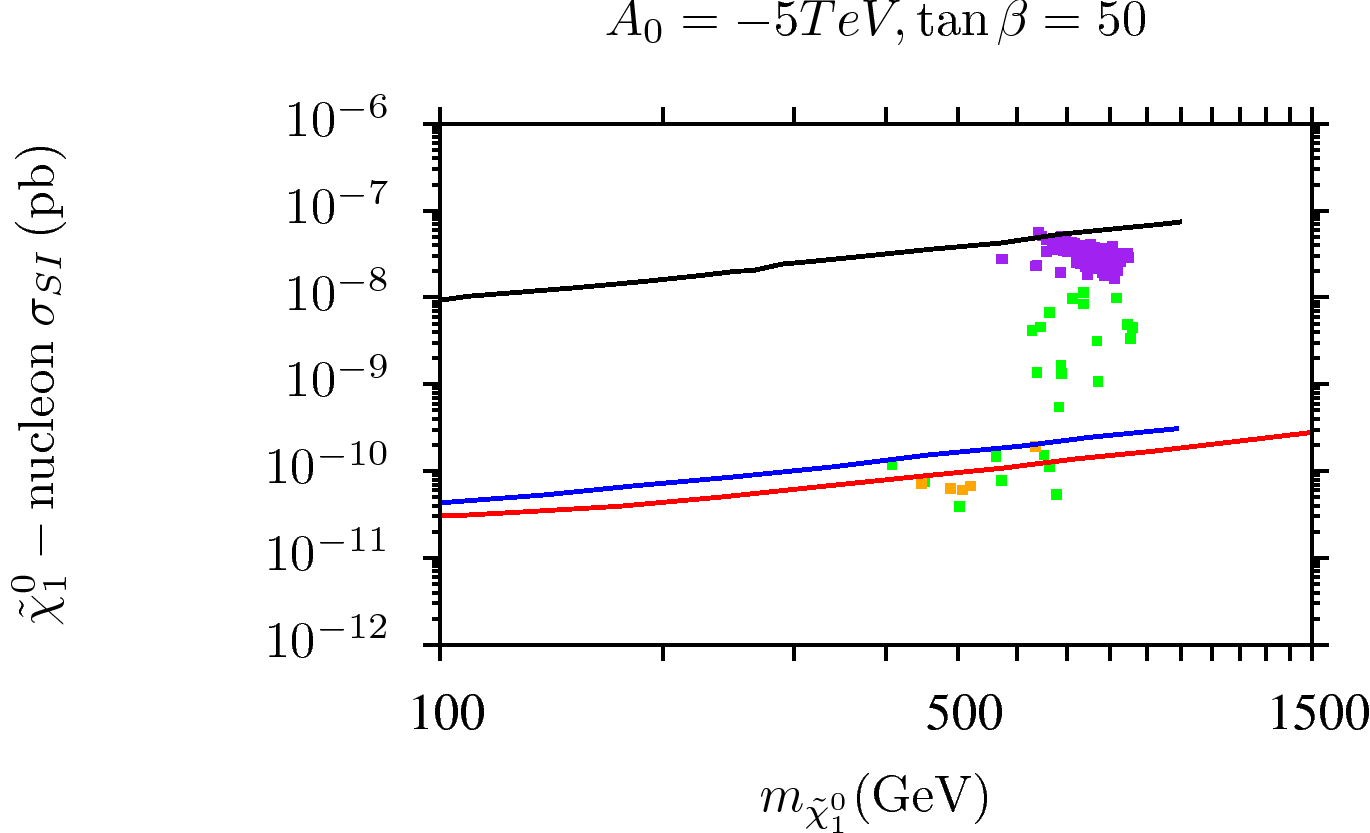}
}
\subfigure{
\includegraphics[width=7cm]{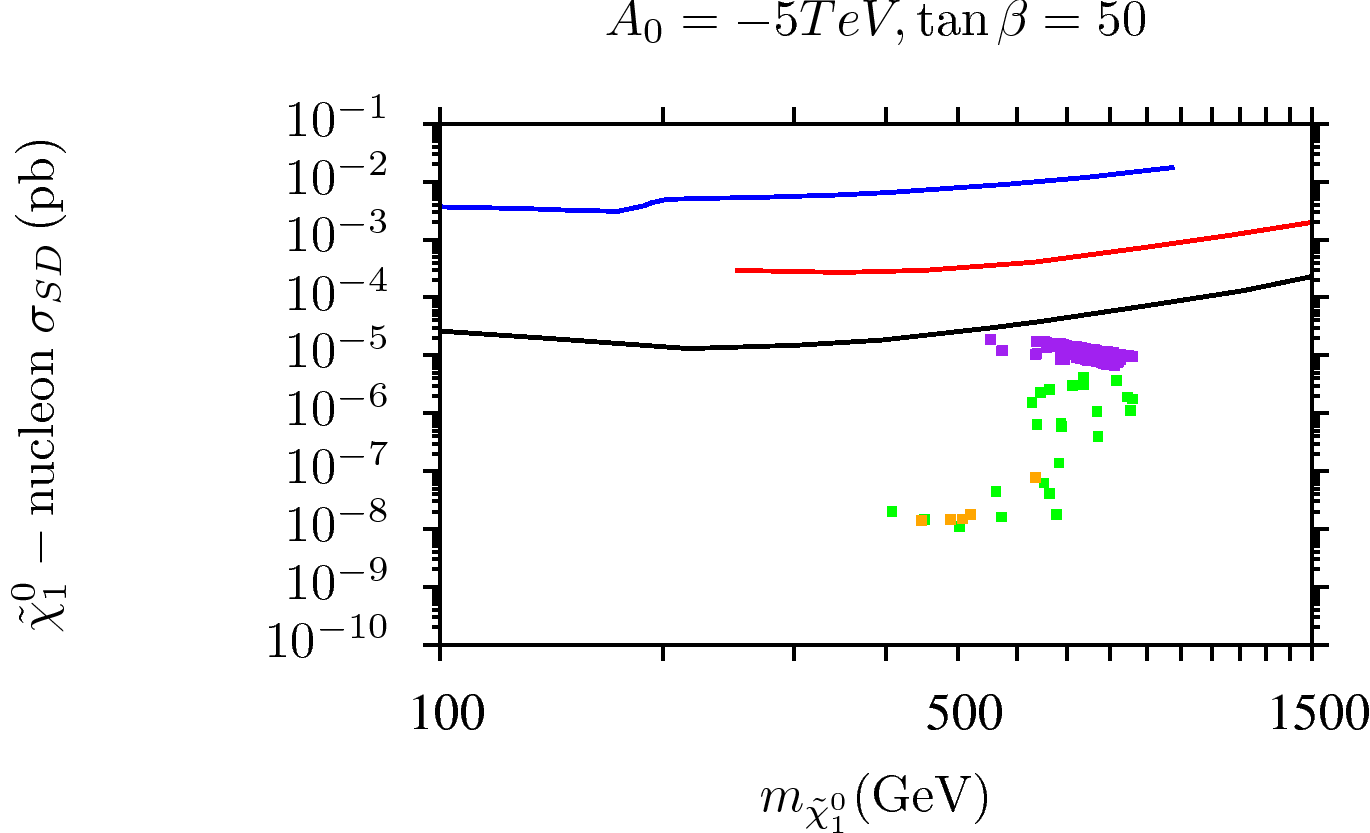}
}
\caption{
Color coding same as in Fig.~\ref{An3TeVXection},
for $A_0=-5$ TeV and $\tan\beta=50$.
}
\label{An5TeVtanb50Xection}
\end{figure}

\begin{figure}
\centering
\subfiguretopcaptrue

\subfigure{
\includegraphics[width=7cm]{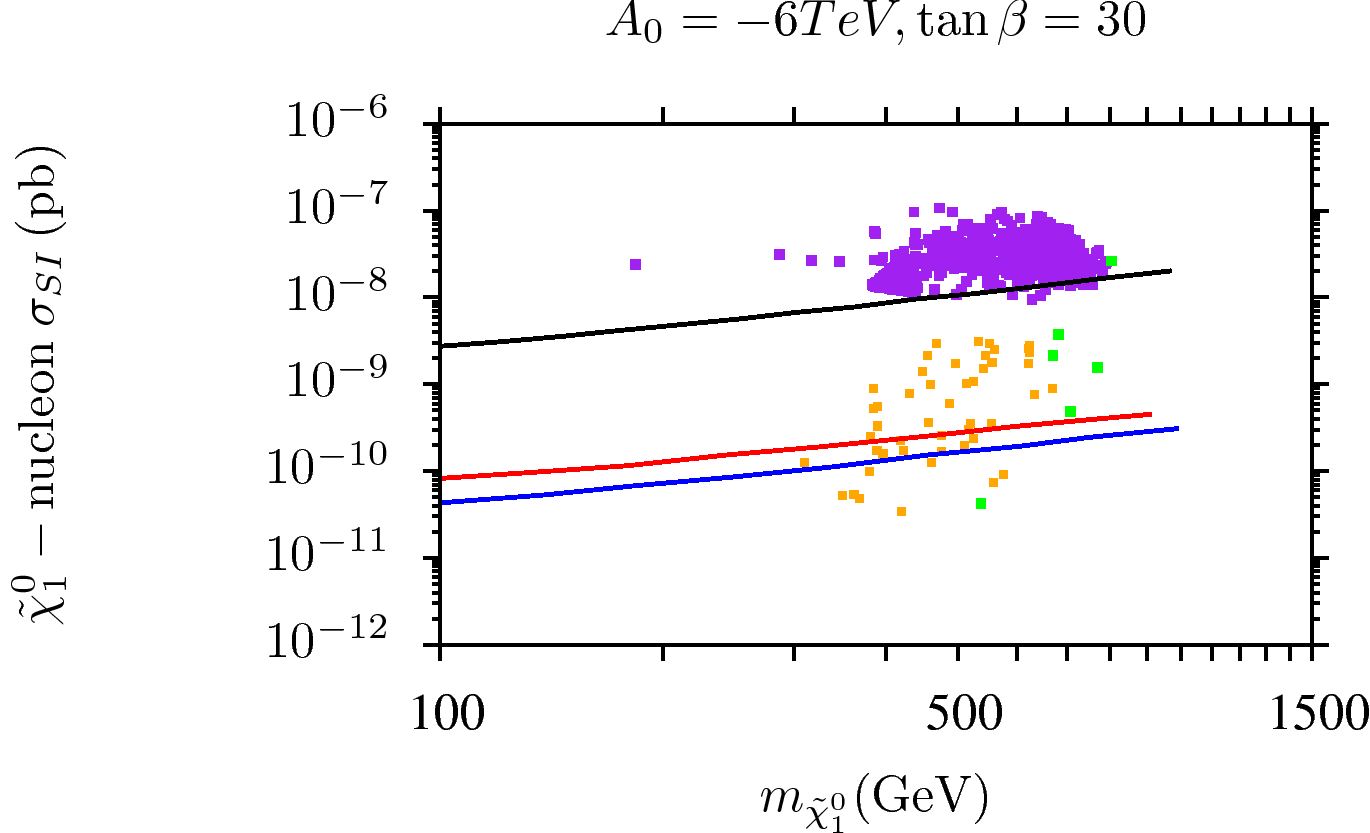}
}
\subfigure{
\includegraphics[width=7cm]{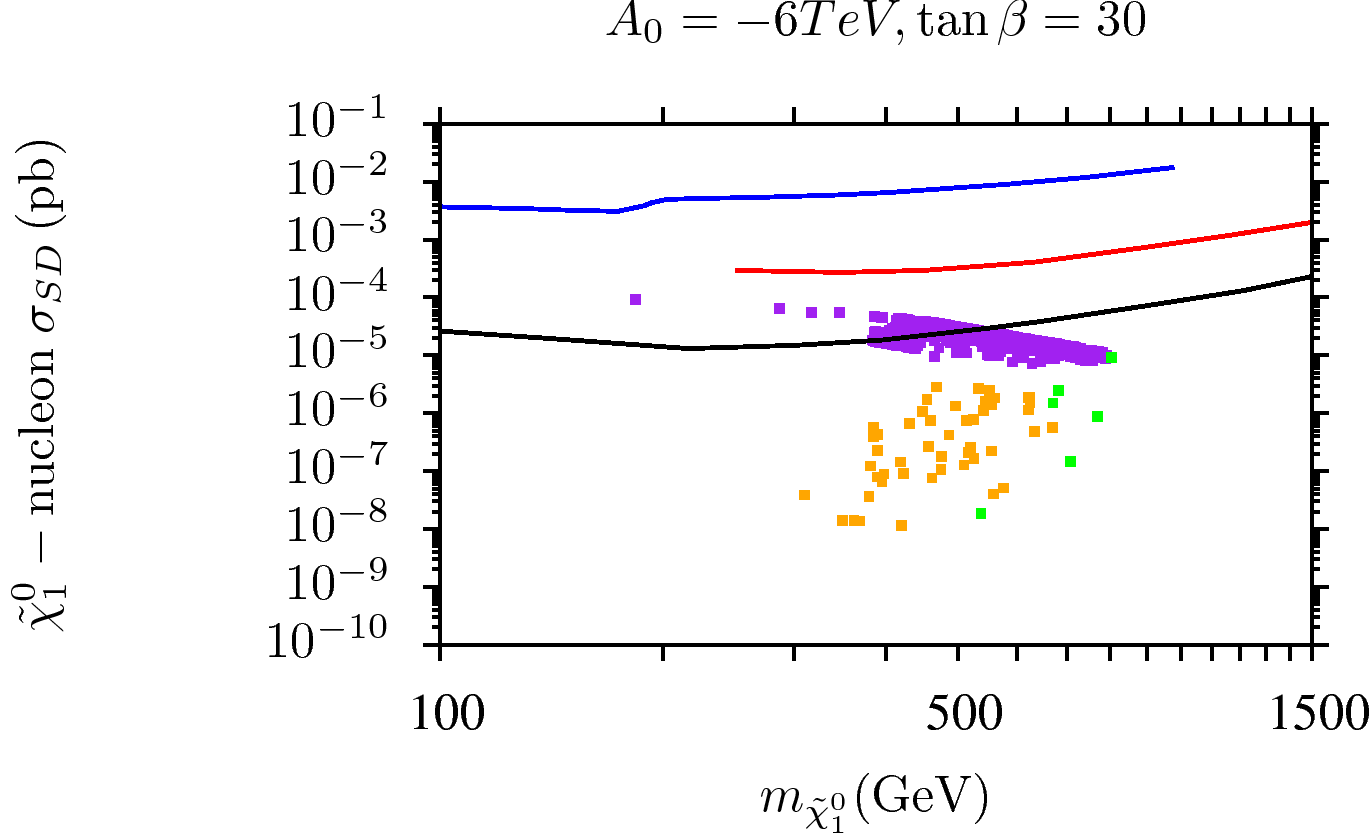}
}

\caption{
Color coding same as in Fig.~\ref{An3TeVXection}, 
for $A_0=-6$ TeV and $\tan\beta=30$.
}
\label{An6TeVXection}
\end{figure}

The cross sections for $A_0$ values of $-4$, $-5$ and $-6$ TeV 
 and $\tan \beta =30$ and $50$ are shown 
 in Figs.~\ref{An4TeVXection}-\ref{An6TeVXection}, respectively. 
Color coding is the same as in Fig.~\ref{An3TeVXection}. 
Interestingly, the resultant SI cross sections are large enough to lie 
within the reach of the future direct dark matter detection experiments 
 (most of the purple points have already been excluded.).

\begin{figure}
\centering
\subfiguretopcaptrue

\subfigure{
\includegraphics[width=7cm]{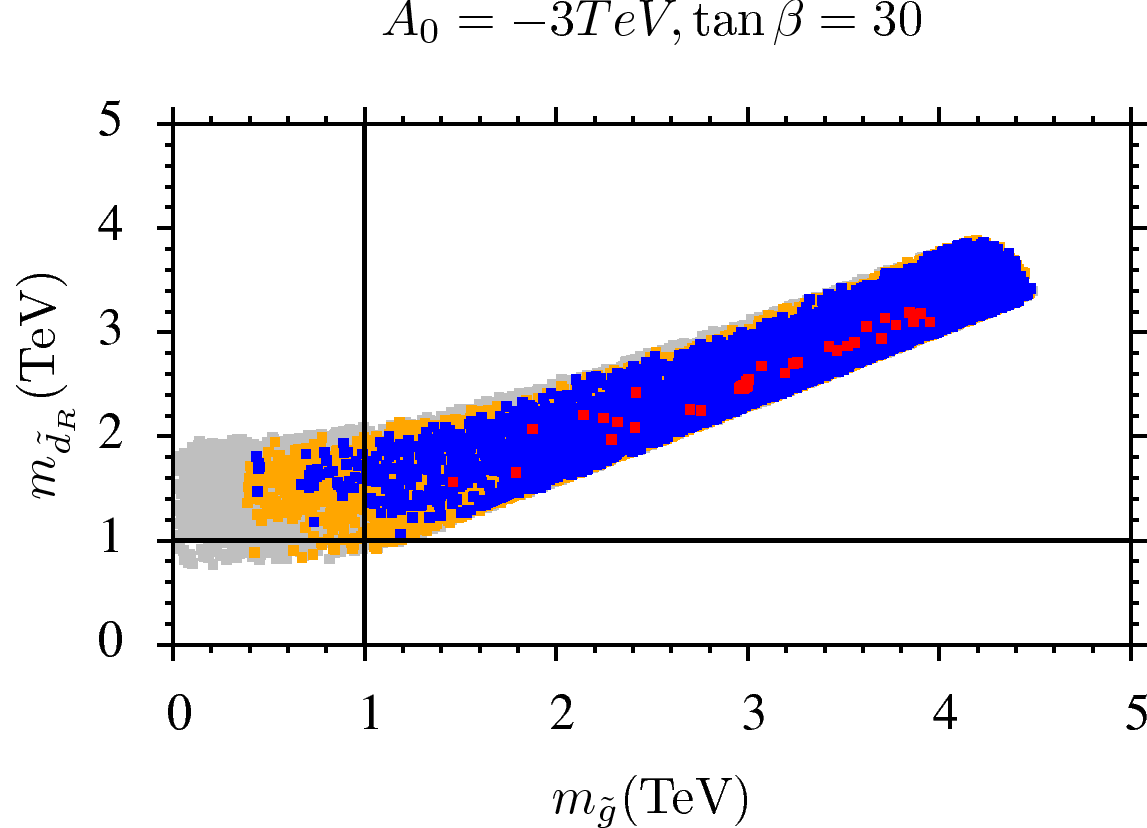}
}
\subfigure{
\includegraphics[width=7cm]{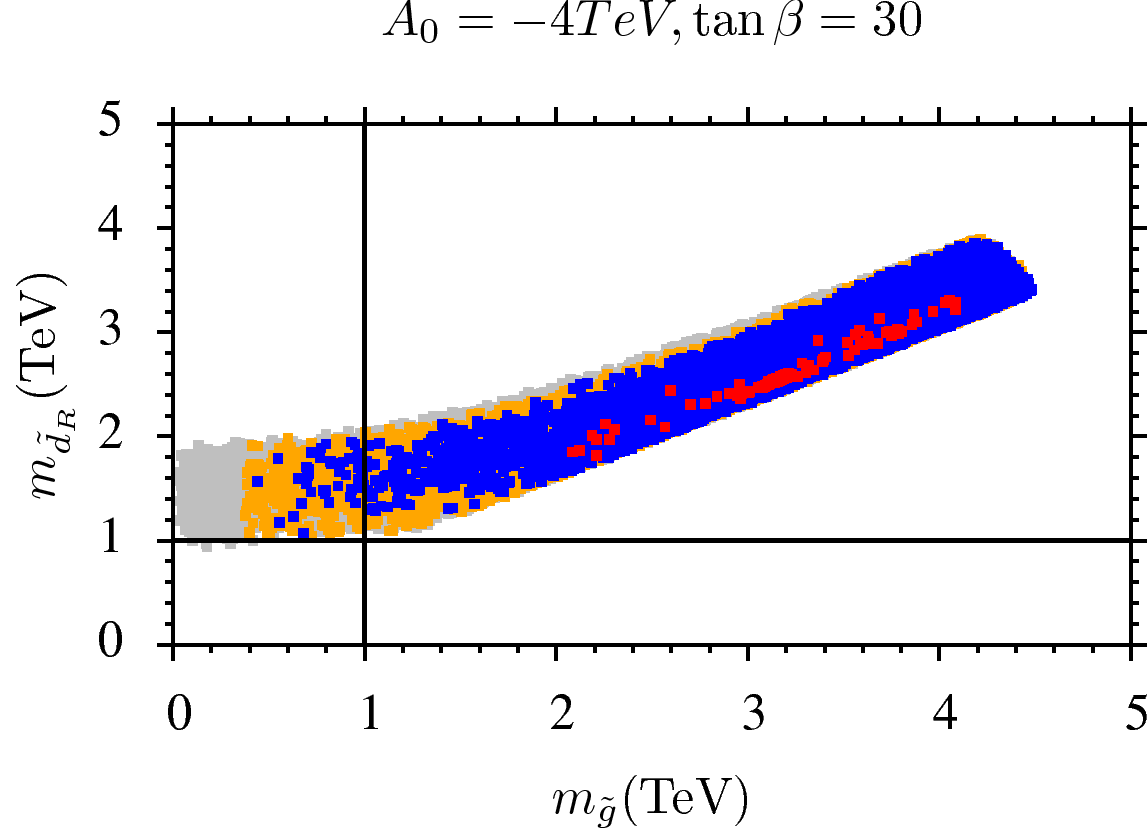}
}
\subfigure{
\includegraphics[width=7cm]{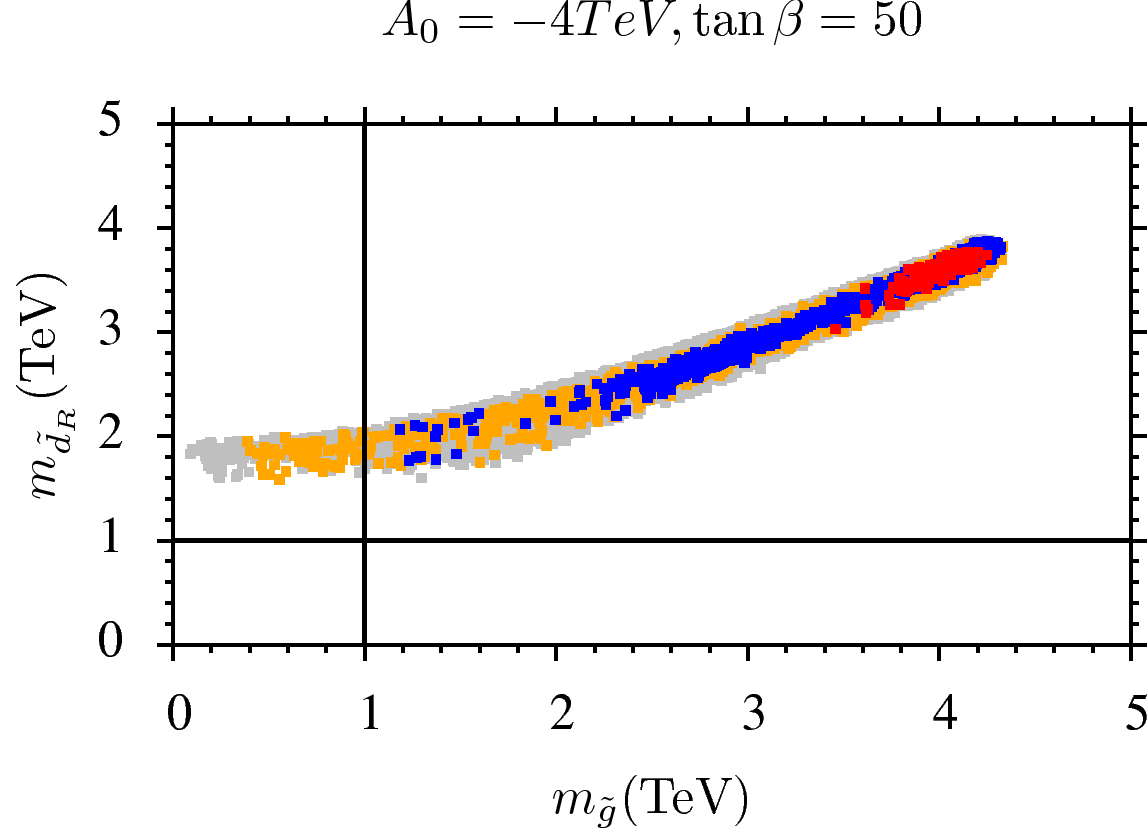}
}
\caption{
Plots in ($m_{\tilde g}$, $m_{\tilde d_R}$)-plane. 
Gray points satisfy the requirements of REWSB and neutralino LSP.
Orange points satisfy mass bounds and B-physics bounds.
Blue points are subset of the orange points and
 represent solutions satisfying the WMAP9 (5$\sigma$) bounds,
 $124 \; {\rm GeV} \leq m_h \leq 127$ GeV,
 $m_{\tilde g}$, $m_{\tilde d_R} \gtrsim 1$ TeV.
Red points are a subset of the blue points and satisfy
 the 3$\sigma$ bounds on $\Delta a_{\mu}$ and current 
 upper bounds set by XENON 100.}

\label{GluDr_1}
\end{figure}

\begin{figure}
\centering
\subfiguretopcaptrue

\subfigure{
\includegraphics[width=7cm]{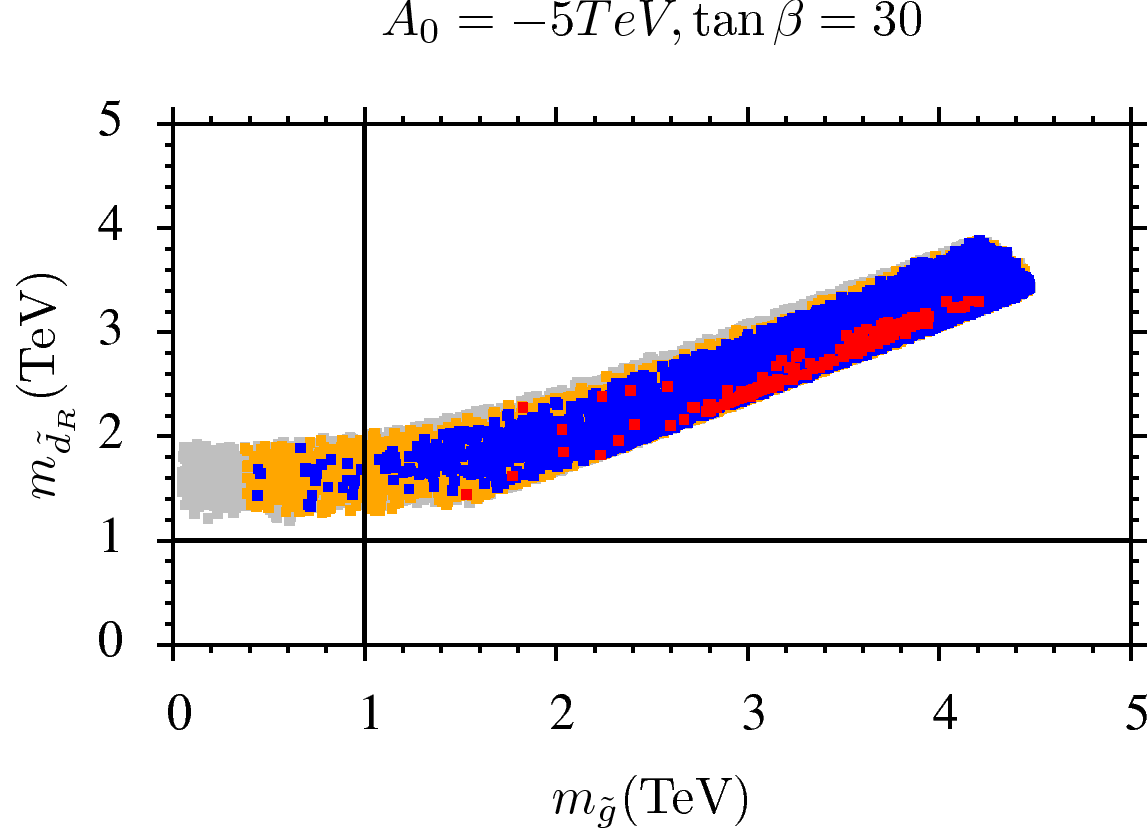}
}
\subfigure{
\includegraphics[width=7cm]{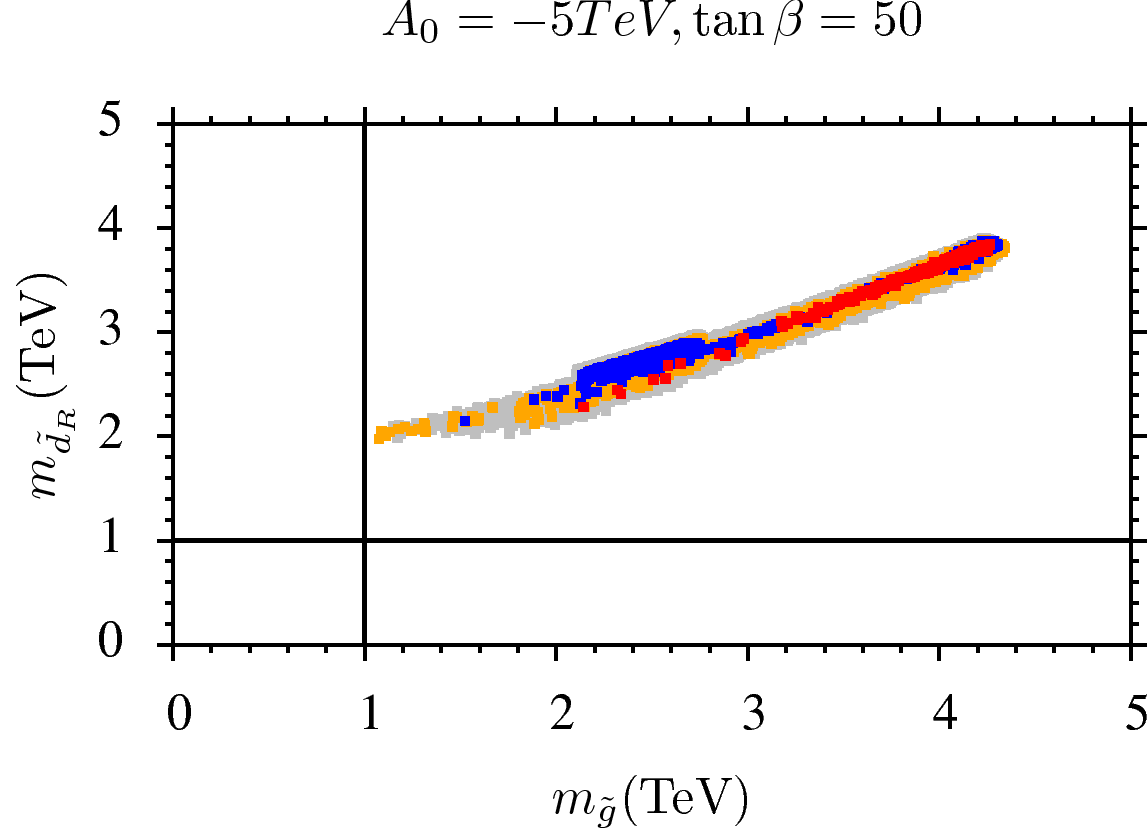}
}\\
\subfigure{
\includegraphics[width=7cm]{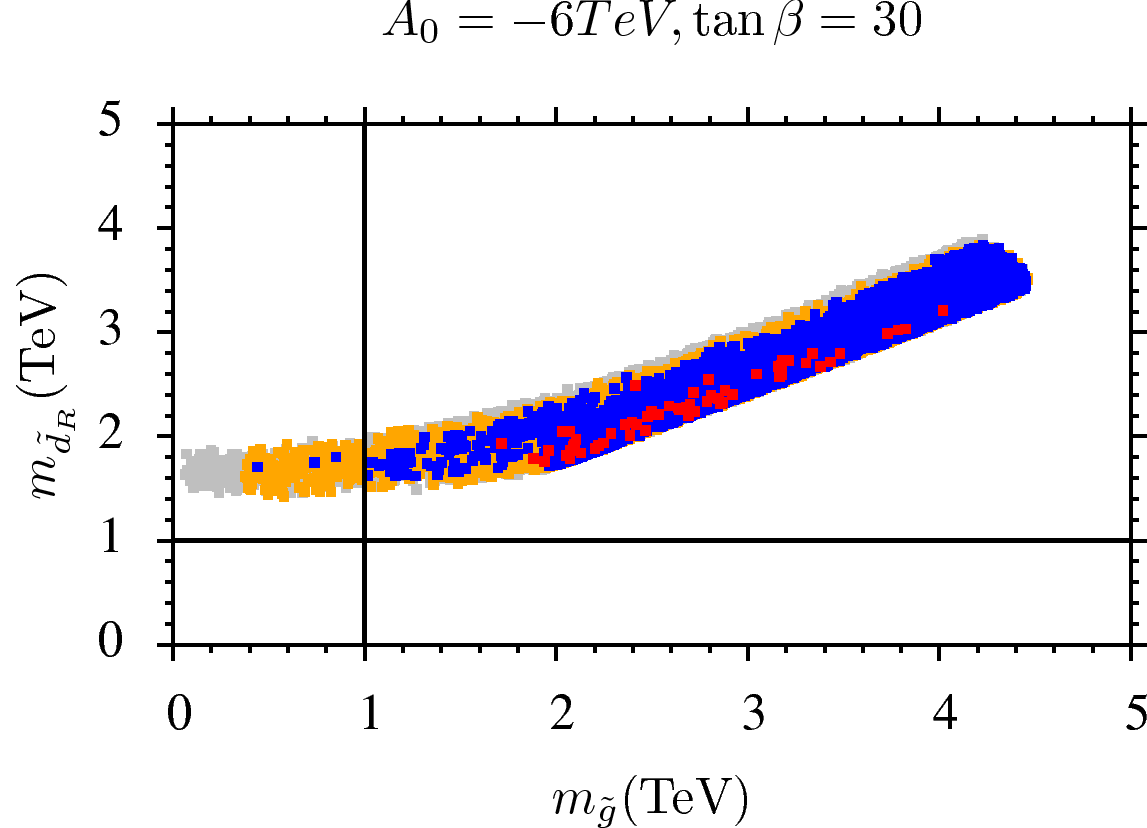}
}
\caption{Color coding same as in Fig.~\ref{GluDr_1}
 .}

\label{GluDr_2}
\end{figure}

In Fig.~\ref{GluDr_1} we present our results 
 in the ($m_{\tilde g}$, $m_{\tilde d_R}$)-plane. 
Gray points satisfy the requirements of REWSB and neutralino LSP.
Orange points satisfy the mass and B-physics bounds.
Blue points are a subset of the orange points and
 represent solutions satisfying the WMAP9 (5$\sigma$) bounds,
 $124 \; {\rm GeV} \leq m_h \leq 127$ GeV,
 $m_{\tilde g}$, $m_{\tilde d_R} \gtrsim 1$ TeV.
Red points are subset of the blue points and satisfy
 the 3$\sigma$ bounds on $\Delta a_{\mu}$, and 
the  current upper bounds set by XENON 100. 
It has been shown in~\cite{Baer:2012vr} that 
 for $m_{\tilde q} \simeq m_{\tilde g}$, 
 the LHC14 with 100 fb$^{-1}$ of data will probe  
 $m_{\tilde g}$ up to $3$ TeV. 
Therefore, some of our results can be tested
 in the near future at the LHC.
Similar results can be seen for other $A_0$ and $\tan\beta$ values in Fig~\ref{GluDr_2}.

Finally, we select four benchmark points 
 from the results with different values of $A_0$ and $\tan \beta$ 
 and show the mass spectrum in Table~\ref{tab:bm}. 
The first, second, third and fourth columns are taken, 
 respectively, from 
 the stau-neutralino coannihilation region, 
 the bino-Higgsino mixed dark matter region, 
 the sneutrino-neutralino coannihilation region, 
 and the A-resonance region.
These points satisfy all phenomenological 
 constraints and will be tested in the near future 
 at the LHC experiments with collider energy upgrade 
 and by the future direct dark matter detection 
 experiments.

\section{Conclusion}
\label{sec:conclude}
The long-sought after Higgs boson of the Standard Model 
 was discovered at the LHC 
 by the ATLAS and CMS experiments  
 through a variety of decay channels. 
Because the SM-like Higgs boson mass is predicted 
 in the MSSM as a function of stop masses, 
 the observed mass of 125-126 GeV has a great impact 
 on the MSSM. In particular, it requires a relatively large stop mass.
In the context of the CMSSM-like parameterization, 
the sfermion masses are found to be as large as $1-10$ TeV, 
 and the discovery of (colored) sparticles at the LHC  
 could be very challenging. 
On the other hand, the muon $g-2$ has been  precisely measured, 
 and the observed data shows more than 3$\sigma$ deviation 
 from the Standard Model prediction. 
This indicates that new physics below the TeV scale might exist, 
such that the associated new particles give rise to additional contributions  
 to the muon $g-2$. 
The MSSM is a prime candidate for this new physics,
 however relatively light sparticles (neutralino, charginos and sleptons) 
 with masses less than 1 TeV are required 
 to account for the muon $g-2$ data.

We have attempted to reconcile 
  these two results in a SU(5) inspired extension  
  of the CMSSM parameterization for the soft SUSY breaking terms. 
In our setup, $m_{\bar 5}$ and $m_{10}$,   
 as well as soft  masses$^2$ of the Higgs doublets,
 are taken as free parameters. 
With this generalization of the CMSSM, 
 we have performed random parameter scans 
 and examined a variety of phenomenological constraints 
 including the observed Higgs boson mass of around 125 GeV, 
 the muon $g-2$, the WMAP9 results for neutralino dark matter 
 abundance, and the observations of rare B-meson decays. 
We have identified parameter regions which satisfy all the constraints. 
In particular, a soft mass$^2$ splitting between 
 ${\bar 5}$ and ${10}$ matter fields, motivated 
 by SU(5) GUT plays a crucial 
 role in accommodating the Higgs boson mass and 
 the muon $g-2$ data. 
We find $m_{\bar 5} < m_{10}$ 
 in the allowed parameter regions. 
The correct neutralino relic abundance is achieved 
 by four different parameter sets, 
namely the sneutrino-neutralino coannihilation region, 
the stau-neutralino coannihilation region,
 the neutralino-Higgsino mixed region 
 and the A-resonance region.
For these regions, we have calculated 
 the spin independent/dependent cross sections 
 for the elastic scattering of neutralino dark matter 
 with a nucleon. 
The resultant cross sections for these
scenarios are shown to be within the reach of future experiments 
 for direct dark matter searches. 
We have highlighted four benchmark points 
 which show characteristic particle mass spectra
 satisfying all the constraints (including the WMAP9 and Planck2013 bounds at 5$\sigma$) and can be tested in the near future at the LHC  
 and by the dark matter direct detection experiments.

\section*{Acknowledgement}
We would like to thank Ilia Gogoladze and Azar Mustafayev for useful discussions.
This work is supported in part by the DOE Grant 
 No. DE-FG02-10ER41714 (N.O.) and 
No. DE-FG02-91ER40626 (S.R and Q.S).
This work used the Extreme Science and Engi-
neering Discovery Environment (XSEDE), which is supported by the National Science
Foundation grant number OCI-1053575




%

%

%

%
%

%
\clearpage



\begin{table}[h!]
\label{table1}
\centering
\begin{tabular}{lcccc}
\hline
\hline
                 & Point 1  &Point 2 & Point 3 & Point 4  \\
\hline
$m_{10}$        & 1046  & 2382  & 2988    & 3788   \\
$m_{\bar 5}$    & 492.8 & 542.7 & 1675    & 1131    \\
$M_{1/2} $      & 1381  & 1471   & 921.2   & 912.2     \\
$A_0$           & -3000 & -4000 & -5000   & -6000   \\
$\tan\beta$     & 30    & 30     &  50    & 30  \\
$\mu$           & 1556 & 697.8   &  1500  & 586.4   \\
$m_{A}$         & 1340  & 1127  &  806.3 & 882.4   \\
\hline
$m_h$          & 124  & 124.5   & 124  &  125 \\
$m_H$          & 1349 & 1134    & 811   & 888  \\
$m_{H^{\pm}}$  & 1351 & 1138    & 817   & 892 \\

\hline
$m_{\tilde{g}}$ & 2970  & 3200  & 2146   & 2145         \\
$m_{\tilde{\chi}^0_{1,2}}$
                & 601,1126  &632, 711   &407, 776    &398, 582  \\
$m_{\tilde{\chi}^0_{3,4}}$
               &1560, 1567  &716, 1218 &1497, 1501  &597, 788   \\

$m_{\tilde{\chi}^{\pm}_{1,2}}$
               &1130,1567   &719, 1201  &776, 1504  &591, 776   \\
\hline $m_{ \tilde{u}_{L,R}}$
               &2882,2809   &3701, 3677  &3483, 3485    &4161, 4229   \\
$m_{\tilde{t}_{1,2}}$
               &1606,2306   &1766, 2779   &1415, 2294    &1404, 2839  \\
\hline $m_{ \tilde{d}_{L,R}}$
               &2884,2595   &3702, 2717  &3484, 2386   &4162, 1931     \\
$m_{\tilde{b}_{1,2}}$
               &2281,2351   &2286, 2781 &1272, 2282    &921, 2848    \\
\hline
$m_{\tilde{\nu}_{1}}$
              & 1076  & 1206  &1808   &   1409      \\
$m_{\tilde{\nu}_{3}}$ & 880  & 822  & 501   & 690       \\
\hline
\hline
$m_{ \tilde{e}_{L,R}}$
              &1084,1077   &1217, 2331 &1810, 2959   &1419, 3682 \\
$m_{\tilde{\tau}_{1,2}}$
             & 617,897   &839, 1963 &507, 1606   &721, 3241 \\
\hline

$\sigma_{SI}({\rm pb})$
             & 3.99$\times 10^{-11}$  & 1.17$\times 10^{-8}$ & 1.08$\times 10^{-10}$ & 2.45$\times 10^{-9}$ \\

$\sigma_{SD}({\rm pb})$
             & 2.10$\times 10^{-8}$  & 1.25$\times 10^{-5}$ & 2.81$\times 10^{-8}$ & 4.11$\times 10^{-6}$ \\

$\Omega_{CDM}h^{2}$
            & 0.11  & 0.11    &0.11 & 0.11 \\
$\Delta a_{\mu}$ & $2.76\times 10^{-10}$ &  $3.40 \times 10^{-10}$   & $3.16 \times 10^{-10}$    & $3.86 \times 10^{-10}$  \\
\hline
\hline
\end{tabular}
\caption{
Benchmark points for $A_0=-3,-4,-5,-6$ TeV and $\tan \beta =30,50$. 
Sparticle and Higgs masses are in GeV.
These benchmark points satisfy all phenomenological constraints. 
Point 1, 2, 3 and 4 are chosen, respectively, from 
 the stau-neutralino coannihilation region, 
 the bino-Higgsino mixed dark matter region, 
 the sneutrino-neutralino coannihilation region, 
 and the A-resonance region. 
\label{tab:bm}}
\end{table}
%
%

\end{document}